\documentclass[10pt,epsfig]{article}

\usepackage{enumerate}
\usepackage{float}
\usepackage{subfig}
\usepackage{color}
\usepackage[table]{xcolor}
\usepackage{slashed}
\usepackage{multirow}
\usepackage{cite}

\let\counterwithin\relax
\usepackage{chngcntr}
\usepackage{amssymb, amsmath,mathrsfs}

\usepackage{graphics}
\usepackage{graphicx}
\usepackage{epsf}
\usepackage{epsfig}
\usepackage{float}
\usepackage{makecell}
\usepackage{multirow}
\usepackage{color}
\usepackage{xcolor}
\usepackage{simplewick}

\usepackage[utf8]{inputenc}
\usepackage{amsmath}
\usepackage{amssymb}
\usepackage{subfig}
\usepackage[normalem]{ulem}

\usepackage{mathtools}
\usepackage{amsmath}
\usepackage{diagbox}

\newcommand\undermat[2]{
	\makebox[0.5pt][l]{$\smash{\underbrace{\phantom{%
					\begin{matrix}#2\end{matrix}}}_{ \let\scriptstyle\textstyle\text{\large $#1$}}}$}#2}
\newcommand\overmat[2]{
	\makebox[-1pt][l]{$\smash{\overbrace{\phantom{%
					\begin{matrix}#2\end{matrix}}}^{ \let\scriptstyle\textstyle\text{\large $#1$}}}$}#2}    
\usepackage{tikz}
\usepackage[vcentermath]{youngtab}
\usepackage{slashed} 
\usepackage[font=small]{caption} 
\usepackage{times} 

\usepackage{bm}       
\usepackage{bbm} 

\usepackage{relsize}  

\usepackage{autobreak}  

\usepackage{makeidx} 
\usepackage{bbding} 
\usepackage{listings} 
\usepackage{ytableau}

\usepackage[colorlinks=true,
linkcolor=blue,
urlcolor=red,
citecolor=red]{hyperref}

\long\def\rpl#1!!#2!!{\textcolor{red}{#1} \textcolor{blue}{#2}}

\usepackage[top=1in, left=0.95in, bottom=1.1in, right=0.95in]{geometry}
\def\baselinestretch{1.27}
\usepackage[toc]{appendix}

\newcommand{\beq}{\begin {equation}}  
\newcommand{\eeq}{\end   {equation}} 
\newcommand{\bea}{\begin {eqnarray}} 
\newcommand{\eea}{\end   {eqnarray}}  
\newcommand{\baa}{\begin {array}   } 
\newcommand{\eaa}{\end   {array}   }     
\newcommand{\bit}{\begin {itemize} }
\newcommand{\eit}{\end   {itemize} }
\newcommand{\be }{\begin {equation}} 
\newcommand{\ee }{\end   {equation}}
\newcommand{\nn }{\nonumber        }

\newcommand{\mc}[1]{\mathcal{#1}}

\newcommand{\vev}[1]{ \left\langle {#1}  \right\rangle }

\newcommand{\eq}[1]{\begin{equation}\begin{split} #1 \end{split}\end{equation}}

\newcommand{\comment}[1]{}

\newcolumntype{M}[1]{>{\centering\arraybackslash}m{#1}}
\newcolumntype{N}{@{}m{0pt}@{}}

\allowdisplaybreaks
\begin{document}

\begin{center}

{\Large \textbf  {Operator Bases in Effective Field Theories with Sterile Neutrinos: $d \leq 9$}}\\[10mm]

Hao-Lin Li$^{a}$\footnote{lihaolin@itp.ac.cn}, Zhe Ren$^{a, b}$\footnote{renzhe@itp.ac.cn}, Ming-Lei Xiao$^{a}$\footnote{mingleix@itp.ac.cn}, Jiang-Hao Yu$^{a, b, c, d, e}$\footnote{jhyu@itp.ac.cn}, Yu-Hui Zheng$^{a, b}$\footnote{zhengyuhui@itp.ac.cn}\\[10mm]

\noindent 
$^a${\em \small CAS Key Laboratory of Theoretical Physics, Institute of Theoretical Physics, Chinese Academy of Sciences,    \\ Beijing 100190, P. R. China}  \\
$^b${\em \small School of Physical Sciences, University of Chinese Academy of Sciences,   Beijing 100049, P.R. China}   \\
$^c${\em \small Center for High Energy Physics, Peking University, Beijing 100871, China} \\
$^d${\em \small School of Fundamental Physics and Mathematical Sciences, Hangzhou Institute for Advanced Study, UCAS, Hangzhou 310024, China} \\
$^e${\em \small International Centre for Theoretical Physics Asia-Pacific, Beijing/Hangzhou, China}\\[10mm]

\date{\today}   
          
\end{center}

\begin{abstract} 

We obtain the complete and independent bases of effective operators at mass dimension 5, 6, 7, 8, 9 in both standard model effective field theory with light sterile right-handed neutrinos ($\nu$SMEFT) and low energy effective field theory with light sterile neutrinos ($\nu$LEFT). These theories provide systematical parametrizations on all possible Lorentz-invariant physical effects involving in the Majorana/Dirac neutrinos, with/without the lepton number violations. In the $\nu$SMEFT, we find that there are 2 (18), 29 (1614), 80 (4206), 323 (20400), 1358 (243944) independent operators with sterile neutrinos included at the dimension 5, 6, 7, 8, 9 for one (three) generation of fermions, while 24, 5223, 3966, 25425, 789426 independent operators in the $\nu$LEFT. 

\end{abstract}

\newpage

\setcounter{tocdepth}{4}
\setcounter{secnumdepth}{4}

\tableofcontents

\setcounter{footnote}{0}

\def\baselinestretch{1.5}
\counterwithin{equation}{section}

\newpage


\section{Introduction}
The Standard Model (SM) is an important achievement of human understanding of the micro-world, which has been examined at various high energy experiments with very high precision. Since that, more and more attention has been drawn to search for new physics beyond the SM but not find any new resonances yet. 
In this case, effective field theory (EFT) providing a systematical framework to parameterize various new physics and describing physical systems below the scale of the new physics, has witnessed great progress in recent years. 
The first and famous achievement of the EFT is the Fermi theory. Nowadays, with only the SM degree of freedom included, the standard model effective field theory (SMEFT)~\cite{Weinberg:1979sa,Buchmuller:1985jz,Grzadkowski:2010es,Lehman:2014jma,Li:2020gnx,Murphy:2020rsh,Li:2020xlh,Liao:2020jmn,Liao:2016hru} provides an EFT framework above the electroweak scale while all SM particles remain massless. The SMEFT contains all the SM fields to construct the effective Lagrangians and respects the SM gauge symmetry $SU(3)_C\times SU(2)_W\times U(1)_Y$.
After electroweak symmetry breaking and integrating out the massive gauge bosons $W^\pm$, $Z$, the top quark $t$ and the Higgs boson $h$, the low energy effective field theory (LEFT)~\cite{Jenkins:2017jig,Liao:2020zyx,Li:2020tsi,Murphy:2020cly} describes all possible physics below the electroweak scale, regarding to the gauge symmetry $SU(3)_C\times U(1)_{\rm EM}$.

The neutrino oscillation experiments have shown evidence that neutrinos are massive, yet the left-handed neutrinos in the SM do not acquire the Dirac mass terms via the Yukawa interactions like other fermions in the SM. One natural solution to this is to add the right-handed neutrinos to the SM and generate the Dirac mass terms after electroweak symmetry breaking, in which the massive neutrinos are Dirac-type and the lepton number is conserved. On the other hand, Majorana mass terms are naturally there since the right-handed neutrinos are the SM gauge singlet and thus sterile to the SM, which violate the lepton number, and the neutrinos are Majorana neutrinos. If the newly-added right-handed neutrinos are heavy, above the TeV scale, for example, then they should be integrated out above the electroweak scale, and the SMEFT and LEFT frameworks are still valid. However, if the right-handed neutrinos are light, such as the sub-GeV or KeV sterile neutrinos, which could be dark matter candidate, see Ref.~\cite{Adhikari:2016bei} for a review, so that they can not be integrated out at the electroweak scale or even at lower energy scale, the standard model effective field theory with right-handed neutrinos ($\nu$SMEFT) and low energy effective field theory with right-handed neutrinos ($\nu$LEFT) are needed to describe various new physics effects. The construction of effective operators involving the right-handed neutrinos in the $\nu$SMEFT has been considered at the dimension 5 to 7 in literature~\cite{delAguila:2008ir,Aparici:2009fh,Bhattacharya:2015vja,Liao:2016qyd}. 
In the $\nu$LEFT, the lepton-number-conserving subset of operators involving the right-handed neutrinos up to dimension 6 has been listed in Ref.~\cite{Chala:2020vqp}, while the lepton-number-violating subset in Ref.~\cite{Li:2020lba} up to dimension 6. 

With the light sterile neutrino included, $\nu$SMEFT operators would modify different kinds of physical processes and at the same time, give rise to new signatures compared to the SMEFT operators. It has been applied to investigate various sterile neutrino production processes at the LHC~\cite{Alcaide:2019pnf,deVries:2020qns}, and also appears in various exotic decay processes, such as the Higgs exotic decays~\cite{Butterworth:2019iff}, sterile neutrino decays~\cite{Duarte:2015iba,Ballett:2016opr}, leptonic radiative decays~\cite{Aparici:2009fh,Zhang:2021tsq}, coherent neutrino scattering experiments and beta decays~\cite{Li:2020lba,Li:2020wxi,Bischer:2019ttk,Han:2020pff}. In particular, in the neutrino-less double beta decay processes, the sterile neutrinos would induce new kinds of long-range neutrino potential~\cite{Dekens:2020ttz}. However, we expect that if the dimension-9 operators in the $\nu$SMEFT dominate, just like the dimensional-9 contributions in the SMEFT, it would induce new types of short-range potential, and thus give rise to additional contribution to both the short-range and long-range processes. Furthermore, in the proton decay processes, if the sterile neutrinos exist, it would induce new types of exotic decay processes as discussed in Ref.~\cite{Helo:2018bgb}. We expect the higher dimension of operators, the lower of the cutoff scale in the proton exotic decay processes. Therefore, the higher dimensional operators in the $\nu$SMEFT which contribute to the proton decay processes, would also be accessible to the collider searches in which the cutoff could be around the TeV scale. Overall, writing down the higher dimensional operators would be useful to investigate various processes involving in sterile neutrino, especially the baryon and lepton number violation processes, in which new physics effect might be reachable by the future colliders. 

To avoid over-counting or miscounting of operators, it's important to work with a complete and independent basis of operators. The traditional method to list the independent operator basis is to consider all possible operators, then to restrict them with equation of motion (EOM), Fierz identities, and integration by parts (IBP)  repeatedly, but the complexity of this method increases exponentially when it comes to higher dimensions.  Compared with the traditional method, our method to obtain the complete bases automatically generate the independent structures~\cite{Li:2020gnx, Li:2020xlh, Li:2020tsi}, without the need of EOM and IBP relations. Furthermore, since spin $\le \frac12$ massive particles do not affect the amplitude-operator correspondence which we used in the massless cases, the method can still be applied in both the $\nu$SMEFT and $\nu$LEFT.

In this work, we introduce a systematical method to list the operator bases of the $\nu$SMEFT and $\nu$LEFT for a given dimension, which has been applied to SMEFT and LEFT~\cite{Li:2020gnx, Li:2020xlh} to guarantee its correctness. Based on the amplitude-operator correspondence, we list the independent Lorentz invariant structure, equivalently, the amplitude basis as functions of spinor-helicity variables, generated via semi-standard Young tableaux (SSYTs) construction. More details can be found in~\cite{Henning:2019enq,Li:2020xlh,Li:2020gnx}. The method still holds when the massive Majorana neutrinos are taken into account, since the Majorana fermions and Dirac fermions have the same wave function in the two-component spinor formalism. Then the gauge structures are also obtained in terms of  invariant group tensors. The Littlewood-Richardson rule allows us to construct a set of singlet Young tableaux of the gauge group indices from the constituting particles, which induces the basis of gauge group factors. Finally, taking the direct product of gauge structures and Lorentz invariant produces the independent flavor-blind amplitudes. Afterwards, the symmetries of the identity particles are expressed by symmetries of flavor indices. We obtain the final results as the 
independent basis of flavor-specified operators with definite permutation symmetries. 

In this paper the operator bases in $\nu$SMEFT and $\nu$LEFT at dimension 5 to 9 are written explicitly as the key results. In $\nu$SMEFT, at dimension 5, the only violation pattern of operators involving sterile neutrinos $N$ is $\left(\Delta B, \Delta L\right)=(0,\pm2)$, while there are $(0,\pm4)$, $(\pm1,\pm1)$ at dimension 6 and $(0,\pm2)$, $(\pm1,\mp1)$ at dimension 7. The violation patterns at dimension 8 are the same as that at dimension 6, and there are additional $(0,\pm6)$, $(\pm1,\pm3)$ at dimension 9 compared to violation patterns at dimension 7. In $\nu$LEFT, the only violation pattern of operators involving sterile neutrinos $N$ is $\left(\Delta B, \Delta L\right)=(0,\pm2)$ at dimension 5, and violation patterns at dimension 6, 7, 8 are the same, which are $(0,\pm2)$, $(0,\pm4)$, $(\pm1,\pm1)$, $(\pm1,\mp1)$. There are additional $(0,\pm6)$, $(\pm1,\pm3)$, $(\pm1,\mp3)$ at dimension 9 compared to violation patterns at dimension 6, 7, 8. Table \ref{tab:sumdim567}-\ref{tab:sumdim9} contain the useful statistics of operators in $\nu$SMEFT and table \ref{tab:sumdim5678L}-\ref{tab:sumdim9L} give the statistics for $\nu$LEFT.

This paper is organized as follows. Section~\ref{sec:pre} first briefly reviews the massive spinor helicity amplitudes and amplitude-operator correspondence, then discusses our notations to construct operators, especially similarities and differences between Dirac neutrinos and Majorana neutrinos. Section~\ref{sec:OpeB} introduces the method of obtaining operator bases. Independent Lorentz and gauge structures can be determined by Young tableaux, and flavor specified operators are obtained after inner product decomposition. Section~\ref{sec:listnuSM} and Section~\ref{sec:listnuL} list high dimension operators involving right-handed neutrinos up to dimension 9 in $\nu$SMEFT and $\nu$LEFT respectively. Section~\ref{sec:con} is our conclusion.

\section{Operators in Spinor-helicity Formalism}\label{sec:pre}

{Amplitude-operator correspondence that connects the local on-shell amplitude  to  the operator producing such amplitude has been extensively used in enumerating operator bases for different kinds of effective field theories}~\cite{Shadmi:2018xan, Ma:2019gtx, Aoude:2019tzn, Durieux:2019eor, Falkowski:2019zdo,Henning:2019enq,Li:2020gnx,Li:2020xlh}.
Since our method of operator construction relies on the amplitude-operator correspondence, we will briefly introduce the spinor-helicity formalism and discuss the amplitude-operator correspondence first in this section. Then  we present the building blocks and operator basis under circumstances where the neutrinos are Dirac or Majorana type.

\subsection{Spinor Helicity Amplitudes}\label{sec:amp}

We start by briefly reviewing relevant notation in the massive spinor-helicity formalism developed recently in Ref.~\cite{Arkani-Hamed:2017jhn}, which could be applied to both the $\nu$SMEFT and $\nu$LEFT in this work. Momenta in the four dimension space-time can always be presented in terms of functions of two spinors, such that
\begin{align}
    p_{\alpha\dot\alpha}=p_{\mu}\sigma^{\mu}_{\alpha\dot\alpha}=\left(\begin{array}{cc}
		E+p_3 & p_1-ip_2\\ p_1+ip_2 & E-p_3
	\end{array}\right)=\lambda^I_{\alpha}\tilde\lambda_{\dot\alpha I},\quad p_{\mu}=\frac12\tilde{\lambda}_{I\dot\alpha}\bar\sigma_{\mu}^{\dot{\alpha}\alpha}\lambda^I_{\alpha},\label{eq:Pchange}
\end{align}
where $\alpha,\dot\alpha$ are the $SU(2)_L\times SU(2)_R$ Lorentz indices, $I$ denotes the $SU(2)$ little group indices for massive particles, which can be omitted for massless particles. The Lorentz invariant brackets of the $SU(2)_L\times SU(2)_R$ group are defined as
\begin{align}
    \langle i^Ij^J\rangle :=\lambda_{i}^{I\alpha}\lambda^J_{j\alpha}=\lambda_{i\alpha}^I\epsilon^{\beta\alpha}\lambda^J_{J\beta}=-\langle j^Ji^I\rangle,\qquad [i^Ij^J]:=\tilde{\lambda}^I_{i\dot\alpha}\tilde{\lambda}^{J\dot\alpha}_j=\tilde{\lambda}^I_{i\dot\alpha}\epsilon^{\dot\alpha \dot\beta}\tilde{\lambda}^{J}_{j\dot\beta}=-[j^Ji^I].
\end{align}
Here the 2-index Levi-Civita symbols are used to raise and lower the indices, defined as
\begin{align}
    \epsilon^{12}=-\epsilon^{21}=\epsilon_{21}=-\epsilon_{12}=1,\quad \lambda^{\alpha I}=\epsilon^{\alpha\beta}\lambda^I_{\beta},\quad \lambda_{\alpha I}=\epsilon_{IJ}\lambda^J_{\alpha},&\quad \tilde\lambda^{\dot\alpha I}=\epsilon^{\dot{\alpha}\dot{\beta}}\tilde\lambda^I_{\dot\beta},\quad \tilde\lambda_{\dot\alpha I}=\epsilon_{IJ}\tilde\lambda^J_{\dot\alpha}.\\
    	\lambda^{I\alpha}\lambda^J_{\alpha}=-m\epsilon^{IJ},\quad \tilde{\lambda}^I_{\dot\alpha}\tilde{\lambda}^{J\dot\alpha}=m\epsilon^{IJ},\quad 
	\lambda^I_{\alpha}\lambda_{\beta I}=m\epsilon_{\alpha\beta},&\quad \tilde{\lambda}^I_{\dot\alpha}\tilde{\lambda}_{\dot\beta I}=m\epsilon_{\dot\alpha\dot\beta}. \label{eq:spin_sum}
\end{align}

For massless fermion and gauge boson, the correspondences between the massless wave function and the helicity spinors follows
\begin{align}
    \epsilon^{\mu}_{i,+}=\frac{\langle \eta|^{\alpha}\sigma^{\mu}_{\alpha\dot\alpha}|i]^{\dot\alpha}}{\sqrt{2}\langle i\eta\rangle},\quad \epsilon^{\mu}_{i,-}=\frac{\langle i|^{\alpha}\sigma^{\mu}_{\alpha\dot\alpha}|\eta]^{\dot\alpha}}{\sqrt{2}[ i\eta]},\quad 	u^{+}=\left(\begin{array}{c}
		0\\\tilde{\lambda}^{\dot\alpha}
	\end{array}\right),\quad  u^{-}=\left(\begin{array}{c}
		\lambda_{\alpha }\\0
	\end{array}\right),\quad \bar{v}^{+}=\left(0,\;\tilde{\lambda}_{\dot{\alpha}} \right),\quad \bar{v}^{-}=\left(\lambda^{\alpha },\; 0 \right),
\label{eq:WFchange}\end{align}
where $\eta$ is the reference spinor parametrizing the gauge redundancy since its value does not affect on the full amplitude.
Notice that we adopt the in-coming momentum convention for all momenta, therefore only $u,\bar{v}$ appear as the in-coming fermions and anti-fermions. $\{\pm\}$ show the helicities of the particles.
For massive fermions, the following forms are taken
\begin{align}
    u^I=\left(\begin{array}{c}
		\lambda_{\alpha }^I\\\tilde{\lambda}^{\dot\alpha I}
	\end{array}\right),\quad \bar{v}^I=\left(\lambda^{\alpha I},\;\tilde{\lambda}_{\dot{\alpha}}^I \right).\label{eq:mWFchange}
\end{align}
With the above replacement rules, the on-shell amplitudes can be written as functions of helicity spinors 
\begin{align}
    \mathcal{A}(\cdots,\epsilon_i,p_i,u_j,p_j,\cdots)=\mathcal{A}(\cdots,\lambda_i,\tilde\lambda_i,\lambda_j,\tilde\lambda_j, \cdots)
\end{align}
Therefore, the on-shell amplitude should contain the following building blocks for massless and massive particles of different helicities  and spins respectively
\begin{align}
	&\text{ helicity }h\text{ massless particle with spinor variables }(\lambda,\tilde\lambda):\quad \mathcal{A} \supset \left\{\begin{array}{ll} \lambda^{r-2h}_{\{\alpha\}}\tilde{\lambda}_{\{\dot\alpha\}}^{r},\quad h\leq 0 \\ 
	\lambda^{r}_{\{\alpha\}}\tilde{\lambda}_{\{\dot\alpha\}}^{r+2h},\quad h\geq 0
	\end{array}\right.\label{eq:LGless} \\
	&\text{ spin }S\text{ massive particle with spinor variables }(\lambda^I,\tilde\lambda^I):\quad \mathcal{A} \supset \left(\lambda^{r+2S-n}_{\{\alpha\}}\tilde{\lambda}^{r+n}_{\{\dot\alpha\}}\right)^{\{I_1\cdots I_{2S}\}},\quad 0\le n\le 2S \label{eq:LGive}
\end{align}
where $\{\cdot\}$ denotes totally symmetric indices.

\subsection{Amplitude-Operator Correspondence}\label{sec:cor}
According to the amplitude-operator correspondence introduced in Ref.~\cite{Shadmi:2018xan, Ma:2019gtx, Aoude:2019tzn, Durieux:2019eor, Falkowski:2019zdo,Henning:2019enq,Li:2020gnx,Li:2020xlh}, it is straightforward to convert any local operators to the on-shell amplitudes using the spinor-helicity formalism introduced above. Firstly, it is necessary to translate the fields in an operator into representations of the $SU(2)_L\times SU(2)_R$ group, following eq.~(\ref{eq:Pchange}) and eq.~(\ref{eq:WFchange}) 
\begin{align}
	\gamma^{\mu}=\left(\begin{array}{cc}
		0&\sigma^{\mu}_{\alpha\dot\alpha}\\\bar{\sigma}^{\mu\dot\alpha\alpha}&0
	\end{array}\right),&\quad \Psi=\left(\begin{array}{c}
		\xi_{\alpha}\\\chi^{\dagger\dot\alpha}
	\end{array}\right),\quad \Psi_M=\left(\begin{array}{c}
		\zeta_{\alpha}\\\zeta^{\dagger\dot\alpha}
	\end{array}\right), \label{eq:fermionfield}\\
	\Psi_{\rm L}=\frac{1-\gamma_5}{2}\Psi=\left(\begin{array}{c}
		\xi_{\alpha}\\ 0
	\end{array}\right)&,\quad \Psi_{\rm R}=\frac{1+\gamma_5}{2}\Psi=\left(\begin{array}{c}
		0\\\chi^{\dagger\dot\alpha}
	\end{array}\right),\\
	D^{\mu}=\frac12 D_{\alpha\dot\alpha}\bar{\sigma}^{\mu\dot\alpha\alpha},\quad F_{\rm{L}\mu\nu}&=\frac14 F_{\rm{L}\alpha\beta}\epsilon_{\dot\alpha\dot\beta}\bar{\sigma}^{\mu\dot\alpha\alpha}\bar{\sigma}^{\nu\dot\beta\beta},\quad F_{\rm{R}\mu\nu}=\frac14 F_{\rm{R}\dot\alpha\dot\beta}\epsilon_{\alpha\beta}\bar{\sigma}^{\mu\dot\alpha\alpha}\bar{\sigma}^{\nu\dot\beta\beta},
\end{align}
where $\xi,\chi$ and $\zeta$ are left-handed Weyl fermions, $\Psi$ and $\Psi_M$ denote Dirac and Majorana fermions respectively. $F_{\rm{L}/\rm{R}}=\frac12 (F\pm i\tilde{F}) $ are the chiral basis of gauge bosons which has definite helicities for the bosons. 
The for the bilinear fermion fields, we take the following rules to obtain the on-shell amplitudes
\begin{align}
	\bar\Psi _1\Psi_2=&\bar\Psi_{{\rm L}1}\Psi_{{\rm R}2}+\bar\Psi_{{\rm R}1}\Psi_{{\rm L}2}\rightarrow\; \bar{v}_1^Iu^J_2=\lambda^{\alpha I}_1\lambda^J_{2\alpha}+\tilde\lambda^I_{1\dot\alpha}\tilde\lambda^{\dot\alpha J}_2=\langle 1^I2^J\rangle+[1^I2^J],\\
	\bar\Psi _1\gamma_5\Psi_2=&\bar\Psi_{{\rm L}1}\Psi_{{\rm R}2}-\bar\Psi_{{\rm R}1}\Psi_{{\rm L}2}\rightarrow\;  \bar{v}_1^I\gamma_5 u^J_2=[1^I2^J]-\langle 1^I2^J\rangle,\\
	\bar\Psi _1\gamma^{\mu}\Psi_2=&\bar\Psi_{{\rm R}1}\gamma^{\mu}\Psi_{{\rm R}2}+\bar\Psi_{{\rm L}1}\gamma^{\mu}\Psi_{{\rm L}2}\rightarrow\; \bar{v}_1^I\gamma^{\mu}u^J_2= \langle 1^I|\sigma^{\mu}|2^J]+[1^I|\bar\sigma ^{\mu}|2^J\rangle,\\
	\bar\Psi _1\gamma^{\mu}\gamma_5\Psi_2=&\bar\Psi_{{\rm R}1}\gamma^{\mu}\Psi_{{\rm R}2}-\bar\Psi_{{\rm L}1}\gamma^{\mu}\Psi_{{\rm L}2}\rightarrow\; \bar{v}_1^I\gamma^{\mu}\gamma_5u^J_2= \langle 1^I|\sigma^{\mu}|2^J]-[1^I|\bar\sigma ^{\mu}|2^J\rangle,\\
	F_{{\rm L}\mu\nu}O^{\mu\nu}=&\frac14 \lambda_{\alpha}\lambda_{\beta}\left(O^{\mu\nu}\sigma_{\nu}\bar{\sigma}_{\mu}\right)^{\alpha\beta},\quad F_{{\rm R}\mu\nu}O^{\mu\nu}=\frac14 \tilde\lambda_{\dot\alpha}\tilde\lambda_{\dot\beta}\left(O^{\mu\nu}\bar\sigma_{\mu}\sigma_{\nu}\right)^{\dot{\alpha}\dot{\beta}},
\end{align}
where eq.~(\ref{eq:WFchange}, \ref{eq:mWFchange}) are used, and
from which the correspondences between a single spinor variable and chiral fermion can be directly identified: 
\begin{align}\label{eq:derivative0}
	\lambda_i^I \rightarrow \psi_i =\Psi_{{\rm L} i},\bar\Psi_{{\rm R} i}\text{ or }\Psi_{M i},\quad \tilde{\lambda}^I_i\rightarrow \psi_i^{\dagger} =\bar{\Psi}_{{\rm L} i},\Psi_{{\rm R} i}\text{ or }\bar{\Psi}_{M i}.
\end{align}
Furthermore,  for a massive particle, when multiple spinors are present, the pair of spinor helicity variables $\lambda^J_i\tilde\lambda_{iJ}$ with contracted little group indices can be translated into derivative acting on particle $i$ yielding the following correspondences: 
\begin{align}\label{eq:derivative1}
	\lambda_i^I\left(\lambda_i^J\tilde\lambda_{iJ}\right)^n \Leftrightarrow D^n\psi_i,&\quad \tilde\lambda_i^I\left(\lambda_i^J\tilde\lambda_{iJ}\right)^n \Leftrightarrow D^n\psi_i^{\dagger},
\end{align}
while for the massless particles, the correspondence remains the same as in  Ref.~\cite{Shadmi:2018xan, Ma:2019gtx, Aoude:2019tzn, Durieux:2019eor, Falkowski:2019zdo,Henning:2019enq,Li:2020gnx,Li:2020xlh,Li:2020tsi}:
\begin{align}\label{eq:derivative2}
	\lambda_i^{n+2}\tilde{\lambda}_i^n \Leftrightarrow D^n F_{{\rm L}i},&\quad \lambda_i^n\tilde\lambda^{n+2} \Leftrightarrow D^n F_{{\rm R}i},\\\label{eq:derivative2}
	\lambda_i^{n+1}\tilde{\lambda}_i^n \Leftrightarrow D^n\psi_i,&\quad \lambda_i^n\tilde{\lambda}^{n+1}_i \Leftrightarrow D^n\psi_i^{\dagger},
\end{align}

There are two comments in order. First, 
the additional interactions with more gauge bosons generated by covariant derivative are not taken into account, because our amplitude operator correspondence applies to local amplitudes only. These vertices are not gauge invariant, and only contribute to parts of  non-local gauge invariant amplitude. 
Second, the IBP relation in operator construction is equivalent to the momentum conservation in on-shell amplitudes. Therefore the IBP redundancy is taken care of via manifesting momentum conservation, which will be treated in the next section.

\subsection{Building Blocks: Majorana verse Dirac Neutrinos}\label{sec:DM}

\begin{table}[h]
	\begin{center}
		\begin{tabular}{|c|cc|ccc|ccc|}
			\hline
			\text{Fields} & $SU(2)_{l}\times SU(2)_{r}$	& $h$ & $SU(3)_{C}$ & $SU(2)_{W}$ & $U(1)_{Y}$ &  Flavor & $B$ & $L$\tabularnewline
			\hline
			$G_{\rm L\alpha\beta}^A$   & $\left(1,0\right)$  & $-1$    & $\boldsymbol{8}$ & $\boldsymbol{1}$ & 0  & $1$ & 0 & 0\tabularnewline
			$W_{\rm L\alpha\beta}^I$   & $\left(1,0\right)$  & $-1$           & $\boldsymbol{1}$ & $\boldsymbol{3}$ & 0  & $1$ & 0 & 0\tabularnewline
			$B_{\rm L\alpha\beta}$   & $\left(1,0\right)$    & $-1$        & $\boldsymbol{1}$ & $\boldsymbol{1}$ & 0  & $1$ & 0 & 0\tabularnewline
			\hline
			$L_{\alpha i}$     & $\left(\frac{1}{2},0\right)$  & $-1/2$  & $\boldsymbol{1}$ & $\boldsymbol{2}$ & $-1/2$  & $n_f$ & 0 & 1\tabularnewline
			$N_{_\mathbb{C}\alpha}$ & $\left(\frac{1}{2},0\right)$ & $-1/2$   & $\boldsymbol{1}$ & $\boldsymbol{1}$ & $0$  & $n_f$ & 0 & $-1$\tabularnewline
			$e_{_\mathbb{C}\alpha}$ & $\left(\frac{1}{2},0\right)$ & $-1/2$   & $\boldsymbol{1}$ & $\boldsymbol{1}$ & $1$  & $n_f$ & 0 & $-1$\tabularnewline
			$Q_{\alpha ai}$     & $\left(\frac{1}{2},0\right)$ & $-1/2$   & $\boldsymbol{3}$ & $\boldsymbol{2}$ & $1/6$  & $n_f$ & $1/3$ & 0\tabularnewline
			$u_{_\mathbb{C}\alpha}^a$ & $\left(\frac{1}{2},0\right)$ & $-1/2$   & $\overline{\boldsymbol{3}}$ & $\boldsymbol{1}$ & $-2/3$  & $n_f$ & $-1/3$ & 0\tabularnewline
			$d_{_\mathbb{C}\alpha}^a$ & $\left(\frac{1}{2},0\right)$ & $-1/2$   & $\overline{\boldsymbol{3}}$ & $\boldsymbol{1}$ & $1/3$  & $n_f$ & $-1/3$ & 0\tabularnewline
			\hline
			$H_i$     & $\left(0,0\right)$&  0     & $\boldsymbol{1}$ & $\boldsymbol{2}$ & $1/2$  & $1$ & 0 & 0\tabularnewline
			\hline
		\end{tabular}
		\caption{\label{tab:SMEFT-field-content}
			The field content of the $\nu$SMEFT, along with their representations under the Lorentz and gauge symmetries, where $N_{_\mathbb{C}}$ denote the right-handed  neutrinos. The representation under Lorentz group is denoted by $(j_l,j_r)$, while the helicity of the field is given by $h = j_r-j_l$ .
			The number of fermion flavors is denoted as $n_f$, which is 3 in the standard model. Their global charges, baryon number $B$ and lepton number $L$ are also listed. All of the fields are accompanied with their Hermitian conjugates that are omitted, $(F_{\rm L \alpha\beta})^\dagger = F_{\rm R \dot\alpha\dot\beta}$ for gauge bosons, $(\psi_\alpha)^\dagger = (\psi^\dagger)_{\dot\alpha}$ for fermions, and $H^\dagger$ for the Higgs, which are under the conjugate representations of all the groups. }
	\end{center}
\end{table}

\begin{table}[t]
	\begin{center}
		\begin{tabular}{|c|cc|cc|ccc|}
			\hline
			\text{Fields} & $SU(2)_{l}\times SU(2)_{r}$	& $h$ & $SU(3)_{C}$ & $U(1)_{\rm EM}$ &  Flavor & $B$ & $L$ \tabularnewline
			\hline
			$G_{\rm L\alpha\beta}^A$   & $\left(1,0\right)$  & $-1$    & $\boldsymbol{8}$ & 0  & $1$ & 0 & 0 \tabularnewline
			$F_{\rm L\alpha\beta}$   & $\left(1,0\right)$    & $-1$        & $\boldsymbol{1}$ & 0  & $1$ & 0 & 0 \tabularnewline
			\hline
			$\nu_{\alpha}$     & $\left(\frac{1}{2},0\right)$  & $-1/2$  & $\boldsymbol{1}$ & $0$  & $n_\nu$ & 0 & $1$ \tabularnewline
			$N_{_\mathbb{C}\alpha}$ & $\left(\frac{1}{2},0\right)$ & $-1/2$   & $\boldsymbol{1}$ & $0$  & $n_\nu$ & 0 & $-1$\tabularnewline
			$e_{\alpha}$ & $\left(\frac{1}{2},0\right)$ & $-1/2$   & $\boldsymbol{1}$ & $-1$  & $n_e$ & 0 & $1$ \tabularnewline
			$e_{_\mathbb{C}\alpha}$ & $\left(\frac{1}{2},0\right)$ & $-1/2$   & $\boldsymbol{1}$ & $1$  & $n_e$ & 0 & $-1$ \tabularnewline
			$u_{\alpha a}$     & $\left(\frac{1}{2},0\right)$ & $-1/2$   & $\boldsymbol{3}$ & $2/3$  & $n_u$ & $1/3$ & 0 \tabularnewline
			$u_{_\mathbb{C}\alpha}^a$ & $\left(\frac{1}{2},0\right)$ & $-1/2$   & $\overline{\boldsymbol{3}}$ & $-2/3$  & $n_u$ & $-1/3$ & 0 \tabularnewline
			$d_{\alpha a}$     & $\left(\frac{1}{2},0\right)$ & $-1/2$   & $\boldsymbol{3}$ & $-1/3$  & $n_d$ & $1/3$ & 0 \tabularnewline
			$d_{_\mathbb{C}\alpha}^a$ & $\left(\frac{1}{2},0\right)$ & $-1/2$   & $\overline{\boldsymbol{3}}$ & $1/3$  & $n_d$ & $-1/3$ & $0$ \tabularnewline
			\hline
		\end{tabular}
		\caption{\label{tab:LEFT-field-content}
			The field content of the $\nu$LEFT.
			The numbers of neutrino flavors, electron flavors, u-type quark flavors and d-type quark flavors are denoted as $n_\nu$, $n_e$, $n_u$ and $n_d$ respectively with $n_\nu=3$, $n_e=3$, $n_u=2$ and $n_d=3$ in $\nu$LEFT.}
	\end{center}
\end{table}

\comment{
Since our building blocks are all two-component spinors, they can be used to construct operators in either cases where neutrinos are Majorana or neutrinos are Dirac, but there remains a subtlety if one want to use these operators, which we will explain in section \ref{sec:DiracMajorana}. If the neutrinos are Majorana fermions, they can be presented as
\begin{eqnarray}
	\nu_M = \left(\begin{array}{c}\nu_{\alpha}\\\nu_{_\mathbb{C}}^{\dot{\alpha}}\end{array}\right), \quad N_M = \left(\begin{array}{c}N_{_\mathbb{C}\alpha}\\N^{\dot{\alpha}}\end{array}\right)
\end{eqnarray}
}

Regarding the building blocks of $\nu$SMEFT, we choose all fields to be left-handed. For example, we choose $N_{_\mathbb{C}\alpha}$ (and its Hermitian conjugate $N_{_\mathbb{C}}^{\dagger\dot{\alpha}}$) instead of $N^{\dot{\alpha}}$ to be the building block. The SM neutrinos $\nu_{\alpha}$ form a $SU(2)_W$ doublet $L_{\alpha}$ along with electrons $e_{\alpha}$, while the right-handed neutrinos are singlets of $SU(2)_W$. Thus, we present the leptons as (chiral) Dirac spinors in our result:
\begin{eqnarray}
	&l_{\rm L}=\left(\begin{array}{c}L_{\alpha}\\0\end{array}\right), \quad  \label{eq:leftneuSM}
	&N_{\rm R}=\left(\begin{array}{c}0\\N_{_\mathbb{C}}^{\dagger\dot{\alpha}}\end{array}\right) \label{eq:rightneuSM},
\end{eqnarray}
where $l_L$ and $N_R$ are four-component chiral Dirac fields with $L_{\alpha} = \left(\begin{array}{c}\nu_{\alpha},e_{\alpha}\end{array}\right)^{T}$  two-component $SU(2)_W$ doublet.
For the building blocks of $\nu$LEFT, where the $SU(2)_W$ symmetry is broken, the notation which writes neutrinos changes to
\begin{eqnarray}
	&\nu_{\rm L}=\left(\begin{array}{c}\nu_{\alpha}\\0\end{array}\right) \label{eq:leftneuL}\\
	&N_{\rm R}=\left(\begin{array}{c}0\\N_{_\mathbb{C}}^{\dagger\dot{\alpha}}\end{array}\right) \label{eq:rightneuL}.
\end{eqnarray}

\comment{
If the neutrinos are Dirac, that is,
\begin{eqnarray}
	\nu_D = \left(\begin{array}{c}\nu_{\alpha}\\N^{\dot{\alpha}}\end{array}\right).
\end{eqnarray}
There are still 2 independent two-component spinors, $\nu_{\alpha}$ and $N^{\dot{\alpha}}$. From the above discussion, it can be seen that our notations relate to two-component spinors only, so eq.~(\ref{eq:leftneuSM}-\ref{eq:rightneuL}) are still valid.

\subsection{Diarc Neutrinos versus Majorana Neutrinos}
}

It is an interesting question as whether the neutrinos are of Dirac type or of Majorana type. Since the neutrinos are neutral after the electroweak symmetry breaking, they are generically Majorana fermions.
All massive fermions consist of two-component spinors with Majorana masses, whilst
a pair of them can be written as a Dirac fermion if and only if there is an $SO(2)$ global symmetry
\eq{
    \mc{L} \supset -m\xi_1\xi_1 - m\xi_2\xi_2 + h.c. ,\qquad
    \begin{pmatrix}
    \xi_1 \\ \xi_2
    \end{pmatrix} \xrightarrow[]{SO(2)} \begin{pmatrix}
    \cos\phi & -\sin\phi \\ \sin\phi & \cos\phi
    \end{pmatrix}\begin{pmatrix}
    \xi_1 \\ \xi_2
    \end{pmatrix}.
}
In our case, the left-handed neutrino and a right-handed counterpart could be their complexification
\eq{
\nu = \xi_1 + i\xi_2 \quad N_{_\mathbb{C}} = \xi_1 - i\xi_2, \qquad 
\mc{L} \supset -m \nu N_{_\mathbb{C}} + h.c. 
}
The $SO(2)$ symmetry becomes an opposite phase shift for $\nu \to e^{i\phi}\nu_L$ and $N_{_\mathbb{C}}\to e^{-i\phi}N_{_\mathbb{C}}$ that can be identified as the lepton number $U(1)_L$\footnote{
Consider the $U(1)^7$ phase symmetry of the field content $H,Q,L,u_{_\mathbb{C}},d_{_\mathbb{C}},e_{_\mathbb{C}},N_{_\mathbb{C}}$ in the $\nu$SMEFT, broken by the 3 Standard Model Yukawa's and an additional Yukawa for neutrinos $HLN_{\mathbb{C}}$ down to $U(1)^3$, one of which gauged to be $U(1)_{\rm EM}$. The remaining two $U(1)'s$ are conventionally divided into Baryon number $U(1)_B$ and Lepton number $U(1)_L$, and by definition, the neutrinos carry no Baryon numbers. Hence the $U(1)$ symmetry for neutrinos has to be the Lepton number $U(1)_L$.
}.
Therefore, from the EFT point of view, any $U(1)_L$ violating effect could result in the breaking of the $SO(2)$ symmetry.
Since the unbroken $SO(2)$ allows us to combine the neutrinos into a Dirac fermion as
\eq{
    \nu_D = \begin{pmatrix}
    \nu_L \\ N_R
    \end{pmatrix} , \qquad \mc{L} \supset -m\overline\nu_D\nu_D,
}
we conclude that the Dirac neutrino is just a notation allowed by the existence of an exact $U(1)_L$ symmetry, hence the question is replaced by whether the $U(1)_L$ violating operators are turned on in the $\nu$SMEFT. 
Once they are turned on, we get the Majorana neutrino, whose four-component form is
\begin{eqnarray}
	\nu_M = \left(\begin{array}{c}\nu_{\alpha}\\\nu_{_\mathbb{C}}^{\dot{\alpha}}\end{array}\right), \quad N_M = \left(\begin{array}{c}N_{_\mathbb{C}\alpha}\\N^{\dot{\alpha}}\end{array}\right).
\end{eqnarray}
We show the complete list of the $U(1)_L$ violating operators in the paper, which should contain the answer. For generality, we stick to the chiral fermion notation eq.~(\ref{eq:leftneuSM}-\ref{eq:rightneuL}) throughout the paper.

\section{Operator Basis}\label{sec:OpeB}

In this section, we will briefly introduce the Young Tensor method to obtain the complete operator bases of $\nu$SMEFT and $\nu$LEFT, then give a specific example of the method. More details can be found in Ref.~\cite{Li:2020gnx, Li:2020xlh, Li:2020tsi}.

\subsection{Operator Construction using Young Tensor}\label{sec:OpeC}

The independent Lorentz structures of a certain class corresponding to  amplitudes of $N$ particles containing $2n$ $\lambda$s and $2\tilde{n}$ $\tilde{\lambda}$s can be presented as $SU(N)$ semi-standard Young tableaux (SSYTs) of a so-called primary Young diagram. The primary Young diagram is as follows,\\
\begin{eqnarray}\label{eq:primary_YD}
	Y_{N,n,\tilde{n}} \quad = \quad \arraycolsep=0pt\def\arraystretch{1}
	\rotatebox[]{90}{\text{$N-2$}} \left\{
	\begin{array}{cccccc}
		\yng(1,1) &\ \ldots{}&\ \yng(1,1)& \overmat{n}{\yng(1,1)&\ \ldots{}\  &\yng(1,1)} \\
		\vdotswithin{}& & \vdotswithin{}&&&\\
		\undermat{\tilde{n}}{\yng(1,1)\ &\ldots{}&\ \yng(1,1)} &&&
	\end{array}
	\right.\ .
	\\
	\nonumber 
\end{eqnarray}
The number of certain indices $i$ to be filled in the primary Young diagram to form SSYTs is determined by $\#i=\tilde{n}-2h_i$, where $h_i$ denotes the helicity of the $i$th particle in the class and the set $\{h_i\}$ are sorted in the order $h_i \leq h_{i+1}$, $i=1,\cdots,N$. The Fock's condition of Young tableaux corresponds to the IBP relation and the Schouten identities of operators, and the SSYTs obtained in this way form a complete and independent amplitude basis that spans a subspace of the representation $Y_{N,n,\tilde{n}}$ of the auxiliary $SU(N)$ group . The correspondence between SSYTs and amplitudes translates columns of SSYTs into brackets using the following rules
\eq{\label{eq:YT_translate}
	\young(i,j) \sim \vev{ij}, \qquad \begin{array}{c} \young({{k_1}},{{k_2}}) \\ \vdots \\ \young({{k_{N-3}}},{{k_{N-2}}}) \end{array} \sim \mc{E}^{k_1\dots k_{N-2}ij}[ij].
}
An concrete example of this correspondence is demonstrated in eq.~\eqref{eq:lexamplet} and eq.~\eqref{eq:lexampleb}.
So far, we manage the massless particles except for the massive Majorana neutrino. 
As discussed in Ref.~\cite{Li:2020tsi}, we show that our algorithm for finding the massive amplitude basis can be directly applied to the theory with massive particle of spin $S=0,1/2$, because the following one to one correspondence between massless and massive amplitudes exists:
\eq{\label{eq:map2mass}\begin{array}{lcr}
	h=0             &\lambda^{r}\tilde{\lambda}^{r}\stackrel{}{\longrightarrow} \left(\lambda^J\tilde{\lambda}_J\right)^r                   & S=0,\ n=0\\
	h=-\frac12      &\lambda^{1+r}\tilde{\lambda}^r\stackrel{}{\longrightarrow} \lambda^I\left(\lambda^J\tilde{\lambda}_J\right)^r          & S=\frac12,\ n=0\\
	h=\frac12       &\lambda^{r}\tilde{\lambda}^{1+r}\stackrel{}{\longrightarrow} \tilde{\lambda}^I\left(\lambda^J\tilde{\lambda}_J\right)^r  & S=\frac12,\ n=1.
\end{array}}
Therefore we treat the massive Majorana neutrino as a massless particle in our method, finding the operator basis for a given dimension. 
Taking into account the amplitude-operator correspondence introduced in eq.~\eqref{eq:derivative0}  to eq.~\eqref{eq:derivative2}, one can further translate the amplitudes to operators as an example showed in eq.~\eqref{eq:lexampleo}.


The independent gauge structures of a certain type contain the Levi-Civita tensors contracted with fundamental gauge group indices of building blocks. Other representations, such as anti-fundamental and adjoint representations, can be converted to the one with fundamental indices only, which corresponds to a particular Young tableau as shown in the following examples, 
\eq{\label{eq:example_T_gluon}
	&\epsilon_{acd}\lambda^A{}_b^d G^A = G_{abc} \sim \young(ab,c)\,,\\
	&\epsilon_{abc}Q^{\dagger,{c}} = Q^\dagger_{ab} \sim \young(a,b)\,,\\
	&\epsilon_{jk}\tau^I{}_i^k W^I = W_{ij}\sim \young(ij) \,,\\
	&\epsilon_{ij}H^{\dagger, j} = H_i \sim \young(i) \,.
}
Then the independent gauge structures are obtained using the modified Littlewood-Richardson rule which applies to the Young tableaux\cite{Li:2020gnx}.

If there are repeated fields in an operator, more redundancies emerge since certain permutation symmetries of flavor indices are supposed to vanish due to permutation symmetries of the Lorentz and gauge structures and spin-statistics of identical particles. To remove these redundancies, we introduce the projectors \cite{Li:2020gnx} to pick out the permutation symmetries of Lorentz and gauge structures and obtain the operators with allowed permutation symmetries of flavor indices through inner product decomposition, since the effect of permuting flavor indices is equivalent to that of permuting gauge and Lorentz structures:
\begin{eqnarray}
	\underbrace{\pi\circ {\cal O}^{\{f_{k},...\}}}_{\rm permute\ flavor} &=& \underbrace{\left(\pi\circ T_{{\rm SU3}}^{\{g_k,...\}}\right)\left(\pi\circ T_{{\rm SU2}}^{\{h_k,...\}}\right)}_{\rm permute\ gauge}\underbrace{\left(\pi\circ{\cal M}^{\{f_k,...\}}_{\{g_{k},...\},\{h_{k},...\}}\right)}_{\rm permute\ Lorentz}.
	\label{eq:tperm}
\end{eqnarray}

After an extra step to simplify the result while keeping track of the permutation symmetries, which we called de-symmetrization, the final result is expressed as $\mc{Y}^{[\lambda]}_x\circ\mc{O}^{\rm (m)}_i$, where $\mc{Y}^{[\lambda]}_x$ is the $x$th Young symmetrizer of the $S_n$ group representation $\lambda$, and $\mc{O}^{\rm (m)}_i$ is a monomial. For example,
\eq{
	&\mc{Y}^{[3]}_1 O^{prs} = \mc{Y}\left[\tiny{\young(prs)}\right] O^{prs} = O^{prs}+O^{rps}+O^{psr}+O^{rsp}+O^{srp}+O^{spr}, \\
	&\mc{Y}^{[2,1]}_1 O^{prs} = \mc{Y}\left[\tiny{\young(pr,s)}\right] O^{prs} = O^{prs}+O^{rps}-O^{srp}-O^{spr}, \\
	&\mc{Y}^{[1^3]}_1 O^{prs} = \mc{Y}\left[\tiny{\young(p,r,s)}\right] O^{prs} = O^{prs}-O^{rps}-O^{psr}+O^{rsp}-O^{srp}+O^{spr}.
}
The action of a Young symmetrizer can also be interpreted as acting on the Wilson coefficient tensor instead of the operator since
\begin{eqnarray}
\sum_{p_i} C_{p_1p_2...p_n}\left( {\cal Y}[\lambda]{\cal O}^{p_1p_2...p_n}\right) = \sum_{p_i}\left( {\cal Y}^{-1}[\lambda] C_{p_1p_2...p_n}\right){\cal O}^{p_1p_2...p_n},
\end{eqnarray}
where ${\cal Y}^{-1}[\lambda]$ means taking inverse of each constituting permutation in ${\cal Y}[\lambda]$. ${\cal Y}^{-1}[\lambda] C_{p_1p_2...p_n}$ will project out the $\lambda$ irreducible representation of $S_n$ group in terms of the Wilson coefficient $ C_{p_1p_2...p_n}$, and the operator is still a monomial in this case. An explicit example will be given in section~\ref{sec:eg}.

\subsection{Procedure and Example}\label{sec:eg}

Before getting started, we should clarify some terminology that will be used in the following content.
\begin{itemize}
    \item Class: A (Lorentz) class is formed by abstract fields that are Lorentz irreducible representations and covariant derivatives.
    \item Type: Substituting the specific fields of $\nu$SMEFT/$\nu$LEFT into each class, the combination of fields (and covariant derivatives) that can form gauge invariant is called a type.
    \item Term: In each type, we organize Lorentz and gauge invariant flavor tensors into different irreducible representations of the symmetric group of flavor indices, and each of the irreducible representations is called a term.
    \item Operator (flavor-specified): Each flavor-specified component of a term viewed as an irreducible flavor tensor is an operator.
\end{itemize}

Let us take the operator type $L N_{_\mathbb{C}}^3 H D^2$ as an example. The independent Lorentz structures are presented by the following SSYTs.
\begin{eqnarray}\label{eq:lexamplet}
    \young(1133,2244,5) \quad \young(1133,2245,4) \quad \young(1134,2245,3) \quad \young(1122,3344,5) \quad \young(1122,3345,4) \nonumber \\
    \young(1124,2335,4) \quad \young(1124,2345,3) \quad \young(1123,2445,3) \quad \young(1123,2344,5) \quad \young(1123,2345,4)
\end{eqnarray}
These SSYTs are independent and complete, and any non-SSYT can be converted to SSTYs with the Fock's conditions~\cite{ma2007group}, thus form the so-called y-basis (Young tableau basis) of this type and can be interpreted as amplitudes using eq.~(\ref{eq:YT_translate})
\begin{eqnarray}\label{eq:lexampleb}
    &\vev{12}\vev{34}^2 [34] \quad -\vev{12}\vev{34}\vev{35}[35] \quad \vev{12}\vev{34}\vev{45}[45] \quad -\vev{13}\vev{24}^2 [24] \quad \vev{13}\vev{24}\vev{25}[25] \nn \\
    &-\vev{13}\vev{23}\vev{45} [35] \quad \vev{13}\vev{24}\vev{45}[45] \quad \vev{14}\vev{24}\vev{35}[45] \quad \vev{13}\vev{24}\vev{34} [34] \quad -\vev{13}\vev{24}\vev{35}[35].\label{eq:ybamp}
\end{eqnarray}
Furthermore, the y-basis can be expressed as Lorentz structures of operators using the amplitude-operator correspondence eq.~(\ref{eq:derivative0})-\eqref{eq:derivative2}, and the basis vectors are
\begin{eqnarray}\label{eq:lexampleo}
\begin{split}
&{\cal M}^{\rm y}_1=L_1^{\alpha} N_{_\mathbb{C}}{}_2{}_{\alpha} \left(D N_{_\mathbb{C}}{}_3\right)^{\beta\gamma}_{\dot{\alpha}} \left(D N_{_\mathbb{C}}{}_4\right)_{\beta\gamma}^{\dot{\alpha}} H_5, \\
&{\cal M}^{\rm y}_2=-L_1^{\alpha} N_{_\mathbb{C}}{}_2{}_{\alpha} \left(D N_{_\mathbb{C}}{}_3\right)^{\beta\gamma}_{\dot{\alpha}}  N_{_\mathbb{C}}{}_4{}_{\beta} \left(D H_5\right)_{\gamma}^{\dot{\alpha}}, \\ 
&{\cal M}^{\rm y}_3=L_1^{\alpha} N_{_\mathbb{C}}{}_2{}_{\alpha}  N_{_\mathbb{C}}{}_3{}^{\beta}  \left(D N_{_\mathbb{C}}{}_4\right)_{\beta\dot{\alpha}}^{\gamma} \left(D H_5\right)_{\gamma}^{\dot{\alpha}}, \\ 
&{\cal M}^{\rm y}_4=-L_1^{\alpha} \left(D N_{_\mathbb{C}}{}_2\right)^{\beta\gamma}_{\dot{\alpha}} N_{_\mathbb{C}}{}_3{}_{\alpha} \left(D N_{_\mathbb{C}}{}_4\right)_{\beta\gamma}^{\dot{\alpha}} H_5, \\
&{\cal M}^{\rm y}_5=L_1^{\alpha} \left(D N_{_\mathbb{C}}{}_2\right)^{\beta\gamma}_{\dot{\alpha}} N_{_\mathbb{C}}{}_3{}_{\alpha}  N_{_\mathbb{C}}{}_4{}_{\beta} \left(D H_5\right)_{\gamma}^{\dot{\alpha}}, \\
&{\cal M}^{\rm y}_6=-L_1^{\alpha}  N_{_\mathbb{C}}{}_2{}^{\beta} \left(D N_{_\mathbb{C}}{}_3\right)_{\alpha\beta\dot{\alpha}}  N_{_\mathbb{C}}{}_4{}^{\gamma} \left(D H_5\right)_{\gamma}^{\dot{\alpha}}, \\
&{\cal M}^{\rm y}_7=L_1^{\alpha}  N_{_\mathbb{C}}{}_2{}^{\beta} N_{_\mathbb{C}}{}_3{}_{\alpha}  \left(D N_{_\mathbb{C}}{}_4\right)^{\gamma}_{\beta\dot{\alpha}} \left(D H_5\right)_{\gamma}^{\dot{\alpha}}, \\
&{\cal M}^{\rm y}_8=L_1^{\alpha}  N_{_\mathbb{C}}{}_2{}^{\beta} N_{_\mathbb{C}}{}_3{}^{\gamma}  \left(D N_{_\mathbb{C}}{}_4\right)_{\alpha\beta\dot{\alpha}} \left(D H_5\right)_{\gamma}^{\dot{\alpha}}, \\
&{\cal M}^{\rm y}_9=L_1^{\alpha}  N_{_\mathbb{C}}{}_2{}^{\beta} \left(D N_{_\mathbb{C}}{}_3\right)^{\gamma}_{\alpha\dot{\alpha}} \left(D N_{_\mathbb{C}}{}_4\right)_{\beta\gamma}^{\dot{\alpha}} H_5, \\
&{\cal M}^{\rm y}_{10}=-L_1^{\alpha}  N_{_\mathbb{C}}{}_2{}^{\beta} \left(D N_{_\mathbb{C}}{}_3\right)^{\gamma}_{\alpha\dot{\alpha}} N_{_\mathbb{C}}{}_4{}_{\beta} \left(D H_5\right)_{\gamma}^{\dot{\alpha}}.
\end{split}
\end{eqnarray}

The m-basis of this type, which independent monomial operators span, can be obtained by converting $D_{\alpha\beta}$ to $D_{\mu}$ and finding the independent monomials, which in this case are
\begin{eqnarray}
\begin{split}
&{\cal M}^{\rm m}_1=\left(L_{pi} N_{_\mathbb{C}}{}_{r}\right) \left(D_{\mu } N_{_\mathbb{C}}{}_{s} D^{\mu } N_{_\mathbb{C}}{}_{t}\right) H_{j}, \\
&{\cal M}^{\rm m}_2=\left(L_{pi} N_{_\mathbb{C}}{}_{r}\right) \left(D_{\mu } N_{_\mathbb{C}}{}_{s} N_{_\mathbb{C}}{}_{t}\right) D^{\mu } H_{j}, \\
&{\cal M}^{\rm m}_3=\left(L_{pi} N_{_\mathbb{C}}{}_{r}\right) \left(N_{_\mathbb{C}}{}_{s} D_{\mu } N_{_\mathbb{C}}{}_{t}\right) D^{\mu } H_{j}, \\
&{\cal M}^{\rm m}_4=\left(L_{pi} N_{_\mathbb{C}}{}_{s}\right) \left(D_{\mu } N_{_\mathbb{C}}{}_{r} D^{\mu } N_{_\mathbb{C}}{}_{t}\right) H_{j}, \\
&{\cal M}^{\rm m}_5=\left(L_{pi} N_{_\mathbb{C}}{}_{s}\right) \left(D_{\mu } N_{_\mathbb{C}}{}_{r} N_{_\mathbb{C}}{}_{t}\right) D^{\mu } H_{j}, \\
&{\cal M}^{\rm m}_6=\left(L_{pi} N_{_\mathbb{C}}{}_{t}\right) \left(N_{_\mathbb{C}}{}_{r} D_{\mu } N_{_\mathbb{C}}{}_{s}\right) D^{\mu } H_{j}, \\
&{\cal M}^{\rm m}_7=i\left(L_{pi} \sigma_{\mu\nu} N_{_\mathbb{C}}{}_{t}\right) \left(N_{_\mathbb{C}}{}_{r} D^{\mu } N_{_\mathbb{C}}{}_{s}\right) D^{\nu } H_{j}, \\
&{\cal M}^{\rm m}_8=\left(L_{pi} N_{_\mathbb{C}}{}_{s}\right) \left(N_{_\mathbb{C}}{}_{r} D_{\mu } N_{_\mathbb{C}}{}_{t}\right) D^{\mu } H_{j}, \\
&{\cal M}^{\rm m}_9=i\left(L_{pi} \sigma_{\mu\nu} N_{_\mathbb{C}}{}_{s}\right) \left(N_{_\mathbb{C}}{}_{r} D^{\mu } N_{_\mathbb{C}}{}_{t}\right) D^{\nu } H_{j}, \\
&{\cal M}^{\rm m}_{10}=i\left(L_{pi} \sigma_{\mu\nu} N_{_\mathbb{C}}{}_{r}\right) \left(D^{\mu } N_{_\mathbb{C}}{}_{s} D^{\nu } N_{_\mathbb{C}}{}_{t}\right) H_{j}.
\end{split}
\end{eqnarray}
After obtaining the m-basis, we can symmetrize the Lorentz structures into irreducible representations with regards to permutation of flavor indices of the repeated fields $N_{_\mathbb{C}}$, which form the so-called p-basis Lorentz structures (symmetric permutation basis) ${\cal M}^{\lambda,\xi}_x$
\begin{eqnarray}
\begin{pmatrix}
{\cal M}^{[3],1}_{1}\\
{\cal M}^{[3],2}_{1}\\
{\cal M}^{[2,1],1}_{1}\\
{\cal M}^{[2,1],1}_{2}\\
{\cal M}^{[2,1],2}_{1}\\
{\cal M}^{[2,1],2}_{2}\\
{\cal M}^{[2,1],3}_{1}\\
{\cal M}^{[2,1],3}_{2}\\
{\cal M}^{[1^3],1}_{1}\\
{\cal M}^{[1^3],2}_{1}
\end{pmatrix}
=\begin{pmatrix}
0 & -\frac{2}{3} & 0 & 0 & \frac{2}{3} & \frac{1}{6} & -\frac{1}{6} & -\frac{1}{2} & \frac{1}{6} & 0\\
0 & 0 & 0 & 0 & 0 & -\frac{1}{2} & \frac{1}{2} & \frac{1}{2} & \frac{1}{2} & 0\\
\frac{2}{3} & 0 & 0 & -\frac{2}{3} & 0 & 0 & 0 & 0 & 0 & 0\\
-\frac{2}{3} & -\frac{2}{3} & 0 & \frac{2}{3} & \frac{2}{3} & -1 & -\frac{1}{3} & \frac{2}{3} & 0 & \frac{2}{3}\\
0 & 0 & \frac{2}{3} & 0 & 0 & -\frac{1}{3} & \frac{1}{3} & \frac{1}{3} & -\frac{1}{3} & 0\\
0 & -\frac{2}{3} & 0 & 0 & 0 & -\frac{1}{3} & -\frac{1}{3} & \frac{1}{3} & \frac{1}{3} & 0\\
0 & \frac{2}{3} & \frac{4}{3} & 0 & -\frac{2}{3} & -\frac{1}{3} & \frac{1}{3} & -\frac{1}{3} & -\frac{1}{3} & 0\\
0 & -\frac{2}{3} & 0 & 0 & -\frac{2}{3} & \frac{2}{3} & 0 & \frac{2}{3} & 0 & 0\\
\frac{2}{3} & -\frac{2}{3} & 0 & \frac{4}{3} & \frac{2}{3} & -1 & -\frac{1}{3} & \frac{2}{3} & 0 & \frac{2}{3}\\
0 & -\frac{2}{3} & -\frac{2}{3} & 0 & 0 & -\frac{1}{2} & -\frac{1}{6} & -\frac{1}{2} & \frac{1}{6} & 0\\
\end{pmatrix}
\begin{pmatrix}
{\cal M}^{\rm m}_{1}\\
{\cal M}^{\rm m}_{2}\\
{\cal M}^{\rm m}_{3}\\
{\cal M}^{\rm m}_{4}\\
{\cal M}^{\rm m}_{5}\\
{\cal M}^{\rm m}_{6}\\
{\cal M}^{\rm m}_{7}\\
{\cal M}^{\rm m}_{8}\\
{\cal M}^{\rm m}_{9}\\
{\cal M}^{\rm m}_{10}
\end{pmatrix},
\end{eqnarray}
where $\lambda$ denotes the representation of the symmetric group, and $x=1,\cdots,d_{\lambda}$. $\xi$ is the multiplicity of the representation. 
As we can see, the symmetrization procedure here is a full-rank conversion matrix that converts m-basis vectors to p-basis vectors, which guarantees the independence and completeness of the p-basis.
Here one subtlety emerges, as we elaborated in Ref.~\cite{Li:2020gnx}, the Grassmann nature of fermions needs to be taken into account, so the final permutation symmetry of Lorentz structure should be $\lambda^{\rm T}$ instead of $\lambda$.

The gauge group structure of this type is simple. The only non-trivial gauge group factor is the $SU(2)_W$ factor $T^m_{SU2}=\epsilon^{ij}$,
\begin{eqnarray}
L_i \sim \young(i)\,, \ H_j \sim \young(j)\,, \quad
\young(i,j)\,.
\end{eqnarray}

At last, the flavor structures of operators are determined by the inner-product decomposition of the Lorentz structures and the gauge structures. We denote the operators with certain flavor symmetries $\lambda$ by ${\cal O}^{\rm (p)}_{(\lambda,x),\xi}$ and find that
\begin{eqnarray}
\begin{pmatrix}
{\cal O}^{\rm (p)}_{([1^3],1),1} \\
{\cal O}^{\rm (p)}_{([1^3],1),2} \\
{\cal O}^{\rm (p)}_{([2,1],1),1} \\
{\cal O}^{\rm (p)}_{([2,1],2),1} \\
{\cal O}^{\rm (p)}_{([2,1],1),2} \\
{\cal O}^{\rm (p)}_{([2,1],2),2} \\
{\cal O}^{\rm (p)}_{([2,1],1),3} \\
{\cal O}^{\rm (p)}_{([2,1],2),3} \\
{\cal O}^{\rm (p)}_{([3],1),1} \\
{\cal O}^{\rm (p)}_{([3],1),2} 
\end{pmatrix}
=
\begin{pmatrix}
0 & -\frac{2}{3} & 0 & 0 & \frac{2}{3} & \frac{1}{6} & -\frac{1}{6} & -\frac{1}{2} & \frac{1}{6} & 0\\
0 & 0 & 0 & 0 & 0 & -\frac{1}{2} & \frac{1}{2} & \frac{1}{2} & \frac{1}{2} & 0\\
\frac{2}{3} & 0 & 0 & -\frac{2}{3} & 0 & 0 & 0 & 0 & 0 & 0\\
-\frac{2}{3} & -\frac{2}{3} & 0 & \frac{2}{3} & \frac{2}{3} & -1 & -\frac{1}{3} & \frac{2}{3} & 0 & \frac{2}{3}\\
0 & 0 & \frac{2}{3} & 0 & 0 & -\frac{1}{3} & \frac{1}{3} & \frac{1}{3} & -\frac{1}{3} & 0\\
0 & -\frac{2}{3} & 0 & 0 & 0 & -\frac{1}{3} & -\frac{1}{3} & \frac{1}{3} & \frac{1}{3} & 0\\
0 & \frac{2}{3} & \frac{4}{3} & 0 & -\frac{2}{3} & -\frac{1}{3} & \frac{1}{3} & -\frac{1}{3} & -\frac{1}{3} & 0\\
0 & -\frac{2}{3} & 0 & 0 & -\frac{2}{3} & \frac{2}{3} & 0 & \frac{2}{3} & 0 & 0\\
\frac{2}{3} & -\frac{2}{3} & 0 & \frac{4}{3} & \frac{2}{3} & -1 & -\frac{1}{3} & \frac{2}{3} & 0 & \frac{2}{3}\\
0 & -\frac{2}{3} & -\frac{2}{3} & 0 & 0 & -\frac{1}{2} & -\frac{1}{6} & -\frac{1}{2} & \frac{1}{6} & 0\\
\end{pmatrix}
\begin{pmatrix}
{\cal M}^{\rm m}_1 T^{\rm m}_{\rm SU2}\\
{\cal M}^{\rm m}_2 T^{\rm m}_{\rm SU2}\\
{\cal M}^{\rm m}_3 T^{\rm m}_{\rm SU2}\\
{\cal M}^{\rm m}_4 T^{\rm m}_{\rm SU2}\\
{\cal M}^{\rm m}_5 T^{\rm m}_{\rm SU2}\\
{\cal M}^{\rm m}_6 T^{\rm m}_{\rm SU2}\\
{\cal M}^{\rm m}_7 T^{\rm m}_{\rm SU2}\\
{\cal M}^{\rm m}_8 T^{\rm m}_{\rm SU2}\\
{\cal M}^{\rm m}_9 T^{\rm m}_{\rm SU2}\\
{\cal M}^{\rm m}_{10} T^{\rm m}_{\rm SU2}
\end{pmatrix}.
\end{eqnarray}
It should be noted that the basis $\{{\cal O}^{\rm (p)}_{(\lambda,x),\xi}\}$ is over complete, as discussed in Ref.~\cite{Li:2020gnx} and \cite{Fonseca:2019yya}, so we only keep the $x=1$ basis vector for each representations. Therefore the complete operator basis of type $W_L L N_{_\mathbb{C}} H D^2$ is 
\begin{eqnarray}
\{{\cal O}^{\rm (p)}_{([1^3],1),1}, \quad
{\cal O}^{\rm (p)}_{([1^3],1),2}, \quad
{\cal O}^{\rm (p)}_{([2,1],1),1}, \quad
{\cal O}^{\rm (p)}_{([2,1],1),2}, \quad
{\cal O}^{\rm (p)}_{([2,1],1),3}, \quad
{\cal O}^{\rm (p)}_{([3],1),1}, \quad
{\cal O}^{\rm (p)}_{([3],1),2}\}.
\end{eqnarray}

The p-basis obtained above is very long. To simplify the form of the result, we apply the desymmetrization procedure \cite{Li:2020xlh} and express the operator basis as monomial flavor tensors with certain flavor symmetry $\lambda$, that is, $\mc{Y}^{[\lambda]}_x\circ\mc{O}^{\rm (m)}_i$, called the p'-basis. In this example, the p'-basis is
\begin{eqnarray}
\begin{split}
{\cal O}^{({\rm p'})}_{([1^3],1),1}&=\mathcal{Y}\left[\tiny{\young(r,s,t)}\right]\epsilon ^{ij} D^{\mu } H_{j} \left(L_{pi} N_{_\mathbb{C}}{}_{r}\right) \left(D_{\mu } N_{_\mathbb{C}}{}_{s} N_{_\mathbb{C}}{}_{t}\right), \\
{\cal O}^{({\rm p'})}_{([1^3],1),2}&=\mathcal{Y}\left[\tiny{\young(r,s,t)}\right]i\epsilon ^{ij} D^{\nu } H_{j} \left(L_{pi} \sigma_{\mu\nu} N_{_\mathbb{C}}{}_{t}\right) \left(N_{_\mathbb{C}}{}_{r} D^{\mu } N_{_\mathbb{C}}{}_{s}\right), \\
{\cal O}^{({\rm p'})}_{([2,1],1),1}&=\mathcal{Y}\left[\tiny{\young(rs,t)}\right]\epsilon ^{ij} H_{j} \left(L_{pi} N_{_\mathbb{C}}{}_{r}\right) \left(D_{\mu } N_{_\mathbb{C}}{}_{s} D^{\mu } N_{_\mathbb{C}}{}_{t}\right), \\
{\cal O}^{({\rm p'})}_{([2,1],1),2}&=\mathcal{Y}\left[\tiny{\young(rs,t)}\right]\epsilon ^{ij} D^{\mu } H_{j} \left(L_{pi} N_{_\mathbb{C}}{}_{r}\right) \left(D_{\mu } N_{_\mathbb{C}}{}_{s} N_{_\mathbb{C}}{}_{t}\right), \\
{\cal O}^{({\rm p'})}_{([2,1],1),3}&=\mathcal{Y}\left[\tiny{\young(rs,t)}\right]\epsilon ^{ij} D^{\mu } H_{j} \left(L_{pi} N_{_\mathbb{C}}{}_{r}\right) \left(N_{_\mathbb{C}}{}_{s} D_{\mu } N_{_\mathbb{C}}{}_{t}\right), \\
{\cal O}^{({\rm p'})}_{([3],1),1}&=\mathcal{Y}\left[\tiny{\young(rst)}\right]\epsilon ^{ij} H_{j} \left(L_{pi} N_{_\mathbb{C}}{}_{r}\right) \left(D_{\mu } N_{_\mathbb{C}}{}_{s} D^{\mu } N_{_\mathbb{C}}{}_{t}\right), \\
{\cal O}^{({\rm p'})}_{([3],1),2}&=\mathcal{Y}\left[\tiny{\young(rst)}\right]\epsilon ^{ij} D^{\mu } H_{j} \left(L_{pi} N_{_\mathbb{C}}{}_{r}\right) \left(D_{\mu } N_{_\mathbb{C}}{}_{s} N_{_\mathbb{C}}{}_{t}\right).
\end{split}
\end{eqnarray}
This is the final form of operators in our result, eq.~(\ref{cl:psiL4hD2l}), after converting two-component spinors to four-component spinors. The Young symmetrizer in front of each monomial operator is interpreted as acting on the Wilson coefficient tensor $C_{prst}$, so that the independent component of the Wilson coefficient tensor that corresponds to each flavor specified operator is constrained. For example, for three generations of fermions, the flavor tensor $C_{prst}$ contains the following independent components for each possible value of $p=1,2,3$: 
\begin{eqnarray}
\young(r,s,t): & \young(1,2,3) \nn \\
\young(rs,t): & \young(11,2) \quad \young(11,3) \quad \young(12,2) \quad \young(12,3) \quad \young(13,2) \quad \young(13,3) \quad \young(22,3) \quad \young(23,3) \\
\young(rst): &\young(111) \quad \young(112) \quad \young(113) \quad \young(122) \quad \young(123) \quad \young(133) \quad \young(222) \quad \young(223) \quad \young(233) \quad \young(333). \nn
\end{eqnarray}
Thus the flavor tensor can be further written as
\begin{eqnarray}
C_{prst} = C_{p \tiny{\young(r,s,t)}} + C_{p \tiny{\young(rs,t)}} + C_{p \tiny{\young(rst)}}.
\end{eqnarray}

\section{Lists of Operators in $\nu$SMEFT}\label{sec:listnuSM}
In this section, we list the complete and independent operator basis in $\nu$SMEFT from dimension 5 to dimension 9, and the statistic results of the operator basis are listed in table \ref{tab:sumdim567}, \ref{tab:sumdim8} and \ref{tab:sumdim9}. It should be noted that the two-component Weyl fermions are used in our building block and converted to four-component (chiral) Dirac fermions in the final result. We give the relations for conversion here for readers' convenience.
\begin{align}
	&q_{\rm{L}}=\begin{pmatrix}Q\\0\end{pmatrix},\quad u_{\rm{R}}=\left(\begin{array}{c}0\\u_{_\mathbb{C}}^{\dagger}\end{array}\right),\quad d_{\rm R}=\left(\begin{array}{c}0\\d_{_\mathbb{C}}^{\dagger}\end{array}\right),\quad l_{\rm L}=\left(\begin{array}{c}L\\0\end{array}\right),\quad e_{\rm R}=\left(\begin{array}{c}0\\e_{_\mathbb{C}}^{\dagger}\end{array}\right),\quad
	N_{\rm R}=\left(\begin{array}{c}0\\N_{_\mathbb{C}}^{\dagger}\end{array}\right).\\
	&\bar{q}_{\rm{L}}=\left(0\,,\,Q^{\dagger} \right),\quad \bar{u}_{\rm{R}}=\left(u_{_\mathbb{C}}\,,\,0 \right),\quad \bar{d}_{\rm R}=\left(d_{_\mathbb{C}}\,,\,0\right),\quad \bar{l}_{\rm L}=\left(0\,,\,L^{\dagger}\right),\quad \bar{e}_{\rm R}=\left(e_{_\mathbb{C}}\,,\,0\right),\quad \bar{N}_{\rm R}=\left(N_{_\mathbb{C}}\,,\,0\right).
\end{align}
Since each four-component fermion is given a unique name here, for simplicity and consistency with other references, the subscripts $\rm{L}$ and $\rm{R}$ are omitted without causing any confusion.

\begin{table}
	\begin{align*}
		\begin{array}{cc|c|c|c|c|c}
			\hline
			\multicolumn{7}{c}{\text{Dim-5 operators}}\\
			\hline
			N & (n,\tilde{n}) & \text{Classes} & \mathcal{N}_{\text{type}} & \mathcal{N}_{\text{term}} & \mathcal{N}_{\text{operator}} & \text{Equations}\\
			\hline
			3 & (2,0) & F_{\rm{L}} \psi{}^2    +h.c. & 0+0+2+0 & 2 & n_f(n_f-1) & \multirow{1}*{(\ref{cl:FLpsiL2f})} \\
			\hline
			4 & (1,0) & \psi{}^2 \phi {}^2    +h.c. & 0+0+2+0 & 2 & n_f(n_f+1) & \multirow{1}*{(\ref{cl:psi2h2f})} \\
			\hline
			\multicolumn{2}{c|}{\text{Total}} & 4 & 0+0+4+0 & 4 & 2 n_f^2 & \\
			\hline\hline
			\multicolumn{7}{c}{\text{Dim-6 operators}}\\
			\hline
			N & (n,\tilde{n}) & \text{Classes} & \mathcal{N}_{\text{type}} & \mathcal{N}_{\text{term}} & \mathcal{N}_{\text{operator}} & \text{Equations}\\
			\hline
			4 & (2,0) & \psi^4    +h.c. & 4+2+0+2 & 14 & \frac{1}{6} n_f^2 (49 n_f^2 -1) & \multirow{1}*{(\ref{cl:psiL4f}-\ref{cl:psiL4l})} \\
			& & F_{\rm{L}} \psi{}^2 \phi    +h.c. & 4+0+0+0 & 4 & 4 n_f^2 & \multirow{1}*{(\ref{cl:FLpsiL2hf}-\ref{cl:FLpsiL2hl})} \\
			\cline{2-7}
			& (1,1) & \psi{}^2 \psi^\dagger{}^2 & 10+2+0+0 & 12 & \frac{1}{4} n_f^2 (41 n_f^2+6 n_f+1) & \multirow{1}*{(\ref{cl:psiL2psiR2f}-\ref{cl:psiL2psiR2l})} \\
			& & \psi \psi^\dagger \phi {}^2 D & 3+0+0+0 & 3 & 3 n_f^2 & \multirow{1}*{(\ref{cl:psiLpsiRh2Df}-\ref{cl:psiLpsiRh2Dl})} \\
			\hline
			5 & (1,0) & \psi{}^2 \phi {}^3    +h.c. & 2+0+0+0 & 2 & 2 n_f^2 & \multirow{1}*{(\ref{cl:psiL2h3f})} \\
			\hline
			\multicolumn{2}{c|}{\text{Total}} & 8 & 23+4+0+2 & 35 & \frac{1}{12} n_f^2 (221 n_f^2+18 n_f+109) & \\
			\hline\hline
			\multicolumn{7}{c}{\text{Dim-7 operators}}\\
			\hline
			N & (n,\tilde{n}) & \text{Classes} & \mathcal{N}_{\text{type}} & \mathcal{N}_{\text{term}} & \mathcal{N}_{\text{operator}} & \text{Equations}\\
			\hline
			4 & (3,0) & F_{\rm{L}}^2 \psi^2    +h.c. & 0+0+6+0 & 6 & 3 n_f (n_f+1) & \multirow{1}*{(\ref{cl:FL2psiL2f}-\ref{cl:FL2psiL2l})} \\
			\cline{2-7}
			& (2,1) & F_{\rm{L}}^2 \psi^\dagger{}^2    +h.c. & 0+0+6+0 & 6 & 3 n_f (n_f+1) & \multirow{1}*{(\ref{cl:FL2psiR2f}-\ref{cl:FL2psiR2l})} \\
			& & \psi^3 \psi^\dagger D   +h.c. & 0+4+20+0 & 24 & \frac{1}{3} n_f^2 (43 n_f^2-15 n_f+2) & \multirow{1}*{(\ref{cl:psiL3psiRDf}-\ref{cl:psiL3psiRDl})} \\
			& & F_{\rm{L}} \psi \psi^\dagger \phi D   +h.c. & 0+0+8+0 & 8 & 8n_f^2 & \multirow{1}*{(\ref{cl:FLpsiLpsiRhDf}-\ref{cl:FLpsiLpsiRhDl})} \\
			& & \psi{}^2 \phi {}^2 D^2   +h.c. & 0+0+4+0 & 6 & 4n_f^2 & \multirow{1}*{(\ref{cl:psiL2h2D2f}-\ref{cl:psiL2h2D2l})} \\
			\hline
			5 & (2,0) & \psi{}^4 \phi    +h.c. & 0+2+10+0 & 24 & 12 n_f^4 & \multirow{1}*{(\ref{cl:psiL4hf}-\ref{cl:psiL4hl})} \\
			& & F_{\rm{L}} \psi{}^2 \phi {}^2   +h.c. & 0+0+6+0 & 6 & 2 n_f (2 n_f-1) & \multirow{1}*{(\ref{cl:FLpsiL2h2f}-\ref{cl:FLpsiL2h2l})} \\
			\cline{2-7}
			& (1,1) &  \psi{}^2 \psi^\dagger{}^2 \phi & 0+4+22+0 & 30 & n_f^3 (23 n_f+3) & \multirow{1}*{(\ref{cl:psiL2psiR2hf}-\ref{cl:psiL2psiR2hl})} \\
			& & \psi \psi^\dagger \phi {}^3 D & 0+0+2+0 & 4 & 4 n_f^2 & \multirow{1}*{(\ref{cl:psiLpsiRh3Df})} \\
			\hline
			6 & (1,0) & \psi{}^2 \phi {}^4    +h.c. & 0+0+2+0 & 2 & n_f(n_f+1) & \multirow{1}*{(\ref{cl:psiL2h4f})} \\
			\hline
			\multicolumn{2}{c|}{\text{Total}} & 18 & 0+10+86+0 & 116 & \frac{1}{3} n_f (148 n_f^3-6 n_f^2+83 n_f+15) & \\
			\hline
		\end{array}
	\end{align*}
	\caption{The complete statistics of dimension 5, 6, 7 $\nu$SMEFT operators.
		$N$ in the leftmost column shows the number of particles. $(n,\tilde{n})$ are the numbers of $\epsilon$
		and $\tilde{\epsilon}$ in the Lorentz structure. $\mathcal{N}_{\rm type}$, $\mathcal{N}_{\rm term}$, and $\mathcal{N}_{\rm operator}$
		show the number of types, terms and Hermitian operators respectively (independent conjugates
		are counted), while the numbers under $\mathcal{N}_{\rm type}$ describe the sum of each possible $|\Delta L|$
		types/operators with $\mathcal{N}=\mathcal{N}(|\Delta L|=0)+\mathcal{N}(|\Delta L|=1)+\mathcal{N}(|\Delta L|=2)+\mathcal{N}(|\Delta L|=4)$. The links in the rightmost column refer to the list(s) of the terms in given
		classes.
	}
	\label{tab:sumdim567}
\end{table}

\begin{table}
	\begin{align*}
		\begin{array}{cc|c|c|c|c|c}
			\hline
			N & (n,\tilde{n}) & \text{Classes} & \mathcal{N}_{\text{type}} & \mathcal{N}_{\text{term}} & \mathcal{N}_{\text{operator}} & \text{Equations}\\
			\hline
			4 & (3,1) & \psi{}^4   D^2    +h.c. & 4+0+2+2 & 22 & \frac{1}{4} n_f (49 n_f^3-6 n_f^2+3 n_f+2) & \multirow{1}*{(\ref{cl:psiL4D2f}-\ref{cl:psiL4D2l})} \\
			& & F_{\rm{L}} \psi{}^2 \phi D^2    +h.c. & 4+0+0+0 & 8 & 8 n_f^2 & \multirow{1}*{(\ref{cl:FLpsiL2hD2f}-\ref{cl:FLpsiL2hD2l})} \\
			\cline{2-7}
			& (2,2) & F_{\rm{L}} F_{\rm{R}} \psi \psi^\dagger   D & 3+0+0+0 & 3 & 3 n_f^2 & \multirow{1}*{(\ref{cl:FLFRpsiLpsiRDf}-\ref{cl:FLFRpsiLpsiRDl})} \\
			& & \psi{}^2 \psi^\dagger{}^2   D^2 & 10+2+0+0 & 24 & \frac{1}{2} n_f^2(41 n_f^2+1) & \multirow{1}*{(\ref{cl:psiL2psiR2D2f}-\ref{cl:psiL2psiR2D2l})}\\
			& & F_{\rm{R}} \psi{}^2 \phi D^2 +h.c. & 4+0+0+0 & 4 & 4 n_f^2 & \multirow{1}*{(\ref{cl:FRpsiL2hD2f}-\ref{cl:FRpsiL2hD2l})}\\
			& & \psi \psi^\dagger \phi {}^2 D^3 & 3+0+0+0 & 4 & 4 n_f^2 & \multirow{1}*{(\ref{cl:psiLpsiRh2D3f}-\ref{cl:psiLpsiRh2D3l})}\\
			\hline
			5 & (3,0) & F_{\rm{L}} \psi{}^4    +h.c. & 10+4+0+2 & 50 & \frac{1}{4} n_f (133 n_f^3+2 n_f^2-n_f+2) & \multirow{1}*{(\ref{cl:FLpsiL4f}-\ref{cl:FLpsiL4l})} \\
			& & F_{\rm{L}}{}^2 \psi{}^2 \phi +h.c. & 8+0+0+0 & 12 & 12 n_f^2 & \multirow{1}*{(\ref{cl:FL2psiL2hf}-\ref{cl:FL2psiL2hl})}\\
			\cline{2-7}
			& (2,1) & F_{\rm{L}} \psi{}^2 \psi^\dagger{}^2 +h.c. & 42+12+0+0 & 58 & \frac{1}{2} n_f^2 (97 n_f^2-1) & \multirow{1}*{(\ref{cl:FLpsiL2psiR2f}-\ref{cl:FLpsiL2psiR2l})} \\
			& & F_{\rm{L}}{}^2 \psi^\dagger{}^2 \phi +h.c. & 8+0+0+0 & 8 & 8 n_f^2 & \multirow{1}*{(\ref{cl:FL2psiR2hf}-\ref{cl:FL2psiR2hl})}\\
			& & \psi{}^3 \psi^\dagger \phi D +h.c. & 24+6+0+2 & 108 & n_f^3 (87 n_f-1) & \multirow{1}*{(\ref{cl:psiL3psiRhDf}-\ref{cl:psiL3psiRhDl})}\\
			& & F_{\rm{L}} \psi \psi^\dagger \phi {}^2 D +h.c. & 12+0+0+0 & 16 & 16 n_f^2 & \multirow{1}*{(\ref{cl:FLpsiLpsiRh2Df}-\ref{cl:FLpsiLpsiRh2Dl})}\\
			& & \psi{}^2 \phi {}^3 D^2 +h.c. & 2+0+0+0 & 12 & 12 n_f^2 & \multirow{1}*{(\ref{cl:psiL2h3D2f})}\\
			\hline
			6 & (2,0) & \psi{}^4 \phi {}^2    +h.c. & 8+2+0+2 & 30 & \frac{5}{6} n_f^2(23 n_f^4+1) & \multirow{1}*{(\ref{cl:psiL4h2f}-\ref{cl:psiL4h2l})} \\
			& & F_{\rm{L}} \psi{}^2 \phi {}^3 +h.c. & 4+0+0+0 & 6 & 6 n_f^2 & \multirow{1}*{(\ref{cl:FLpsiL2h3f}-\ref{cl:FLpsiL2h3l})}\\
			\cline{2-7}
			& (1,1) & \psi{}^2 \psi^\dagger{}^2 \phi {}^2 & 16+4+0+2 & 28 & \frac{1}{4} n_f^2 (91 n_f^2+2 n_f+3) & \multirow{1}*{(\ref{cl:psiL2psiR2h2f}-\ref{cl:psiL2psiR2h2l})} \\
			& & \psi \psi^\dagger \phi {}^4 D & 3+0+0+0 & 3 & 3 n_f^2 & \multirow{1}*{(\ref{cl:psiLpsiRh4Df}-\ref{cl:psiLpsiRh4Dl})}\\
			\hline
			7 & (1,0) & \psi{}^2 \phi {}^5    +h.c. & 2+0+0+0 & 2 & 2 n_f^2 & \multirow{1}*{(\ref{cl:psiL2h5f})} \\
			\hline
			\multicolumn{2}{c|}{\text{Total}} & 31 & 167+30+2+10 & 398 & \frac{1}{12} n_f (2921 n_f^3-18 n_f^2+961 n_f+12) & \\
			\hline
		\end{array}
	\end{align*}
	\caption{The complete statistics of dimension 8 $\nu$SMEFT operators.
		The numbers under $\mathcal{N}_{\rm type}$ describe the sum of each possible $|\Delta L|$
		types/operators with $\mathcal{N}=\mathcal{N}(|\Delta L|=0)+\mathcal{N}(|\Delta L|=1)+\mathcal{N}(|\Delta L|=2)+\mathcal{N}(|\Delta L|=4)$.
	}
	\label{tab:sumdim8}
\end{table}

\begin{table}
	\begin{align*}
		\begin{array}{cc|c|c|c|c|c}
			\hline
			N & (n,\tilde{n}) & \text{Classes} & \mathcal{N}_{\text{type}} & \mathcal{N}_{\text{term}} & \mathcal{N}_{\text{operator}} & \text{Equations}\\
			\hline
			4 & (4,1) & F_{\rm{L}}{}^2 \psi{}^2   D^2    +h.c. & 0+6+0+0 & 12 & 6 n_f (n_f+1) & \multirow{1}*{(\ref{cl:FL2psiL2D2f}-\ref{cl:FL2psiL2D2l})} \\
			\cline{2-7}
			& (3,2) & F_{\rm{L}} F_{\rm{R}} \psi{}^2   D^2 +h.c. & 0+6+0+0 & 6 & 3 n_f (n_f+1) & \multirow{1}*{(\ref{cl:FLFRpsiL2D2f}-\ref{cl:FLFRpsiL2D2l})} \\
			& & F_{\rm{L}}{}^2 \psi^\dagger{}^2   D^2 +h.c. & 0+6+0+0 & 6 & 3 n_f (n_f+1) & \multirow{1}*{(\ref{cl:FL2psiR2D2f}-\ref{cl:FL2psiR2D2l})}\\
			& &  \psi{}^3 \psi^\dagger   D^3 +h.c. & 4+20+0+0 & 46 & \frac{2}{3} n_f^2 (43 n_f^2-1) & \multirow{1}*{(\ref{cl:psiL3psiRD3f}-\ref{cl:psiL3psiRD3l})}\\
			& & F_{\rm{L}} \psi \psi^\dagger \phi D^3 +h.c. & 0+8+0+0 & 16 & 16 n_f^2 & \multirow{1}*{(\ref{cl:FLpsiLpsiRhD3f}-\ref{cl:FLpsiLpsiRhD3l})}\\
			& & \psi{}^2 \phi {}^2 D^4 +h.c. & 0+4+0+0 & 8 & n_f (5 n_f+1) & \multirow{1}*{(\ref{cl:psiL2h2D4f}-\ref{cl:psiL2h2D4l})}\\
			\hline
			5 & (4,0) & F_{\rm{L}}{}^3 \psi{}^2    +h.c. & 0+10+0+0 & 16 & 4 n_f (2 n_f-1) & \multirow{1}*{(\ref{cl:FL3psiL2f}-\ref{cl:FL3psiL2l})} \\
			\cline{2-7}
			& (3,1) & F_{\rm{L}}{}^3 \psi^\dagger{}^2 +h.c. & 0+4+0+0 & 4 & 2 n_f (n_f+1) & \multirow{1}*{(\ref{cl:FL3psiR2f}-\ref{cl:FL3psiR2l})} \\
			& & F_{\rm{L}} \psi{}^3 \psi^\dagger   D +h.c. & 10+42+0+0 & 222 & \frac{2}{3} n_f^2(212 n_f^2+1) & \multirow{1}*{(\ref{cl:FLpsiL3psiRDf}-\ref{cl:FLpsiL3psiRDl})}\\
			& & F_{\rm{L}}{}^2 \psi \psi^\dagger \phi D +h.c. & 0+16+0+0 & 32 & 32 n_f^2 & \multirow{1}*{(\ref{cl:FL2psiLpsiRhDf}-\ref{cl:FL2psiLpsiRhDl})}\\
			& & \psi{}^4 \phi D^2 +h.c. & 2+10+0+0 & 120 & 60 n_f^4 & \multirow{1}*{(\ref{cl:psiL4hD2f}-\ref{cl:psiL4hD2l})}\\
			& & F_{\rm{L}} \psi{}^2 \phi {}^2 D^2 +h.c. & 0+8+0+0 & 42 & 2 n_f (14 n_f-1) & \multirow{1}*{(\ref{cl:FLpsiL2h2D2f}-\ref{cl:FLpsiL2h2D2l})}\\
			\cline{2-7}
			& (2,2) & F_{\rm{L}} F_{\rm{R}}{}^2 \psi{}^2 +h.c. & 0+12+0+0 & 12 & 6 n_f (n_f-1) & \multirow{1}*{(\ref{cl:FLFR2psiL2f}-\ref{cl:FLFR2psiL2l})} \\
			& & F_{\rm{R}} \psi{}^3 \psi^\dagger   D +h.c. & 10+42+0+0 & 166 & 2 n_f^3 (53 n_f-5) & \multirow{1}*{(\ref{cl:FRpsiL3psiRDf}-\ref{cl:FRpsiL3psiRDl})}\\
			& & F_{\rm{L}} F_{\rm{R}} \psi \psi^\dagger \phi D & 0+10+0+0 & 24 & 24 n_f^2 & \multirow{1}*{(\ref{cl:FLFRpsiLpsiRhDf}-\ref{cl:FLFRpsiLpsiRhDl})}\\
			& & \psi{}^2 \psi^\dagger{}^2 \phi D^2 & 4+22+0+0 & 210 & n_f^3 (161 n_f-3) & \multirow{1}*{(\ref{cl:psiL2psiR2hD2f}-\ref{cl:psiL2psiR2hD2l})}\\
			& & F_{\rm{R}} \psi{}^2 \phi {}^2 D^2 +h.c. & 0+8+0+0 & 24 & 4 n_f (4 n_f-1) & \multirow{1}*{(\ref{cl:FRpsiL2h2D2f}-\ref{cl:FRpsiL2h2D2l})}\\
			& & \psi \psi^\dagger \phi {}^3 D^3 & 0+2+0+0 & 20 & 20 n_f^2 & \multirow{1}*{(\ref{cl:psiLpsiRh3D3f})}\\
			\hline
			6 & (3,0) & \psi{}^6    +h.c. & 6+10+6+2 & 130 & \frac{1}{72} n_f^2 (1921 n_f^4-219 n_f^3-335 n_f^2+75 n_f-2) & \multirow{1}*{(\ref{cl:psiL6f}-\ref{cl:psiL6l})} \\
			& & F_{\rm{L}} \psi{}^4 \phi +h.c. & 6+26+0+0 & 110 & n_f^3 (53 n_f-9) & \multirow{1}*{(\ref{cl:FLpsiL4hf}-\ref{cl:FLpsiL4hl})}\\
			& & F_{\rm{L}}{}^2 \psi{}^2 \phi {}^2 +h.c. & 0+12+0+0 & 18 & 2 n_f (6 n_f+1) & \multirow{1}*{(\ref{cl:FL2psiL2h2f}-\ref{cl:FL2psiL2h2l})}\\
			\cline{2-7}
			& (2,1) & \psi{}^4 \psi^\dagger{}^2 +h.c. & 40+106+14+0 & 474 & \frac{1}{12} n_f^3 (2455 n_f^3+91 n_f^2-91 n_f+65) & \multirow{1}*{(\ref{cl:psiL4psiR2f}-\ref{cl:psiL4psiR2l})} \\
			& & F_{\rm{L}} \psi{}^2 \psi^\dagger{}^2 \phi +h.c. & 24+116+0+0 & 176 & 138 n_f^4 & \multirow{1}*{(\ref{cl:FLpsiL2psiR2hf}-\ref{cl:FLpsiL2psiR2hl})}\\
			& & F_{\rm{L}}{}^2 \psi^\dagger{}^2 \phi {}^2 +h.c. & 0+10+0+0 & 10 & 2 n_f (3 n_f+2) & \multirow{1}*{(\ref{cl:FL2psiR2h2f}-\ref{cl:FL2psiR2h2l})}\\
			& & \psi{}^3 \psi^\dagger \phi {}^2 D +h.c. & 10+44+0+0 & 268 & n_f^2 (181 n_f^2-7 n_f-2) & \multirow{1}*{(\ref{cl:psiL3psiRh2Df}-\ref{cl:psiL3psiRh2Dl})}\\
			& & F_{\rm{L}} \psi \psi^\dagger \phi {}^3 D +h.c. & 0+8+0+0 & 32 & 32 n_f^2 & \multirow{1}*{(\ref{cl:FLpsiLpsiRh3Df}-\ref{cl:FLpsiLpsiRh3Dl})}\\
			& & \psi{}^2 \phi {}^4 D^2 +h.c. & 0+4+0+0 & 20 & 2 n_f (7 n_f+1) & \multirow{1}*{(\ref{cl:psiL2h4D2f}-\ref{cl:psiL2h4D2l})}\\
			\hline
			7 & (2,0) & \psi{}^4 \phi {}^3    +h.c. & 2+12+0+0 & 28 & \frac{4}{3} n_f^2 (10 n_f^2-1) & \multirow{1}*{(\ref{cl:psiL4h3f}-\ref{cl:psiL4h3l})} \\
			& & F_{\rm{L}} \psi{}^2 \phi {}^4 +h.c. & 0+6+0+0 & 6 & 2 n_f (2 n_f-1) & \multirow{1}*{(\ref{cl:FLpsiL2h4f}-\ref{cl:FLpsiL2h4l})}\\
			\cline{2-7}
			& (1,1) & \psi{}^2 \psi^\dagger{}^2 \phi {}^3 & 4+22+0+0 & 34 & 2 n_f^3 (13 n_f+2) & \multirow{1}*{(\ref{cl:psiL2psiR2h3f}-\ref{cl:psiL2psiR2h3l})} \\
			& & \psi \psi^\dagger \phi {}^5 D & 0+2+0+0 & 4 & 4 n_f^2 & \multirow{1}*{(\ref{cl:psiLpsiRh5Df})}\\
			\hline
			8 & (1,0) & \psi{}^2 \phi {}^6    +h.c. & 0+2+0+0 & 2 & n_f (n_f+1) & \multirow{1}*{(\ref{cl:psiL2h6f})} \\
			\hline
			\multicolumn{2}{c|}{\multirow{2}*{\text{Total}}} & \multirow{2}*{59} & \multirow{2}*{122+616+20+2} & \multirow{2}*{2298} & \frac{1}{72} n_f (16651 n_f^5+327 n_f^4+64519 n_f^3 \\ & & & & & -1335 n_f^2+17182 n_f+432) \\
			\hline
		\end{array}
	\end{align*}
	\caption{The complete statistics of dimension 9 $\nu$SMEFT operators.
		The numbers under $\mathcal{N}_{\rm type}$ describe the sum of each possible $|\Delta L|$
		types/operators with $\mathcal{N}=\mathcal{N}(|\Delta L|=1)+\mathcal{N}(|\Delta L|=2)+\mathcal{N}(|\Delta L|=3)+\mathcal{N}(|\Delta L|=6)$.
	}
	\label{tab:sumdim9}
\end{table}

Here, we give some conversions between four-component spinors and two-component spinors for readers' convenience. For four-component bilinears formed by
\begin{align}
	\Psi=\left(\begin{array}{c} \xi_{\alpha}\\\chi^{\dagger\dot{\alpha}} \end{array} \right),\quad \bar{\Psi}=\Psi^{\dagger}\gamma^0=\left(\chi^{\alpha},\;\xi^{\dagger}_{\dot{\alpha}} \right)\;,
\end{align}
the following relations are useful to convert the four-component bilinears to the two-component spinors.
\eq{\label{eq:bilinear}
	\bar{\Psi}_1\Psi_2=&\chi_1^{\alpha}\xi_{2\alpha}+\xi^{\dagger}_{1\dot{\alpha}}\chi^{\dagger\dot{\alpha}}_2\;,\\
	\bar{\Psi}_1\gamma^{\mu}\Psi_2=&\chi_1^{\alpha}\sigma^{\mu}_{\alpha\dot{\alpha}}\chi^{\dagger\dot{\alpha}}_2+\xi^{\dagger}_{1\dot{\alpha}}\bar{\sigma}^{\mu\dot{\alpha}\alpha}\xi_{2\alpha}\;,\\
	\bar{\Psi}_1\sigma^{\mu\nu}\Psi_2=&\chi_1^{\alpha}\left(\sigma^{\mu\nu}\right)_\alpha{}^\beta\xi_{2\beta}+\xi^{\dagger}_{1\dot{\alpha}}\left(\bar{\sigma}^{\mu\nu}\right)^{\dot{\alpha}}{}_{\dot{\beta}}\chi^{\dagger\dot{\beta}}_{2}\;,\\
	\Psi^{\rm{T}}_1C\Psi_2=&\xi_1^{\alpha}\xi_{2\alpha}+\chi^{\dagger}_{1\dot{\alpha}}\chi^{\dagger\dot{\alpha}}_2\;,\\
	\Psi^{\rm{T}}_1C\gamma^{\mu}\Psi_2=&\xi_1^{\alpha}\sigma^{\mu}_{\alpha\dot{\alpha}}\chi^{\dagger\dot{\alpha}}_2+\chi^{\dagger}_{1\dot{\alpha}}\bar{\sigma}^{\mu\dot{\alpha}\alpha}\xi_{2\alpha}\;,\\
	\Psi^{\rm{T}}_1C\sigma^{\mu\nu}\Psi_2=&\xi_1^{\alpha}\left(\sigma^{\mu\nu}\right)_\alpha{}^\beta\xi_{2\beta}+\chi^{\dagger}_{1\dot{\alpha}}\left(\bar{\sigma}^{\mu\nu}\right)^{\dot{\alpha}}{}_{\dot{\beta}}\chi^{\dagger\dot{\beta}}_2\;,\\
	\bar{\Psi}_1C\bar{\Psi}_2^{\rm{T}}=&\xi^{\dagger}_{1\dot{\alpha}}\xi^{\dagger\dot{\alpha}}_2+\chi^{\alpha}_1\chi_{2\alpha}\;,\\
	\bar{\Psi}_1\gamma^{\mu}C\bar{\Psi}_2^{\rm{T}}=&\chi_1^{\alpha}\sigma^{\mu}_{\alpha\dot{\alpha}}\xi^{\dagger\dot{\alpha}}_2+\xi^{\dagger}_{1\dot{\alpha}}\bar{\sigma}^{\mu\dot{\alpha}\alpha}\chi_{2\alpha}\;,\\
	\bar{\Psi}_1\sigma^{\mu\nu}C\bar{\Psi}_2^{\rm{T}}=&\xi^{\dagger}_{1\dot{\alpha}}\left(\bar{\sigma}^{\mu\nu}\right)^{\dot{\alpha}}{}_{\dot{\beta}}\xi^{\dagger\dot{\beta}}_2+\chi^{\alpha}_1\left(\sigma^{\mu\nu}\right)_\alpha{}^\beta\chi_{2\beta}\;,
}
where $\gamma^{\mu}=\begin{pmatrix}
0&\sigma^{\mu}_{\alpha\dot{\beta}}\\\bar{\sigma}^{\mu\dot{\alpha}\beta}&0
\end{pmatrix}$ and $\sigma^{\mu\nu}=\dfrac{i}{2}[\gamma^\mu,\gamma^\nu]=\begin{pmatrix}
\left(\sigma^{\mu\nu}\right)_\alpha{}^\beta&0\\0&\left(\bar{\sigma}^{\mu\nu}\right)^{\dot{\alpha}}{}_{\dot{\beta}}
\end{pmatrix}$. In our notation, the Dirac charge conjugation matrix $C=i\gamma^0\gamma^2=\begin{pmatrix} \epsilon_{\alpha\beta}&0\\0&\epsilon^{\dot{\alpha}\dot{\beta}}\end{pmatrix}=\begin{pmatrix} -\epsilon^{\alpha\beta}&0\\0&-\epsilon_{\dot{\alpha}\dot{\beta}}\end{pmatrix}=-C^{-1}$ and the following relations are needed to compare with the literature, 
\begin{eqnarray}\label{eq:psiC}
    \Psi^c=C\overline{\Psi}^{\rm T}, \quad 
    \overline{\Psi^c} = -\Psi^{\rm T} C^{-1} = \Psi^{\rm T} C.
\end{eqnarray}
It should be noted that the transpose symbol $\rm{T}$ is omitted in our result. Here we give some examples of the conversion
\begin{eqnarray}
    \epsilon _{abc} \left(\overline{d}_{p}{}^{a} C \overline{N}_{s}\right) \left(\overline{d}_{r}{}^{b} C \overline{u}_{t}{}^{c}\right) &=& \epsilon _{abc} \left(d_{_\mathbb{C}}{}_{p}{}^{a} N_{_\mathbb{C}}{}_{s}\right) \left(d_{_\mathbb{C}}{}_{r}{}^{b} u_{_\mathbb{C}}{}_{t}{}^{c}\right), \\
    i H_{i} D^{\mu } H^{\dagger}{}^{i} \left(\overline{N}_{p} \gamma _{\mu } N_{r}\right) &=& i H_{i} D^{\mu } H^{\dagger}{}^{i} \left(N_{_\mathbb{C}}{}_{p} \sigma _{\mu } N^{\dagger}_{_\mathbb{C}}{}_{r}\right), \\
    i B_{\rm{L}}{}^{\mu }{}^{\nu } \left(\overline{N}_{p} \sigma _{\mu }{}_{\nu } C \overline{N}_{r}\right) &=& i B_{\rm{L}}{}^{\mu }{}^{\nu } \left(N_{_\mathbb{C}}{}_{p} \sigma _{\mu }{}_{\nu } N_{_\mathbb{C}}{}_{r}\right).
\end{eqnarray}
The Hermitian conjugates of the fermion chains are
\begin{eqnarray}\label{eq:hermit1}
    \left[\bar\Psi_1\Psi_2 \right]^{\dagger}=\bar\Psi_2\Psi_1,\qquad \left[\bar\Psi_1\gamma^{\mu}\Psi_2 \right]^{\dagger}=&\bar\Psi_2\gamma^{\mu}\Psi_1,\qquad \left[\bar\Psi_1\sigma^{\mu\nu}\Psi_2 \right]^{\dagger}=\bar\Psi_2\sigma^{\mu\nu}\Psi_1,\\
    \left[\Psi_1C\Psi_2 \right]^{\dagger}=\bar\Psi_2C\bar\Psi_1,\quad \left[\Psi_1C\gamma^{\mu}\Psi_2 \right]^{\dagger}=&\bar\Psi_2\gamma^{\mu}C\bar\Psi_1,\quad \left[\Psi_1C\sigma^{\mu\nu}\Psi_2 \right]^{\dagger}=\bar\Psi_2\sigma^{\mu\nu}C\bar\Psi_1.\label{eq:hermit2}
\end{eqnarray}

In the following content that lists the operator basis, we should clarify that each $\psi$ ($\psi^{\dagger}$) in an operator class means a two-component left-handed (right handed) spinor in this class. The baryon number and lepton number violation pattern of each operator type are presented next to the type as $(\Delta B, \Delta L)$. The subscripts and superscripts $\{p,r,s,t,u,v\}$, $\{i,j,k,l,m,n\}$, $\{I,J,K,L\}$, $\{a,b,c,d,e,f\}$ and $\{A,B,C,D\}$ denote flavor indices, $SU(2)_W$ group (anti)fundamental representation indices, $SU(2)_W$ group adjoint representation indices, $SU(3)_C$ group (anti)fundamental representation indices and $SU(3)_C$ group adjoint representation indices respectively.

\subsection{Lists of the Dim-5 Operators}

\indent \indent \underline{Class $F_{\rm{L}} \psi{}^2    $}: 1 type

\begin{align}

&\begin{array}{c|l}\label{cl:FLpsiL2f}

\mathcal{O}_{B_{\rm{L}} \bar{N}^2    }(0,-2)

&\mathcal{Y}\left[\tiny{\young(p,r)}\right]i B_{\rm{L}}{}^{\mu }{}^{\nu } \left(\overline{N}_{p} \sigma _{\mu }{}_{\nu } C \overline{N}_{r}\right)

\end{array}

\end{align}

\underline{Class $  \psi{}^2 \phi {}^2  $}: 1 type

\begin{align}

&\begin{array}{c|l}\label{cl:psi2h2f}

\mathcal{O}_{  \bar{N}^2 H H^{\dagger}   }(0,-2)

&\mathcal{Y}\left[\tiny{\young(pr)}\right]H_{i} H^{\dagger}{}^{i} \left(\overline{N}_{p} C \overline{N}_{r}\right)

\end{array}

\end{align}
These operators contribute to the sterile neutrino masses and right-handed neutrino magnetic moments, listed and discussed in Ref.~\cite{delAguila:2008ir,Aparici:2009fh}. Here $(0,-2)$ denotes the baryon and lepton numbers and each of the $\psi$s or $\psi^{\dagger}$s in the class name means a left-handed two-component spinor or a right-handed two-component spinor in the class respectively.

We take these two operators as an example to discuss the connection between our results and those in other literature. In Ref.~\cite{delAguila:2008ir}, the above two operators are written as $\left(\overline{N} \sigma^{\mu\nu} N^c\right) B_{\mu\nu}$ and $\left(\overline{N} N^c\right) \left(\phi^{\dagger} \phi\right)$. These operators are exactly the same as our result after some conversions between notations. For gauge bosons contracting with $\sigma$ matrices, the following relations are useful,
\begin{eqnarray}\label{eq:FLRtilde1}
&&F_{\rm L}{}^{\mu\nu} \left(\sigma_{\mu\nu}\right)_{\alpha}{}^{\beta}=F^{\mu\nu} \left(\sigma_{\mu\nu}\right)_{\alpha}{}^{\beta}=-i\tilde{F}^{\mu\nu} \left(\sigma_{\mu\nu}\right)_{\alpha}{}^{\beta}, \quad F_{\rm R}{}^{\mu\nu} \left(\sigma_{\mu\nu}\right)_{\alpha}{}^{\beta}=0, \\
&&F_{\rm{R}}{}^{\mu\nu}\left(\bar{\sigma}_{\mu\nu}\right)^{\dot{\alpha}}{}_{\dot{\beta}}=F^{\mu\nu}\left(\bar{\sigma}_{\mu\nu}\right)^{\dot{\alpha}}{}_{\dot{\beta}}=i\tilde{F}^{\mu\nu}\left(\bar{\sigma}_{\mu\nu}\right)^{\dot{\alpha}}{}_{\dot{\beta}}, \quad F_{\rm{L}}{}^{\mu\nu}\left(\bar{\sigma}_{\mu\nu}\right)^{\dot{\alpha}}{}_{\dot{\beta}}=0,\label{eq:FLRtilde2}
\end{eqnarray}
where $F$ can be the field strength of any gauge boson in the $\nu$SMEFT or $\nu$LEFT and $F_{\rm L/R}=\dfrac{1}{2} \left(F \mp i\tilde{F}\right)$.
Taking account of eq.~(\ref{eq:psiC}) and eq.~(\ref{eq:FLRtilde1}), eq.~(\ref{cl:FLpsiL2f}) and eq.~(\ref{cl:psi2h2f}) become
\begin{eqnarray}\label{eq:BN2}
    &&\mathcal{Y}\left[\tiny{\young(p,r)}\right]i B^{\mu }{}^{\nu } \left(\overline{N}_{p} \sigma _{\mu }{}_{\nu } N^c_{r}\right), \\
    &&\mathcal{Y}\left[\tiny{\young(pr)}\right]H_{i} H^{\dagger}{}^{i} \left(\overline{N}_{p} N^c_{r}\right).\label{eq:N2H2}
\end{eqnarray}
Now we can see that these two operators are the same operators as these in Ref.~\cite{delAguila:2008ir}. Besides, we also explicitly show the flavor symmetries of the operators. For example, the flavor structure of the two right-handed neutrinos in eq.~(\ref{eq:BN2}) must be antisymmetric, while the flavor structure of the two right-handed neutrinos in eq.~(\ref{eq:N2H2}) must be symmetric.

\comment{ With these conversions, eq.~(\ref{cl:FLpsiL2f}) and eq.~(\ref{cl:psi2h2f}) can be expressed with two-component spinors as
\begin{eqnarray}
    &&\mathcal{Y}\left[\tiny{\young(p,r)}\right]i B_{\rm L}{}^{\mu }{}^{\nu } \left[N_{_\mathbb{C}}{}_{p}{}^{\alpha} \left(\sigma _{\mu }{}_{\nu }\right)_{\alpha}{}^{\beta} N_{_\mathbb{C}}{}_{r}_{\beta}\right], \\
    &&\mathcal{Y}\left[\tiny{\young(pr)}\right]H_{i} H^{\dagger}{}^{i} \left(N_{_\mathbb{C}}{}_{p}{}^{\alpha} N_{_\mathbb{C}}{}_{r}{}_{\alpha}\right).
\end{eqnarray}
}

\subsection{Lists of the Dim-6 Operators}

These operators were listed and discussed in Ref.~\cite{Bhattacharya:2015vja,Liao:2016qyd}. We will take type $\bar{d}^2 \bar{N} \bar{u}$ as an example and compare our result with that of Ref.~\cite{Liao:2016qyd}. To do that, first we find the Hermitian conjugate of operators eq.~(\ref{ty:bd2bNbu}), which are
\begin{eqnarray}\label{eq:bd2bNbu1}
          \epsilon ^{abc} \left(N_{s} C d_{p}{}_{a}\right) \left(u_{t}{}_{c} C d_{r}{}_{b}\right)=\left[\epsilon _{abc} \left(\overline{d}_{p}{}^{a} C \overline{N}_{s}\right) \left(\overline{d}_{r}{}^{b} C \overline{u}_{t}{}^{c}\right)\right]^{\dagger},\\
		\epsilon ^{abc} \left(d_{r}{}_{b} C d_{p}{}_{a}\right) \left(u_{t}{}_{c} C N_{s}\right)=\left[\epsilon _{abc} \left(\overline{d}_{p}{}^{a} C \overline{d}_{r}{}^{b}\right) \left(\overline{N}_{s} C \overline{u}_{t}{}^{c}\right)\right]^{\dagger},\label{eq:bd2bNbu2}
\end{eqnarray}
derived from eq.~\eqref{eq:hermit2}.
Afterwards, taking account of the flavor structure of eq.~(\ref{eq:bd2bNbu2}), we find
\begin{eqnarray}
    \epsilon ^{abc} \left(d_{r}{}_{b} C d_{p}{}_{a}\right) \left(u_{t}{}_{c} C N_{s}\right) &=& -\epsilon ^{abc} \left(N_{s} C d_{p}{}_{a}\right) \left(u_{t}{}_{c} C d_{r}{}_{b}\right) + \epsilon ^{abc} \left(N_{s} C d_{r}{}_{a}\right) \left(u_{t}{}_{c} C d_{p}{}_{b}\right) \\
    &=& - \mathcal{Y}\left[\tiny{\young(p,r)}\right]\epsilon ^{abc} \left(N_{s} C d_{p}{}_{a}\right) \left(u_{t}{}_{c} C d_{r}{}_{b}\right) \\
    &\propto& \mathcal{Y}\left[\tiny{\young(p,r)}\right]\epsilon ^{abc} \left(d_{r}{}_{b} C d_{p}{}_{a}\right) \left(u_{t}{}_{c} C N_{s}\right),
\end{eqnarray}
where we used Schouten identity in the first equality. After these conversions, we can write our result as
\begin{eqnarray}
    &&\mathcal{Y}\left[\tiny{\young(pr)}\right]\epsilon ^{abc} \left(N_{s} C d_{p}{}_{a}\right) \left(u_{t}{}_{c} C d_{r}{}_{b}\right), \\
    &&\mathcal{Y}\left[\tiny{\young(p,r)}\right]\epsilon ^{abc} \left(N_{s} C d_{p}{}_{a}\right) \left(u_{t}{}_{c} C d_{r}{}_{b}\right),
\end{eqnarray}
which can be reorganized into one term $\epsilon ^{abc} \left(N_{s} C d_{p}{}_{a}\right) \left(u_{t}{}_{c} C d_{r}{}_{b}\right)$ and is the same as operator $\mathcal{O}_{uddN}$ in Ref.~\cite{Liao:2016qyd}.\\

\indent \underline{Class $  \psi{}^4    $}: 4 types

\begin{align}
	
	&
\begin{array}{c|l}\label{cl:psiL4f}
		
		\multirow{2}*{$\mathcal{O}_{  \bar{d} l \bar{N} q    }^{(1,2)}(0,0)$}
		
		& \epsilon ^{ij} \left(\overline{N}_{s} q_{ta j}\right) \left(\overline{d}_{p}{}^{a} l_{ri}\right)
		
		\\& \epsilon ^{ij} \left(\overline{d}_{p}{}^{a} C \overline{N}_{s}\right) \left(l_{ri} C q_{ta j}\right)
		
	\end{array}\\
	
	&\begin{array}{c|l}\label{ty:bd2bNbu}
		
		\multirow{2}*{$\mathcal{O}_{  \bar{d}^2 \bar{N} \bar{u}    }^{(1,2)}(-1,-1)$}
		
		&\mathcal{Y}\left[\tiny{\young(pr)}\right]\epsilon _{abc} \left(\overline{d}_{p}{}^{a} C \overline{N}_{s}\right) \left(\overline{d}_{r}{}^{b} C \overline{u}_{t}{}^{c}\right)
		
		\\&\mathcal{Y}\left[\tiny{\young(p,r)}\right]\epsilon _{abc} \left(\overline{d}_{p}{}^{a} C \overline{d}_{r}{}^{b}\right) \left(\overline{N}_{s} C \overline{u}_{t}{}^{c}\right)
		
	\end{array}\\
	
	&\begin{array}{c|l}
		
		\multirow{2}*{$\mathcal{O}_{  \bar{e} l^2 \bar{N}    }^{(1,2)}(0,0)$}
		
		&\mathcal{Y}\left[\tiny{\young(rs)}\right]\epsilon ^{ij} \left(\overline{e}_{p} l_{ri}\right) \left(\overline{N}_{t} l_{sj}\right)
		
		\\&\mathcal{Y}\left[\tiny{\young(r,s)}\right]\epsilon ^{ij} \left(\overline{e}_{p} l_{ri}\right) \left(\overline{N}_{t} l_{sj}\right)
		
	\end{array}\\
	
	&\begin{array}{c|l}
		
		\mathcal{O}_{  \bar{N}^4    }(0,-4)
		
		&\mathcal{Y}\left[\tiny{\young(pr,st)}\right]\left(\overline{N}_{p} C \overline{N}_{r}\right) \left(\overline{N}_{s} C \overline{N}_{t}\right)
		
	\end{array}\label{cl:psiL4l}
	
\end{align}

\underline{Class $F_{\rm{L}} \psi{}^2 \phi  $}: 2 types

\begin{align}
	
	&\begin{array}{c|l}\label{cl:FLpsiL2hf}
		
		\mathcal{O}_{W_{\rm{L}} l \bar{N} H  }(0,0)
		
		& i \left(\tau ^I\right)_k^i \epsilon ^{jk} H_{j} W_{\rm{L}}^{I}{}^{\mu }{}^{\nu } \left(\overline{N}_{r} \sigma _{\mu }{}_{\nu } l_{pi}\right)
		
	\end{array}\\
	
	&\begin{array}{c|l}
		
		\mathcal{O}_{B_{\rm{L}} l \bar{N} H  }(0,0)
		
		& i \epsilon ^{ij} H_{j} B_{\rm{L}}{}^{\mu }{}^{\nu } \left(\overline{N}_{r} \sigma _{\mu }{}_{\nu } l_{pi}\right)
		
	\end{array}\label{cl:FLpsiL2hl}
	
\end{align}

\underline{Class $  \psi{}^2 \psi^\dagger{}^2    $}: 9 types

\begin{align}
	
	&\begin{array}{c|l}\label{cl:psiL2psiR2f}
		
		\mathcal{O}_{  d  N  q^2    }(1,1)
		
		&\mathcal{Y}\left[\tiny{\young(pr)}\right]\epsilon ^{abc} \epsilon ^{ij} \left(d_{sc} C N_{t}\right) \left(q_{pa i} C q_{rb j}\right)
		
	\end{array}\\
	
	&\begin{array}{c|l}
		
		\mathcal{O}_{  \bar{l}  N  q \bar{u}    }(0,0)
		
		& \left(\overline{l}_{s}{}^{i} N_{t}\right) \left(\overline{u}_{r}{}^{a} q_{pa i}\right)
		
	\end{array}\\
	
	&\begin{array}{c|l}
		
		\mathcal{O}_{  \bar{N} N  q \bar{q}     }(0,0)
		
		& \left(\overline{N}_{p} q_{ra i}\right) \left(\overline{q}_{t}^{a i} N_{s}\right)
		
	\end{array}\\
	
	&\begin{array}{c|l}
		
		\mathcal{O}_{  d  \bar{e} N  \bar{u}    }(0,0)
		
		& \left(d_{sa} C N_{t}\right) \left(\overline{e}_{p} C \overline{u}_{r}{}^{a}\right)
		
	\end{array}\\
	
	&\begin{array}{c|l}
		
		\mathcal{O}_{  \bar{N} N  \bar{u} u     }(0,0)
		
		& \left(N_{s} C u_{ta}\right) \left(\overline{N}_{p} C \overline{u}_{r}{}^{a}\right)
		
	\end{array}\\
	
	&\begin{array}{c|l}
		
		\mathcal{O}_{  \bar{d} d  \bar{N} N     }(0,0)
		
		& \left(d_{sa} C N_{t}\right) \left(\overline{d}_{p}{}^{a} C \overline{N}_{r}\right)
		
	\end{array}\\
	
	&\begin{array}{c|l}
		
		\mathcal{O}_{  l \bar{l}  \bar{N} N     }(0,0)
		
		& \left(\overline{N}_{r} l_{pi}\right) \left(\overline{l}_{s}{}^{i} N_{t}\right)
		
	\end{array}\\
	
	&\begin{array}{c|l}
		
		\mathcal{O}_{  \bar{e} e  \bar{N} N     }(0,0)
		
		& \left(e_{s} C N_{t}\right) \left(\overline{e}_{p} C \overline{N}_{r}\right)
		
	\end{array}\\
	
	&\begin{array}{c|l}
		
		\mathcal{O}_{  \bar{N}^2 N ^2    }(0,0)
		
		&\mathcal{Y}\left[\tiny{\young(pr)},\tiny{\young(st)}\right]\left(N_{s} C N_{t}\right) \left(\overline{N}_{p} C \overline{N}_{r}\right)
		
	\end{array}\label{cl:psiL2psiR2l}
	
\end{align}

\underline{Class $  \psi \psi^\dagger \phi {}^2 D$}: 2 types

\begin{align}
	
	&\begin{array}{c|l}\label{cl:psiLpsiRh2Df}
		
		\mathcal{O}_{  \bar{e} N  H^{\dagger} {}^2 D}(0,0)
		
		& i \epsilon _{ij} H^{\dagger}{}^{i} D^{\mu } H^{\dagger}{}^{j} \left(\overline{e}_{p} \gamma _{\mu } N_{r}\right)
		
	\end{array}\\
	
	&\begin{array}{c|l}
		
		\mathcal{O}_{  \bar{N} N  H H^{\dagger}  D}(0,0)
		
		& i H_{i} D^{\mu } H^{\dagger}{}^{i} \left(\overline{N}_{p} \gamma _{\mu } N_{r}\right)
		
	\end{array}\label{cl:psiLpsiRh2Dl}
	
\end{align}

\underline{Class $  \psi{}^2 \phi {}^3  $}: 1 type

\begin{align}
	
	&\begin{array}{c|l}\label{cl:psiL2h3f}
		
		\mathcal{O}_{  l \bar{N} H{}^2 H^{\dagger}   }(0,0)
		
		& \epsilon ^{ik} H_{j} H_{k} H^{\dagger}{}^{j} \left(\overline{N}_{r} l_{pi}\right)
		
	\end{array}
	
\end{align}

\subsection{Lists of the Dim-7 Operators}

\indent \indent \underline{Class $F_{\rm{L}}{}^2 \psi{}^2    $}: 3 types

\begin{align}
	
	&\begin{array}{c|l}\label{cl:FL2psiL2f}
		
		\mathcal{O}_{G_{\rm{L}}{}^2 \bar{N}^2    }(0,-2)
		
		&\mathcal{Y}\left[\tiny{\young(pr)}\right]\left(\overline{N}_{p} C \overline{N}_{r}\right) G_{\rm{L}}^{A}{}_{\mu }{}_{\nu } G_{\rm{L}}^{A}{}^{\mu }{}^{\nu }
		
	\end{array}\\
	
	&\begin{array}{c|l}
		
		\mathcal{O}_{W_{\rm{L}}{}^2 \bar{N}^2    }(0,-2)
		
		&\mathcal{Y}\left[\tiny{\young(pr)}\right]\left(\overline{N}_{p} C \overline{N}_{r}\right) W_{\rm{L}}^{I}{}_{\mu }{}_{\nu } W_{\rm{L}}^{I}{}^{\mu }{}^{\nu }
		
	\end{array}\\
	
	&\begin{array}{c|l}
		
		\mathcal{O}_{B_{\rm{L}}{}^2 \bar{N}^2    }(0,-2)
		
		&\mathcal{Y}\left[\tiny{\young(pr)}\right]B_{\rm{L}}{}_{\mu }{}_{\nu } B_{\rm{L}}{}^{\mu }{}^{\nu } \left(\overline{N}_{p} C \overline{N}_{r}\right)
		
	\end{array}\label{cl:FL2psiL2l}
	
\end{align}

\underline{Class $F_{\rm{L}}{}^2 \psi^\dagger{}^2    $}: 3 types

\begin{align}
	
	&\begin{array}{c|l}\label{cl:FL2psiR2f}
		
		\mathcal{O}_{G_{\rm{L}}{}^2 N ^2    }(0,2)
		
		&\mathcal{Y}\left[\tiny{\young(pr)}\right]\left(N_{p} C N_{r}\right) G_{\rm{L}}^{A}{}_{\mu }{}_{\nu } G_{\rm{L}}^{A}{}^{\mu }{}^{\nu }
		
	\end{array}\\
	
	&\begin{array}{c|l}
		
		\mathcal{O}_{W_{\rm{L}}{}^2 N ^2    }(0,2)
		
		&\mathcal{Y}\left[\tiny{\young(pr)}\right]\left(N_{p} C N_{r}\right) W_{\rm{L}}^{I}{}_{\mu }{}_{\nu } W_{\rm{L}}^{I}{}^{\mu }{}^{\nu }
		
	\end{array}\\
	
	&\begin{array}{c|l}
		
		\mathcal{O}_{B_{\rm{L}}{}^2 N ^2    }(0,2)
		
		&\mathcal{Y}\left[\tiny{\young(pr)}\right]B_{\rm{L}}{}_{\mu }{}_{\nu } B_{\rm{L}}{}^{\mu }{}^{\nu } \left(N_{p} C N_{r}\right)
		
	\end{array}\label{cl:FL2psiR2l}
	
\end{align}

\underline{Class $  \psi{}^3 \psi^\dagger   D$}: 12 types

\begin{align}
	
	&\begin{array}{c|l}\label{cl:psiL3psiRDf}
		
		\mathcal{O}_{  d  \bar{N} q^2   D}(1,-1)
		
		&\mathcal{Y}\left[\tiny{\young(r,s)}\right]i \epsilon ^{abc} \epsilon ^{ij} \left(\overline{N}_{p} \gamma _{\mu } d_{tc}\right) \left(q_{ra i} C D^{\mu } q_{sb j}\right)
		
	\end{array}\\
	
	&\begin{array}{c|l}
		
		\mathcal{O}_{  \bar{l}  \bar{N} q \bar{u}   D}(0,-2)
		
		& i \left(D^{\mu } \overline{u}_{s}{}^{a} q_{ra i}\right) \left(\overline{N}_{p} \gamma _{\mu } C \overline{l}_{t}{}^{i}\right)
		
	\end{array}\\
	
	&\begin{array}{c|l}
		
		\mathcal{O}_{  \bar{d} l N  q   D}(0,2)
		
		& i \epsilon ^{ij} \left(\overline{d}_{p}{}^{a} \gamma _{\mu } N_{t}\right) \left(l_{ri} C D^{\mu } q_{sa j}\right)
		
	\end{array}\\
	
	&\begin{array}{c|l}
		
		\mathcal{O}_{  \bar{N}^2 q \bar{q}    D}(0,-2)
		
		&\mathcal{Y}\left[\tiny{\young(p,r)}\right]i \left(\overline{N}_{r} D^{\mu } q_{sa i}\right) \left(\overline{N}_{p} \gamma _{\mu } C \overline{q}_{t}^{a i}\right)
		
	\end{array}\\
	
	&\begin{array}{c|l}
		
		\mathcal{O}_{  \bar{d}^2 N  \bar{u}   D}(-1,1)
		
		&\mathcal{Y}\left[\tiny{\young(pr)}\right]i \epsilon _{abc} \left(\overline{d}_{p}{}^{a} \gamma _{\mu } N_{t}\right) \left(\overline{d}_{r}{}^{b} C D^{\mu } \overline{u}_{s}{}^{c}\right)
		
	\end{array}\\
	
	&\begin{array}{c|l}
		
		\mathcal{O}_{  d  \bar{e} \bar{N} \bar{u}   D}(0,-2)
		
		& i \left(\overline{e}_{p} \gamma _{\mu } d_{ta}\right) \left(\overline{N}_{r} C D^{\mu } \overline{u}_{s}{}^{a}\right)
		
	\end{array}\\
	
	&\begin{array}{c|l}
		
		\mathcal{O}_{  \bar{N}^2 \bar{u} u    D}(0,-2)
		
		&\mathcal{Y}\left[\tiny{\young(p,r)}\right]i \left(\overline{N}_{p} \gamma _{\mu } u_{ta}\right) \left(\overline{N}_{r} C D^{\mu } \overline{u}_{s}{}^{a}\right)
		
	\end{array}\\
	
	&\begin{array}{c|l}
		
		\mathcal{O}_{  \bar{d} d  \bar{N}^2   D}(0,-2)
		
		&\mathcal{Y}\left[\tiny{\young(r,s)}\right]i \left(\overline{d}_{p}{}^{a} \gamma _{\mu } d_{ta}\right) \left(\overline{N}_{r} C D^{\mu } \overline{N}_{s}\right)
		
	\end{array}\\
	
	&\begin{array}{c|l}
		
		\mathcal{O}_{  \bar{e} l^2 N    D}(0,2)
		
		&\mathcal{Y}\left[\tiny{\young(rs)}\right]i \epsilon ^{ij} \left(\overline{e}_{p} \gamma _{\mu } N_{t}\right) \left(l_{ri} C D^{\mu } l_{sj}\right)
		
	\end{array}\\
	
	&\begin{array}{c|l}
		
		\mathcal{O}_{  l \bar{l}  \bar{N}^2   D}(0,-2)
		
		&\mathcal{Y}\left[\tiny{\young(r,s)}\right]i \left(\overline{N}_{r} C D^{\mu } \overline{N}_{s}\right) \left(\overline{l}_{t}{}^{i} \gamma _{\mu } l_{pi}\right)
		
	\end{array}\\
	
	&\begin{array}{c|l}
		
		\mathcal{O}_{  \bar{e} e  \bar{N}^2   D}(0,-2)
		
		&\mathcal{Y}\left[\tiny{\young(r,s)}\right]i \left(\overline{e}_{p} \gamma _{\mu } e_{t}\right) \left(\overline{N}_{r} C D^{\mu } \overline{N}_{s}\right)
		
	\end{array}\\
	
	&\begin{array}{c|l}
		
		\mathcal{O}_{  \bar{N}^3 N    D}(0,-2)
		
		&\mathcal{Y}\left[\tiny{\young(p,r,s)}\right]i \left(\overline{N}_{p} \gamma _{\mu } N_{t}\right) \left(\overline{N}_{r} C D^{\mu } \overline{N}_{s}\right)
		
	\end{array}\label{cl:psiL3psiRDl}
	
\end{align}

\underline{Class $F_{\rm{L}} \psi \psi^\dagger \phi D$}: 4 types

\begin{align}
	
	&\begin{array}{c|l}\label{cl:FLpsiLpsiRhDf}
		
		\mathcal{O}_{W_{\rm{L}} l N  H D}(0,2)
		
		& i \left(\tau ^I\right)_k^i \epsilon ^{jk} D_{\mu } H_{j} W_{\rm{L}}^{I}{}^{\mu }{}{}_{\nu } \left(l_{pi} C \gamma ^{\nu } N_{r}\right)
		
	\end{array}\\
	
	&\begin{array}{c|l}
		
		\mathcal{O}_{W_{\rm{L}} \bar{l}  \bar{N} H^{\dagger}  D}(0,-2)
		
		& i \epsilon _{jk} \left(\tau ^I\right)_i^k D_{\mu } H^{\dagger}{}^{i} W_{\rm{L}}^{I}{}^{\mu }{}{}_{\nu } \left(\overline{N}_{p} \gamma ^{\nu } C \overline{l}_{r}{}^{j}\right)
		
	\end{array}\\
	
	&\begin{array}{c|l}
		
		\mathcal{O}_{B_{\rm{L}} l N  H D}(0,2)
		
		& i \epsilon ^{ij} B_{\rm{L}}{}^{\mu }{}{}_{\nu } D_{\mu } H_{j} \left(l_{pi} C \gamma ^{\nu } N_{r}\right)
		
	\end{array}\\
	
	&\begin{array}{c|l}
		
		\mathcal{O}_{B_{\rm{L}} \bar{l}  \bar{N} H^{\dagger}  D}(0,-2)
		
		& i \epsilon _{ij} B_{\rm{L}}{}^{\mu }{}{}_{\nu } D_{\mu } H^{\dagger}{}^{i} \left(\overline{N}_{p} \gamma ^{\nu } C \overline{l}_{r}{}^{j}\right)
		
	\end{array}\label{cl:FLpsiLpsiRhDl}
	
\end{align}

\underline{Class $  \psi{}^2 \phi {}^2 D^2$}: 2 types

\begin{align}
	
	&\begin{array}{c|l}\label{cl:psiL2h2D2f}
		
		\mathcal{O}_{  \bar{e} \bar{N} H^{\dagger} {}^2 D^2}(0,-2)
		
		& i \epsilon _{ij} D_{\mu } H^{\dagger}{}^{i} D_{\nu } H^{\dagger}{}^{j} \left(\overline{e}_{p} \sigma ^{\mu }{}^{\nu } C \overline{N}_{r}\right)
		
	\end{array}\\
	
	&\begin{array}{c|l}
		
		\multirow{2}*{$\mathcal{O}_{  \bar{N}^2 H H^{\dagger}  D^2}^{(1,2)}(0,-2)$}
		
		&\mathcal{Y}\left[\tiny{\young(p,r)}\right]i D_{\mu } H_{i} D_{\nu } H^{\dagger}{}^{i} \left(\overline{N}_{p} \sigma ^{\mu }{}^{\nu } C \overline{N}_{r}\right)
		
		\\&\mathcal{Y}\left[\tiny{\young(pr)}\right]D_{\mu } H_{i} \left(\overline{N}_{p} C \overline{N}_{r}\right) D^{\mu } H^{\dagger}{}^{i}
		
	\end{array}\label{cl:psiL2h2D2l}
	
\end{align}

\underline{Class $  \psi{}^4 \phi  $}: 6 types

\begin{align}
	
	&\begin{array}{c|l}\label{cl:psiL4hf}
		
		\multirow{3}*{$\mathcal{O}_{  \bar{N} q^3 H^{\dagger}   }^{(1\sim 3)}(1,-1)$}
		
		&\mathcal{Y}\left[\tiny{\young(rst)}\right]\epsilon ^{abc} \epsilon ^{ik} H^{\dagger}{}^{j} \left(\overline{N}_{p} q_{ra i}\right) \left(q_{sb j} C q_{tc k}\right)
		
		\\&\mathcal{Y}\left[\tiny{\young(rs,t)}\right]\epsilon ^{abc} \epsilon ^{ik} H^{\dagger}{}^{j} \left(\overline{N}_{p} q_{ra i}\right) \left(q_{sb j} C q_{tc k}\right)
		
		\\&\mathcal{Y}\left[\tiny{\young(r,s,t)}\right]\epsilon ^{abc} \epsilon ^{ik} H^{\dagger}{}^{j} \left(\overline{N}_{p} q_{ra i}\right) \left(q_{sb j} C q_{tc k}\right)
		
	\end{array}\\
	
	&\begin{array}{c|l}
		
		\multirow{2}*{$\mathcal{O}_{  \bar{e} \bar{N} q \bar{u} H^{\dagger}   }^{(1,2)}(0,-2)$}
		
		& H^{\dagger}{}^{i} \left(\overline{e}_{p} C \overline{N}_{r}\right) \left(\overline{u}_{t}{}^{a} q_{sa i}\right)
		
		\\& H^{\dagger}{}^{i} \left(\overline{e}_{p} q_{sa i}\right) \left(\overline{N}_{r} C \overline{u}_{t}{}^{a}\right)
		
	\end{array}\\
	
	&\begin{array}{c|l}
		
		\multirow{2}*{$\mathcal{O}_{  \bar{N}^2 q \bar{u} H  }^{(1,2)}(0,-2)$}
		
		&\mathcal{Y}\left[\tiny{\young(p,r)}\right]\epsilon ^{ij} H_{j} \left(\overline{N}_{p} q_{sa i}\right) \left(\overline{N}_{r} C \overline{u}_{t}{}^{a}\right)
		
		\\&\mathcal{Y}\left[\tiny{\young(pr)}\right]\epsilon ^{ij} H_{j} \left(\overline{N}_{p} C \overline{N}_{r}\right) \left(\overline{u}_{t}{}^{a} q_{sa i}\right)
		
	\end{array}\\
	
	&\begin{array}{c|l}
		
		\multirow{2}*{$\mathcal{O}_{  \bar{d} \bar{N}^2 q H^{\dagger}   }^{(1,2)}(0,-2)$}
		
		&\mathcal{Y}\left[\tiny{\young(r,s)}\right]H^{\dagger}{}^{i} \left(\overline{N}_{s} q_{ta i}\right) \left(\overline{d}_{p}{}^{a} C \overline{N}_{r}\right)
		
		\\&\mathcal{Y}\left[\tiny{\young(rs)}\right]H^{\dagger}{}^{i} \left(\overline{N}_{s} q_{ta i}\right) \left(\overline{d}_{p}{}^{a} C \overline{N}_{r}\right)
		
	\end{array}\\
	
	&\begin{array}{c|l}
		
		\multirow{2}*{$\mathcal{O}_{  \bar{e} l \bar{N}^2 H^{\dagger}   }^{(1,2)}(0,-2)$}
		
		&\mathcal{Y}\left[\tiny{\young(s,t)}\right]H^{\dagger}{}^{i} \left(\overline{e}_{p} C \overline{N}_{s}\right) \left(\overline{N}_{t} l_{ri}\right)
		
		\\&\mathcal{Y}\left[\tiny{\young(st)}\right]H^{\dagger}{}^{i} \left(\overline{N}_{s} C \overline{N}_{t}\right) \left(\overline{e}_{p} l_{ri}\right)
		
	\end{array}\\
	
	&\begin{array}{c|l}
		
		\mathcal{O}_{  l \bar{N}^3 H  }(0,-2)
		
		&\mathcal{Y}\left[\tiny{\young(rs,t)}\right]\epsilon ^{ij} H_{j} \left(\overline{N}_{s} C \overline{N}_{t}\right) \left(\overline{N}_{r} l_{pi}\right)
		
	\end{array}\label{cl:psiL4hl}
	
\end{align}

\underline{Class $F_{\rm{L}} \psi{}^2 \phi {}^2  $}: 3 types

\begin{align}
	
	&\begin{array}{c|l}\label{cl:FLpsiL2h2f}
		
		\mathcal{O}_{W_{\rm{L}} \bar{e} \bar{N} H^{\dagger} {}^2  }(0,-2)
		
		& i \epsilon _{jk} \left(\tau ^I\right)_i^k H^{\dagger}{}^{i} H^{\dagger}{}^{j} W_{\rm{L}}^{I}{}^{\mu }{}^{\nu } \left(\overline{e}_{p} \sigma _{\mu }{}_{\nu } C \overline{N}_{r}\right)
		
	\end{array}\\
	
	&\begin{array}{c|l}
		
		\mathcal{O}_{W_{\rm{L}} \bar{N}^2 H H^{\dagger}   }(0,-2)
		
		&\mathcal{Y}\left[\tiny{\young(p,r)}\right]i \left(\tau ^I\right)_j^i H_{i} H^{\dagger}{}^{j} W_{\rm{L}}^{I}{}^{\mu }{}^{\nu } \left(\overline{N}_{p} \sigma _{\mu }{}_{\nu } C \overline{N}_{r}\right)
		
	\end{array}\\
	
	&\begin{array}{c|l}
		
		\mathcal{O}_{B_{\rm{L}} \bar{N}^2 H H^{\dagger}   }(0,-2)
		
		&\mathcal{Y}\left[\tiny{\young(p,r)}\right]i H_{i} H^{\dagger}{}^{i} B_{\rm{L}}{}^{\mu }{}^{\nu } \left(\overline{N}_{p} \sigma _{\mu }{}_{\nu } C \overline{N}_{r}\right)
		
	\end{array}\label{cl:FLpsiL2h2l}
	
\end{align}

\underline{Class $  \psi{}^2 \psi^\dagger{}^2 \phi  $}: 13 types

\begin{align}
	
	&\begin{array}{c|l}\label{cl:psiL2psiR2hf}
		
		\mathcal{O}_{  N ^2 q \bar{u} H  }(0,2)
		
		&\mathcal{Y}\left[\tiny{\young(st)}\right]\epsilon ^{ij} H_{j} \left(N_{s} C N_{t}\right) \left(\overline{u}_{r}{}^{a} q_{pa i}\right)
		
	\end{array}\\
	
	&\begin{array}{c|l}
		
		\mathcal{O}_{  \bar{d} e  N  q H  }(0,2)
		
		& \epsilon ^{ij} H_{j} \left(e_{s} C N_{t}\right) \left(\overline{d}_{p}{}^{a} q_{ra i}\right)
		
	\end{array}\\
	
	&\begin{array}{c|l}
		
		\mathcal{O}_{  \bar{d} N ^2 q H^{\dagger}   }(0,2)
		
		&\mathcal{Y}\left[\tiny{\young(st)}\right]H^{\dagger}{}^{i} \left(N_{s} C N_{t}\right) \left(\overline{d}_{p}{}^{a} q_{ra i}\right)
		
	\end{array}\\
	
	&\begin{array}{c|l}
		
		\multirow{2}*{$\mathcal{O}_{  l N  q \bar{q}  H  }^{(1,2)}(0,2)$}
		
		& \epsilon ^{ik} H_{k} \left(\overline{q}_{t}^{a j} N_{s}\right) \left(l_{pi} C q_{ra j}\right)
		
		\\& \epsilon ^{ij} H_{k} \left(\overline{q}_{t}^{a k} N_{s}\right) \left(l_{pi} C q_{ra j}\right)
		
	\end{array}\\
	
	&\begin{array}{c|l}
		
		\mathcal{O}_{  d ^2 \bar{N} q H  }(1,-1)
		
		&\mathcal{Y}\left[\tiny{\young(s,t)}\right]\epsilon ^{abc} \epsilon ^{ij} H_{j} \left(d_{sb} C d_{tc}\right) \left(\overline{N}_{p} q_{ra i}\right)
		
	\end{array}\\
	
	&\begin{array}{c|l}
		
		\mathcal{O}_{  d  \bar{N} q u  H^{\dagger}   }(1,-1)
		
		& \epsilon ^{abc} H^{\dagger}{}^{i} \left(\overline{N}_{p} q_{ra i}\right) \left(d_{sb} C u_{tc}\right)
		
	\end{array}\\
	
	&\begin{array}{c|l}
		
		\mathcal{O}_{  l N  \bar{u} u  H  }(0,2)
		
		& \epsilon ^{ij} H_{j} \left(N_{s} C u_{ta}\right) \left(\overline{u}_{r}{}^{a} l_{pi}\right)
		
	\end{array}\\
	
	&\begin{array}{c|l}
		
		\mathcal{O}_{  d  \bar{l}  \bar{N} \bar{u} H  }(0,-2)
		
		& H_{i} \left(\overline{N}_{p} C \overline{u}_{r}{}^{a}\right) \left(\overline{l}_{t}{}^{i} d_{sa}\right)
		
	\end{array}\\
	
	&\begin{array}{c|l}
		
		\mathcal{O}_{  \bar{d} d  l N  H  }(0,2)
		
		& \epsilon ^{ij} H_{j} \left(d_{sa} C N_{t}\right) \left(\overline{d}_{p}{}^{a} l_{ri}\right)
		
	\end{array}\\
	
	&\begin{array}{c|l}
		
		\multirow{2}*{$\mathcal{O}_{  l^2 \bar{l}  N  H  }^{(1,2)}(0,2)$}
		
		&\mathcal{Y}\left[\tiny{\young(pr)}\right]\epsilon ^{ik} H_{k} \left(l_{pi} C l_{rj}\right) \left(\overline{l}_{s}{}^{j} N_{t}\right)
		
		\\&\mathcal{Y}\left[\tiny{\young(p,r)}\right]\epsilon ^{ik} H_{k} \left(l_{pi} C l_{rj}\right) \left(\overline{l}_{s}{}^{j} N_{t}\right)
		
	\end{array}\\
	
	&\begin{array}{c|l}
		
		\mathcal{O}_{  \bar{e} e  l N  H  }(0,2)
		
		& \epsilon ^{ij} H_{j} \left(e_{s} C N_{t}\right) \left(\overline{e}_{p} l_{ri}\right)
		
	\end{array}\\
	
	&\begin{array}{c|l}
		
		\mathcal{O}_{  \bar{e} l N ^2 H^{\dagger}   }(0,2)
		
		&\mathcal{Y}\left[\tiny{\young(st)}\right]H^{\dagger}{}^{i} \left(N_{s} C N_{t}\right) \left(\overline{e}_{p} l_{ri}\right)
		
	\end{array}\\
	
	&\begin{array}{c|l}
		
		\mathcal{O}_{  l \bar{N} N ^2 H  }(0,2)
		
		&\mathcal{Y}\left[\tiny{\young(st)}\right]\epsilon ^{ij} H_{j} \left(N_{s} C N_{t}\right) \left(\overline{N}_{r} l_{pi}\right)
		
	\end{array}\label{cl:psiL2psiR2hl}
	
\end{align}

\underline{Class $  \psi \psi^\dagger \phi {}^3 D$}: 1 type

\begin{align}
	
	&\begin{array}{c|l}\label{cl:psiLpsiRh3Df}
		
		\multirow{2}*{$\mathcal{O}_{  l N  H{}^2 H^{\dagger}  D}^{(1,2)}(0,2)$}
		
		& i \epsilon ^{ik} H_{j} H_{k} D^{\mu } H^{\dagger}{}^{j} \left(l_{pi} C \gamma _{\mu } N_{r}\right)
		
		\\& i \epsilon ^{ik} H_{j} H^{\dagger}{}^{j} D^{\mu } H_{k} \left(l_{pi} C \gamma _{\mu } N_{r}\right)
		
	\end{array}
	
\end{align}

\underline{Class $  \psi{}^2 \phi {}^4  $}: 1 type

\begin{align}
	
	&\begin{array}{c|l}\label{cl:psiL2h4f}
		
		\mathcal{O}_{  \bar{N}^2 H{}^2 H^{\dagger} {}^2  }(0,-2)
		
		&\mathcal{Y}\left[\tiny{\young(pr)}\right]H_{i} H_{j} H^{\dagger}{}^{i} H^{\dagger}{}^{j} \left(\overline{N}_{p} C \overline{N}_{r}\right)
		
	\end{array}
	
\end{align}

\subsection{Lists of the Dim-8 Operators}

\indent \indent \underline{Class $  \psi{}^4   D^2$}: 4 types

\begin{align}
	
	&\begin{array}{c|l}\label{cl:psiL4D2f}
		
		\multirow{3}*{$\mathcal{O}_{  \bar{d} l \bar{N} q   D^2}^{(1\sim 3)}(0,0)$}
		
		& \epsilon ^{ij} \left(\overline{d}_{p}{}^{a} l_{ri}\right) \left(D_{\mu } \overline{N}_{s} D^{\mu } q_{ta j}\right)
		
		\\& \epsilon ^{ij} \left(\overline{d}_{p}{}^{a} C \overline{N}_{s}\right) \left(D_{\mu } l_{ri} C D^{\mu } q_{ta j}\right)
		
		\\& i \epsilon ^{ij} \left(D_{\mu } \overline{N}_{s} D_{\nu } q_{ta j}\right) \left(\overline{d}_{p}{}^{a} \sigma ^{\mu }{}^{\nu } l_{ri}\right)
		
	\end{array}\\
	
	&\begin{array}{c|l}
		
		\multirow{3}*{$\mathcal{O}_{  \bar{d}^2 \bar{N} \bar{u}   D^2}^{(1\sim 3)}(-1,-1)$}
		
		&\mathcal{Y}\left[\tiny{\young(pr)}\right]\epsilon _{abc} \left(\overline{d}_{p}{}^{a} C \overline{N}_{s}\right) \left(D_{\mu } \overline{d}_{r}{}^{b} C D^{\mu } \overline{u}_{t}{}^{c}\right)
		
		\\&\mathcal{Y}\left[\tiny{\young(p,r)}\right]\epsilon _{abc} \left(\overline{d}_{p}{}^{a} C \overline{d}_{r}{}^{b}\right) \left(D_{\mu } \overline{N}_{s} C D^{\mu } \overline{u}_{t}{}^{c}\right)
		
		\\&\mathcal{Y}\left[\tiny{\young(p,r)}\right]\epsilon _{abc} \left(\overline{d}_{p}{}^{a} C \overline{N}_{s}\right) \left(D_{\mu } \overline{d}_{r}{}^{b} C D^{\mu } \overline{u}_{t}{}^{c}\right)
		
	\end{array}\\
	
	&\begin{array}{c|l}
		
		\multirow{3}*{$\mathcal{O}_{  \bar{e} l^2 \bar{N}   D^2}^{(1\sim 3)}(0,0)$}
		
		&\mathcal{Y}\left[\tiny{\young(rs)}\right]\epsilon ^{ij} \left(\overline{e}_{p} l_{ri}\right) \left(D^{\mu } \overline{N}_{t} D_{\mu } l_{sj}\right)
		
		\\&\mathcal{Y}\left[\tiny{\young(r,s)}\right]\epsilon ^{ij} \left(\overline{e}_{p} l_{ri}\right) \left(D^{\mu } \overline{N}_{t} D_{\mu } l_{sj}\right)
		
		\\&\mathcal{Y}\left[\tiny{\young(r,s)}\right]i \epsilon ^{ij} \left(D_{\nu } \overline{N}_{t} D_{\mu } l_{sj}\right) \left(\overline{e}_{p} \sigma ^{\mu }{}^{\nu } l_{ri}\right)
		
	\end{array}\\
	
	&\begin{array}{c|l}
		
		\multirow{2}*{$\mathcal{O}_{  \bar{N}^4   D^2}^{(1,2)}(0,-4)$}
		
		&\mathcal{Y}\left[\tiny{\young(pr,st)}\right]\left(\overline{N}_{p} C \overline{N}_{r}\right) \left(D_{\mu } \overline{N}_{s} C D^{\mu } \overline{N}_{t}\right)
		
		\\&\mathcal{Y}\left[\tiny{\young(prst)}\right]\left(\overline{N}_{p} C \overline{N}_{r}\right) \left(D_{\mu } \overline{N}_{s} C D^{\mu } \overline{N}_{t}\right)
		
	\end{array}\label{cl:psiL4D2l}
	
\end{align}

\underline{Class $F_{\rm{L}} \psi{}^2 \phi D^2$}: 2 types

\begin{align}
	
	&\begin{array}{c|l}\label{cl:FLpsiL2hD2f}
		
		\multirow{2}*{$\mathcal{O}_{W_{\rm{L}} l \bar{N} H D^2}^{(1,2)}(0,0)$}
		
		& i \left(\tau ^I\right)_k^i \epsilon ^{jk} D_{\nu } H_{j} W_{\rm{L}}^{I}{}_{\lambda }{}^{\mu } \left(D_{\mu } \overline{N}_{r} \sigma ^{\lambda }{}^{\nu } l_{pi}\right)
		
		\\& \left(\tau ^I\right)_k^i \epsilon ^{jk} D_{\nu } H_{j} W_{\rm{L}}^{I}{}^{\mu }{}^{\nu } \left(D_{\mu } \overline{N}_{r} l_{pi}\right)
		
	\end{array}\\
	
	&\begin{array}{c|l}
		
		\multirow{2}*{$\mathcal{O}_{B_{\rm{L}} l \bar{N} H D^2}^{(1,2)}(0,0)$}
		
		& i \epsilon ^{ij} B_{\rm{L}}{}_{\lambda }{}{}^{\mu } D_{\nu } H_{j} \left(D_{\mu } \overline{N}_{r} \sigma ^{\lambda }{}^{\nu } l_{pi}\right)
		
		\\& \epsilon ^{ij} B_{\rm{L}}{}^{\mu }{}^{\nu } D_{\nu } H_{j} \left(D_{\mu } \overline{N}_{r} l_{pi}\right)
		
	\end{array}\label{cl:FLpsiL2hD2l}
	
\end{align}

\underline{Class $F_{\rm{L}} F_{\rm{R}} \psi \psi^\dagger   D$}: 3 types

\begin{align}
	
	&\begin{array}{c|l}\label{cl:FLFRpsiLpsiRDf}
		
		\mathcal{O}_{G_{\rm{L}} G_{\rm{R}} \bar{N} N    D}(0,0)
		
		& i G_{\rm{L}}^{A}{}^{\mu }{}{}_{\nu } G_{\rm{R}}^{A}{}^{\nu }{}{}_{\lambda } \left(\overline{N}_{p} \gamma ^{\lambda } D_{\mu } N_{r}\right)
		
	\end{array}\\
	
	&\begin{array}{c|l}
		
		\mathcal{O}_{W_{\rm{L}} W_{\rm{R}} \bar{N} N    D}(0,0)
		
		& i W_{\rm{L}}^{I}{}^{\mu }{}{}_{\nu } W_{\rm{R}}^{I}{}^{\nu }{}{}_{\lambda } \left(\overline{N}_{p} \gamma ^{\lambda } D_{\mu } N_{r}\right)
		
	\end{array}\\
	
	&\begin{array}{c|l}
		
		\mathcal{O}_{B_{\rm{L}} B_{\rm{R}} \bar{N} N    D}(0,0)
		
		& i B_{\rm{L}}{}^{\mu }{}{}_{\nu } B_{\rm{R}}{}^{\nu }{}{}_{\lambda } \left(\overline{N}_{p} \gamma ^{\lambda } D_{\mu } N_{r}\right)
		
	\end{array}\label{cl:FLFRpsiLpsiRDl}
	
\end{align}

\underline{Class $  \psi{}^2 \psi^\dagger{}^2   D^2$}: 9 types

\begin{align}
	
	&\begin{array}{c|l}\label{cl:psiL2psiR2D2f}
		
		\multirow{2}*{$\mathcal{O}_{  d  N  q^2   D^2}^{(1,2)}(1,1)$}
		
		&\mathcal{Y}\left[\tiny{\young(p,r)}\right]i \epsilon ^{abc} \epsilon ^{ij} \left(D_{\mu } d_{sc} C D_{\nu } N_{t}\right) \left(q_{pa i} C \sigma ^{\mu }{}^{\nu } q_{rb j}\right)
		
		\\&\mathcal{Y}\left[\tiny{\young(pr)}\right]\epsilon ^{abc} \epsilon ^{ij} \left(q_{pa i} C q_{rb j}\right) \left(D_{\mu } d_{sc} C D^{\mu } N_{t}\right)
		
	\end{array}\\
	
	&\begin{array}{c|l}
		
		\multirow{2}*{$\mathcal{O}_{  \bar{l}  N  q \bar{u}   D^2}^{(1,2)}(0,0)$}
		
		& \left(\overline{u}_{r}{}^{a} q_{pa i}\right) \left(D_{\mu } \overline{l}_{s}{}^{i} D^{\mu } N_{t}\right)
		
		\\& i \left(D_{\mu } \overline{l}_{s}{}^{i} D_{\nu } N_{t}\right) \left(\overline{u}_{r}{}^{a} \sigma ^{\mu }{}^{\nu } q_{pa i}\right)
		
	\end{array}\\
	
	&\begin{array}{c|l}
		
		\multirow{2}*{$\mathcal{O}_{  \bar{N} N  q \bar{q}    D^2}^{(1,2)}(0,0)$}
		
		& \left(\overline{N}_{p} q_{ra i}\right) \left(D^{\mu } \overline{q}_{t}^{a i} D_{\mu } N_{s}\right)
		
		\\& i \left(D_{\nu } \overline{q}_{t}^{a i} D_{\mu } N_{s}\right) \left(\overline{N}_{p} \sigma ^{\mu }{}^{\nu } q_{ra i}\right)
		
	\end{array}\\
	
	&\begin{array}{c|l}
		
		\multirow{2}*{$\mathcal{O}_{  d  \bar{e} N  \bar{u}   D^2}^{(1,2)}(0,0)$}
		
		& \left(\overline{e}_{p} C \overline{u}_{r}{}^{a}\right) \left(D_{\mu } d_{sa} C D^{\mu } N_{t}\right)
		
		\\& i \left(D_{\mu } d_{sa} C D_{\nu } N_{t}\right) \left(\overline{e}_{p} \sigma ^{\mu }{}^{\nu } C \overline{u}_{r}{}^{a}\right)
		
	\end{array}\\
	
	&\begin{array}{c|l}
		
		\multirow{2}*{$\mathcal{O}_{  \bar{N} N  \bar{u} u    D^2}^{(1,2)}(0,0)$}
		
		& \left(\overline{N}_{p} C \overline{u}_{r}{}^{a}\right) \left(D_{\mu } N_{s} C D^{\mu } u_{ta}\right)
		
		\\& i \left(D_{\mu } N_{s} C D_{\nu } u_{ta}\right) \left(\overline{N}_{p} \sigma ^{\mu }{}^{\nu } C \overline{u}_{r}{}^{a}\right)
		
	\end{array}\\
	
	&\begin{array}{c|l}
		
		\multirow{2}*{$\mathcal{O}_{  \bar{d} d  \bar{N} N    D^2}^{(1,2)}(0,0)$}
		
		& \left(\overline{d}_{p}{}^{a} C \overline{N}_{r}\right) \left(D_{\mu } d_{sa} C D^{\mu } N_{t}\right)
		
		\\& i \left(D_{\mu } d_{sa} C D_{\nu } N_{t}\right) \left(\overline{d}_{p}{}^{a} \sigma ^{\mu }{}^{\nu } C \overline{N}_{r}\right)
		
	\end{array}\\
	
	&\begin{array}{c|l}
		
		\multirow{2}*{$\mathcal{O}_{  l \bar{l}  \bar{N} N    D^2}^{(1,2)}(0,0)$}
		
		& \left(\overline{N}_{r} l_{pi}\right) \left(D_{\mu } \overline{l}_{s}{}^{i} D^{\mu } N_{t}\right)
		
		\\& i \left(D_{\mu } \overline{l}_{s}{}^{i} D_{\nu } N_{t}\right) \left(\overline{N}_{r} \sigma ^{\mu }{}^{\nu } l_{pi}\right)
		
	\end{array}\\
	
	&\begin{array}{c|l}
		
		\multirow{2}*{$\mathcal{O}_{  \bar{e} e  \bar{N} N    D^2}^{(1,2)}(0,0)$}
		
		& \left(\overline{e}_{p} C \overline{N}_{r}\right) \left(D_{\mu } e_{s} C D^{\mu } N_{t}\right)
		
		\\& i \left(D_{\mu } e_{s} C D_{\nu } N_{t}\right) \left(\overline{e}_{p} \sigma ^{\mu }{}^{\nu } C \overline{N}_{r}\right)
		
	\end{array}\\
	
	&\begin{array}{c|l}
		
		\multirow{2}*{$\mathcal{O}_{  \bar{N}^2 N ^2   D^2}^{(1,2)}(0,0)$}
		
		&\mathcal{Y}\left[\tiny{\young(p,r)},\tiny{\young(s,t)}\right]i \left(D_{\mu } N_{s} C D_{\nu } N_{t}\right) \left(\overline{N}_{p} \sigma ^{\mu }{}^{\nu } C \overline{N}_{r}\right)
		
		\\&\mathcal{Y}\left[\tiny{\young(pr)},\tiny{\young(st)}\right]\left(\overline{N}_{p} C \overline{N}_{r}\right) \left(D_{\mu } N_{s} C D^{\mu } N_{t}\right)
		
	\end{array}\label{cl:psiL2psiR2D2l}
	
\end{align}

\underline{Class $F_{\rm{R}} \psi{}^2 \phi D^2$}: 2 types

\begin{align}
	
	&\begin{array}{c|l}\label{cl:FRpsiL2hD2f}
		
		\mathcal{O}_{W_{\rm{R}} l \bar{N} H D^2}(0,0)
		
		& i \left(\tau ^I\right)_k^j \epsilon ^{ik} D_{\mu } D_{\nu } H_{j} W_{\rm{R}}^{I}{}^{\mu }{}{}_{\lambda } \left(\overline{N}_{r} \sigma ^{\nu }{}^{\lambda } l_{pi}\right)
		
	\end{array}\\
	
	&\begin{array}{c|l}
		
		\mathcal{O}_{B_{\rm{R}} l \bar{N} H D^2}(0,0)
		
		& i \epsilon ^{ij} B_{\rm{R}}{}^{\mu }{}{}_{\lambda } D_{\mu } D_{\nu } H_{j} \left(\overline{N}_{r} \sigma ^{\nu }{}^{\lambda } l_{pi}\right)
		
	\end{array}\label{cl:FRpsiL2hD2l}
	
\end{align}

\underline{Class $  \psi \psi^\dagger \phi {}^2 D^3$}: 2 types

\begin{align}
	
	&\begin{array}{c|l}\label{cl:psiLpsiRh2D3f}
		
		\mathcal{O}_{  \bar{e} N  H^{\dagger} {}^2 D^3}(0,0)
		
		& i \epsilon _{ij} H^{\dagger}{}^{i} D_{\mu } D_{\nu } H^{\dagger}{}^{j} \left(\overline{e}_{p} \gamma ^{\nu } D^{\mu } N_{r}\right)
		
	\end{array}\\
	
	&\begin{array}{c|l}
		
		\multirow{2}*{$\mathcal{O}_{  \bar{N} N  H H^{\dagger}  D^3}^{(1,2)}(0,0)$}
		
		& i H_{i} D_{\mu } D_{\nu } H^{\dagger}{}^{i} \left(\overline{N}_{p} \gamma ^{\nu } D^{\mu } N_{r}\right)
		
		\\& i D_{\nu } H_{i} D^{\mu } H^{\dagger}{}^{i} \left(\overline{N}_{p} \gamma _{\mu } D^{\nu } N_{r}\right)
		
	\end{array}\label{cl:psiLpsiRh2D3l}
	
\end{align}

\underline{Class $F_{\rm{L}} \psi{}^4    $}: 8 types

\begin{align}
	
	&\begin{array}{c|l}\label{cl:FLpsiL4f}
		
		\multirow{3}*{$\mathcal{O}_{G_{\rm{L}} \bar{d} l \bar{N} q    }^{(1\sim 3)}(0,0)$}
		
		& i \left(\lambda ^A\right)_a^b \epsilon ^{ij} G_{\rm{L}}^{A}{}_{\mu }{}_{\nu } \left(\overline{N}_{s} q_{tb j}\right) \left(\overline{d}_{p}{}^{a} \sigma ^{\mu }{}^{\nu } l_{ri}\right)
		
		\\& i \left(\lambda ^A\right)_a^b \epsilon ^{ij} G_{\rm{L}}^{A}{}^{\mu }{}^{\nu } \left(l_{ri} C q_{tb j}\right) \left(\overline{d}_{p}{}^{a} \sigma _{\mu }{}_{\nu } C \overline{N}_{s}\right)
		
		\\& i \left(\lambda ^A\right)_a^b \epsilon ^{ij} G_{\rm{L}}^{A}{}_{\mu }{}_{\nu } \left(\overline{d}_{p}{}^{a} q_{tb j}\right) \left(\overline{N}_{s} \sigma ^{\mu }{}^{\nu } l_{ri}\right)
		
	\end{array}\\
	
	&\begin{array}{c|l}
		
		\multirow{6}*{$\mathcal{O}_{G_{\rm{L}} \bar{d}^2 \bar{N} \bar{u}    }^{(1\sim 6)}(-1,-1)$}
		
		&\mathcal{Y}\left[\tiny{\young(p,r)}\right]i \epsilon _{ace} \left(\lambda ^A\right)_b^e G_{\rm{L}}^{A}{}_{\mu }{}_{\nu } \left(\overline{N}_{s} C \overline{u}_{t}{}^{c}\right) \left(\overline{d}_{p}{}^{a} \sigma ^{\mu }{}^{\nu } C \overline{d}_{r}{}^{b}\right)
		
		\\&\mathcal{Y}\left[\tiny{\young(p,r)}\right]i \epsilon _{ace} \left(\lambda ^A\right)_b^e G_{\rm{L}}^{A}{}^{\mu }{}^{\nu } \left(\overline{d}_{r}{}^{b} C \overline{u}_{t}{}^{c}\right) \left(\overline{d}_{p}{}^{a} \sigma _{\mu }{}_{\nu } C \overline{N}_{s}\right)
		
		\\&\mathcal{Y}\left[\tiny{\young(p,r)}\right]i \epsilon _{abe} \left(\lambda ^A\right)_c^e G_{\rm{L}}^{A}{}^{\mu }{}^{\nu } \left(\overline{d}_{r}{}^{b} C \overline{u}_{t}{}^{c}\right) \left(\overline{d}_{p}{}^{a} \sigma _{\mu }{}_{\nu } C \overline{N}_{s}\right)
		
		\\&\mathcal{Y}\left[\tiny{\young(pr)}\right]i \epsilon _{ace} \left(\lambda ^A\right)_b^e G_{\rm{L}}^{A}{}_{\mu }{}_{\nu } \left(\overline{N}_{s} C \overline{u}_{t}{}^{c}\right) \left(\overline{d}_{p}{}^{a} \sigma ^{\mu }{}^{\nu } C \overline{d}_{r}{}^{b}\right)
		
		\\&\mathcal{Y}\left[\tiny{\young(pr)}\right]i \epsilon _{ace} \left(\lambda ^A\right)_b^e G_{\rm{L}}^{A}{}^{\mu }{}^{\nu } \left(\overline{d}_{r}{}^{b} C \overline{u}_{t}{}^{c}\right) \left(\overline{d}_{p}{}^{a} \sigma _{\mu }{}_{\nu } C \overline{N}_{s}\right)
		
		\\&\mathcal{Y}\left[\tiny{\young(pr)}\right]i \epsilon _{abe} \left(\lambda ^A\right)_c^e G_{\rm{L}}^{A}{}^{\mu }{}^{\nu } \left(\overline{d}_{r}{}^{b} C \overline{u}_{t}{}^{c}\right) \left(\overline{d}_{p}{}^{a} \sigma _{\mu }{}_{\nu } C \overline{N}_{s}\right)
		
	\end{array}\\
	
	&\begin{array}{c|l}
		
		\multirow{3}*{$\mathcal{O}_{W_{\rm{L}} \bar{d} l \bar{N} q    }^{(1\sim 3)}(0,0)$}
		
		& i \left(\tau ^I\right)_k^i \epsilon ^{jk} W_{\rm{L}}^{I}{}_{\mu }{}_{\nu } \left(\overline{N}_{s} q_{ta j}\right) \left(\overline{d}_{p}{}^{a} \sigma ^{\mu }{}^{\nu } l_{ri}\right)
		
		\\& i \left(\tau ^I\right)_k^i \epsilon ^{jk} W_{\rm{L}}^{I}{}^{\mu }{}^{\nu } \left(l_{ri} C q_{ta j}\right) \left(\overline{d}_{p}{}^{a} \sigma _{\mu }{}_{\nu } C \overline{N}_{s}\right)
		
		\\& i \left(\tau ^I\right)_k^i \epsilon ^{jk} W_{\rm{L}}^{I}{}_{\mu }{}_{\nu } \left(\overline{d}_{p}{}^{a} q_{ta j}\right) \left(\overline{N}_{s} \sigma ^{\mu }{}^{\nu } l_{ri}\right)
		
	\end{array}\\
	
	&\begin{array}{c|l}
		
		\multirow{3}*{$\mathcal{O}_{W_{\rm{L}} \bar{e} l^2 \bar{N}    }^{(1\sim 3)}(0,0)$}
		
		&\mathcal{Y}\left[\tiny{\young(r,s)}\right]i \left(\tau ^I\right)_k^i \epsilon ^{jk} W_{\rm{L}}^{I}{}_{\mu }{}_{\nu } \left(\overline{N}_{t} l_{sj}\right) \left(\overline{e}_{p} \sigma ^{\mu }{}^{\nu } l_{ri}\right)
		
		\\&\mathcal{Y}\left[\tiny{\young(r,s)}\right]i \left(\tau ^I\right)_k^i \epsilon ^{jk} W_{\rm{L}}^{I}{}_{\mu }{}_{\nu } \left(\overline{e}_{p} C \overline{N}_{t}\right) \left(l_{ri} C \sigma ^{\mu }{}^{\nu } l_{sj}\right)
		
		\\&\mathcal{Y}\left[\tiny{\young(rs)}\right]i \left(\tau ^I\right)_k^i \epsilon ^{jk} W_{\rm{L}}^{I}{}_{\mu }{}_{\nu } \left(\overline{N}_{t} l_{sj}\right) \left(\overline{e}_{p} \sigma ^{\mu }{}^{\nu } l_{ri}\right)
		
	\end{array}\\
	
	&\begin{array}{c|l}
		
		\multirow{3}*{$\mathcal{O}_{B_{\rm{L}} \bar{d} l \bar{N} q    }^{(1\sim 3)}(0,0)$}
		
		& i \epsilon ^{ij} B_{\rm{L}}{}_{\mu }{}_{\nu } \left(\overline{N}_{s} q_{ta j}\right) \left(\overline{d}_{p}{}^{a} \sigma ^{\mu }{}^{\nu } l_{ri}\right)
		
		\\& i \epsilon ^{ij} B_{\rm{L}}{}^{\mu }{}^{\nu } \left(l_{ri} C q_{ta j}\right) \left(\overline{d}_{p}{}^{a} \sigma _{\mu }{}_{\nu } C \overline{N}_{s}\right)
		
		\\& i \epsilon ^{ij} B_{\rm{L}}{}_{\mu }{}_{\nu } \left(\overline{d}_{p}{}^{a} q_{ta j}\right) \left(\overline{N}_{s} \sigma ^{\mu }{}^{\nu } l_{ri}\right)
		
	\end{array}\\
	
	&\begin{array}{c|l}
		
		\multirow{3}*{$\mathcal{O}_{B_{\rm{L}} \bar{d}^2 \bar{N} \bar{u}    }^{(1\sim 3)}(-1,-1)$}
		
		&\mathcal{Y}\left[\tiny{\young(pr)}\right]i \epsilon _{abc} B_{\rm{L}}{}_{\mu }{}_{\nu } \left(\overline{N}_{s} C \overline{u}_{t}{}^{c}\right) \left(\overline{d}_{p}{}^{a} \sigma ^{\mu }{}^{\nu } C \overline{d}_{r}{}^{b}\right)
		
		\\&\mathcal{Y}\left[\tiny{\young(pr)}\right]i \epsilon _{abc} B_{\rm{L}}{}^{\mu }{}^{\nu } \left(\overline{d}_{r}{}^{b} C \overline{u}_{t}{}^{c}\right) \left(\overline{d}_{p}{}^{a} \sigma _{\mu }{}_{\nu } C \overline{N}_{s}\right)
		
		\\&\mathcal{Y}\left[\tiny{\young(p,r)}\right]i \epsilon _{abc} B_{\rm{L}}{}^{\mu }{}^{\nu } \left(\overline{d}_{r}{}^{b} C \overline{u}_{t}{}^{c}\right) \left(\overline{d}_{p}{}^{a} \sigma _{\mu }{}_{\nu } C \overline{N}_{s}\right)
		
	\end{array}\\
	
	&\begin{array}{c|l}
		
		\multirow{3}*{$\mathcal{O}_{B_{\rm{L}} \bar{e} l^2 \bar{N}    }^{(1\sim 3)}(0,0)$}
		
		&\mathcal{Y}\left[\tiny{\young(rs)}\right]i \epsilon ^{ij} B_{\rm{L}}{}_{\mu }{}_{\nu } \left(\overline{N}_{t} l_{sj}\right) \left(\overline{e}_{p} \sigma ^{\mu }{}^{\nu } l_{ri}\right)
		
		\\&\mathcal{Y}\left[\tiny{\young(rs)}\right]i \epsilon ^{ij} B_{\rm{L}}{}_{\mu }{}_{\nu } \left(\overline{e}_{p} C \overline{N}_{t}\right) \left(l_{ri} C \sigma ^{\mu }{}^{\nu } l_{sj}\right)
		
		\\&\mathcal{Y}\left[\tiny{\young(r,s)}\right]i \epsilon ^{ij} B_{\rm{L}}{}_{\mu }{}_{\nu } \left(\overline{N}_{t} l_{sj}\right) \left(\overline{e}_{p} \sigma ^{\mu }{}^{\nu } l_{ri}\right)
		
	\end{array}\\
	
	&\begin{array}{c|l}
		
		\mathcal{O}_{B_{\rm{L}} \bar{N}^4    }(0,-4)
		
		&\mathcal{Y}\left[\tiny{\young(pr,s,t)}\right]i B_{\rm{L}}{}^{\mu }{}^{\nu } \left(\overline{N}_{r} C \overline{N}_{t}\right) \left(\overline{N}_{p} \sigma _{\mu }{}_{\nu } C \overline{N}_{s}\right)
		
	\end{array}\label{cl:FLpsiL4l}
	
\end{align}

\underline{Class $F_{\rm{L}}{}^2 \psi{}^2 \phi  $}: 4 types

\begin{align}
	
	&\begin{array}{c|l}\label{cl:FL2psiL2hf}
		
		\mathcal{O}_{G_{\rm{L}}{}^2 l \bar{N} H  }(0,0)
		
		& \epsilon ^{ij} H_{j} G_{\rm{L}}^{A}{}_{\mu }{}_{\nu } G_{\rm{L}}^{A}{}^{\mu }{}^{\nu } \left(\overline{N}_{r} l_{pi}\right)
		
	\end{array}\\
	
	&\begin{array}{c|l}
		
		\multirow{2}*{$\mathcal{O}_{W_{\rm{L}}{}^2 l \bar{N} H  }^{(1,2)}(0,0)$}
		
		& \epsilon ^{ij} H_{j} W_{\rm{L}}^{I}{}_{\mu }{}_{\nu } W_{\rm{L}}^{I}{}^{\mu }{}^{\nu } \left(\overline{N}_{r} l_{pi}\right)
		
		\\& i \left(\tau ^K\right)_k^i \epsilon ^{IJK} \epsilon ^{jk} H_{j} W_{\rm{L}}^{I}{}_{\mu }{}_{\nu } W_{\rm{L}}^{J}{}^{\mu }{}{}_{\lambda } \left(\overline{N}_{r} \sigma ^{\nu }{}^{\lambda } l_{pi}\right)
		
	\end{array}\\
	
	&\begin{array}{c|l}
		
		\multirow{2}*{$\mathcal{O}_{B_{\rm{L}} W_{\rm{L}} l \bar{N} H  }^{(1,2)}(0,0)$}
		
		& \left(\tau ^I\right)_k^i \epsilon ^{jk} H_{j} B_{\rm{L}}{}_{\mu }{}_{\nu } W_{\rm{L}}^{I}{}^{\mu }{}^{\nu } \left(\overline{N}_{r} l_{pi}\right)
		
		\\& i \left(\tau ^I\right)_k^i \epsilon ^{jk} H_{j} B_{\rm{L}}{}_{\mu }{}_{\nu } W_{\rm{L}}^{I}{}^{\mu }{}{}_{\lambda } \left(\overline{N}_{r} \sigma ^{\nu }{}^{\lambda } l_{pi}\right)
		
	\end{array}\\
	
	&\begin{array}{c|l}
		
		\mathcal{O}_{B_{\rm{L}}{}^2 l \bar{N} H  }(0,0)
		
		& \epsilon ^{ij} H_{j} B_{\rm{L}}{}_{\mu }{}_{\nu } B_{\rm{L}}{}^{\mu }{}^{\nu } \left(\overline{N}_{r} l_{pi}\right)
		
	\end{array}\label{cl:FL2psiL2hl}
	
\end{align}

\underline{Class $F_{\rm{L}} \psi{}^2 \psi^\dagger{}^2    $}: 27 types

\begin{align}
	
	&\begin{array}{c|l}\label{cl:FLpsiL2psiR2f}
		
		\multirow{2}*{$\mathcal{O}_{G_{\rm{L}} d  N  q^2    }^{(1,2)}(1,1)$}
		
		&\mathcal{Y}\left[\tiny{\young(pr)}\right]i \left(\lambda ^A\right)_e^b \epsilon ^{ace} \epsilon ^{ij} G_{\rm{L}}^{A}{}_{\mu }{}_{\nu } \left(d_{sc} C N_{t}\right) \left(q_{pa i} C \sigma ^{\mu }{}^{\nu } q_{rb j}\right)
		
		\\&\mathcal{Y}\left[\tiny{\young(p,r)}\right]i \left(\lambda ^A\right)_e^b \epsilon ^{ace} \epsilon ^{ij} G_{\rm{L}}^{A}{}_{\mu }{}_{\nu } \left(d_{sc} C N_{t}\right) \left(q_{pa i} C \sigma ^{\mu }{}^{\nu } q_{rb j}\right)
		
	\end{array}\\
	
	&\begin{array}{c|l}
		
		\mathcal{O}_{G_{\rm{L}} \bar{l}  N  q \bar{u}    }(0,0)
		
		& i \left(\lambda ^A\right)_b^a G_{\rm{L}}^{A}{}_{\mu }{}_{\nu } \left(\overline{l}_{s}{}^{i} N_{t}\right) \left(\overline{u}_{r}{}^{b} \sigma ^{\mu }{}^{\nu } q_{pa i}\right)
		
	\end{array}\\
	
	&\begin{array}{c|l}
		
		\mathcal{O}_{G_{\rm{L}} \bar{N} N  q \bar{q}     }(0,0)
		
		& i \left(\lambda ^A\right)_b^a G_{\rm{L}}^{A}{}_{\mu }{}_{\nu } \left(\overline{q}_{t}^{b i} N_{s}\right) \left(\overline{N}_{p} \sigma ^{\mu }{}^{\nu } q_{ra i}\right)
		
	\end{array}\\
	
	&\begin{array}{c|l}
		
		\mathcal{O}_{G_{\rm{L}} d  \bar{e} N  \bar{u}    }(0,0)
		
		& i \left(\lambda ^A\right)_a^b G_{\rm{L}}^{A}{}_{\mu }{}_{\nu } \left(d_{sb} C N_{t}\right) \left(\overline{e}_{p} \sigma ^{\mu }{}^{\nu } C \overline{u}_{r}{}^{a}\right)
		
	\end{array}\\
	
	&\begin{array}{c|l}
		
		\mathcal{O}_{G_{\rm{L}} \bar{N} N  \bar{u} u     }(0,0)
		
		& i \left(\lambda ^A\right)_a^b G_{\rm{L}}^{A}{}_{\mu }{}_{\nu } \left(N_{s} C u_{tb}\right) \left(\overline{N}_{p} \sigma ^{\mu }{}^{\nu } C \overline{u}_{r}{}^{a}\right)
		
	\end{array}\\
	
	&\begin{array}{c|l}
		
		\multirow{2}*{$\mathcal{O}_{G_{\rm{L}} \bar{d} \bar{N} \bar{q} ^2    }^{(1,2)}(-1,-1)$}
		
		&\mathcal{Y}\left[\tiny{\young(s,t)}\right]i \epsilon _{ace} \epsilon _{ij} \left(\lambda ^A\right)_b^e G_{\rm{L}}^{A}{}_{\mu }{}_{\nu } \left(\overline{q}_{s}^{b i} C \overline{q}_{t}^{c j}\right) \left(\overline{d}_{p}{}^{a} \sigma ^{\mu }{}^{\nu } C \overline{N}_{r}\right)
		
		\\&\mathcal{Y}\left[\tiny{\young(st)}\right]i \epsilon _{ace} \epsilon _{ij} \left(\lambda ^A\right)_b^e G_{\rm{L}}^{A}{}_{\mu }{}_{\nu } \left(\overline{q}_{s}^{b i} C \overline{q}_{t}^{c j}\right) \left(\overline{d}_{p}{}^{a} \sigma ^{\mu }{}^{\nu } C \overline{N}_{r}\right)
		
	\end{array}\\
	
	&\begin{array}{c|l}
		
		\mathcal{O}_{G_{\rm{L}} \bar{d} e  \bar{N} u     }(0,0)
		
		& i \left(\lambda ^A\right)_a^b G_{\rm{L}}^{A}{}_{\mu }{}_{\nu } \left(e_{s} C u_{tb}\right) \left(\overline{d}_{p}{}^{a} \sigma ^{\mu }{}^{\nu } C \overline{N}_{r}\right)
		
	\end{array}\\
	
	&\begin{array}{c|l}
		
		\mathcal{O}_{G_{\rm{L}} \bar{d} d  \bar{N} N     }(0,0)
		
		& i \left(\lambda ^A\right)_a^b G_{\rm{L}}^{A}{}_{\mu }{}_{\nu } \left(d_{sb} C N_{t}\right) \left(\overline{d}_{p}{}^{a} \sigma ^{\mu }{}^{\nu } C \overline{N}_{r}\right)
		
	\end{array}\\
	
	&\begin{array}{c|l}
		
		\mathcal{O}_{G_{\rm{L}} l \bar{N} \bar{q}  u     }(0,0)
		
		& i \left(\lambda ^A\right)_a^b G_{\rm{L}}^{A}{}_{\mu }{}_{\nu } \left(\overline{q}_{s}^{a i} u_{tb}\right) \left(\overline{N}_{r} \sigma ^{\mu }{}^{\nu } l_{pi}\right)
		
	\end{array}\\
	
	&\begin{array}{c|l}
		
		\mathcal{O}_{W_{\rm{L}} d  N  q^2    }(1,1)
		
		&\mathcal{Y}\left[\tiny{\young(pr)}\right]i \left(\tau ^I\right)_k^i \epsilon ^{abc} \epsilon ^{jk} W_{\rm{L}}^{I}{}_{\mu }{}_{\nu } \left(d_{sc} C N_{t}\right) \left(q_{pa i} C \sigma ^{\mu }{}^{\nu } q_{rb j}\right)
		
	\end{array}\\
	
	&\begin{array}{c|l}
		
		\mathcal{O}_{W_{\rm{L}} \bar{l}  N  q \bar{u}    }(0,0)
		
		& i \left(\tau ^I\right)_j^i W_{\rm{L}}^{I}{}_{\mu }{}_{\nu } \left(\overline{l}_{s}{}^{j} N_{t}\right) \left(\overline{u}_{r}{}^{a} \sigma ^{\mu }{}^{\nu } q_{pa i}\right)
		
	\end{array}\\
	
	&\begin{array}{c|l}
		
		\mathcal{O}_{W_{\rm{L}} \bar{N} N  q \bar{q}     }(0,0)
		
		& i \left(\tau ^I\right)_j^i W_{\rm{L}}^{I}{}_{\mu }{}_{\nu } \left(\overline{q}_{t}^{a j} N_{s}\right) \left(\overline{N}_{p} \sigma ^{\mu }{}^{\nu } q_{ra i}\right)
		
	\end{array}\\
	
	&\begin{array}{c|l}
		
		\mathcal{O}_{W_{\rm{L}} \bar{d} \bar{N} \bar{q} ^2    }(-1,-1)
		
		&\mathcal{Y}\left[\tiny{\young(s,t)}\right]i \epsilon _{abc} \epsilon _{jk} \left(\tau ^I\right)_i^k W_{\rm{L}}^{I}{}_{\mu }{}_{\nu } \left(\overline{q}_{s}^{b i} C \overline{q}_{t}^{c j}\right) \left(\overline{d}_{p}{}^{a} \sigma ^{\mu }{}^{\nu } C \overline{N}_{r}\right)
		
	\end{array}\\
	
	&\begin{array}{c|l}
		
		\mathcal{O}_{W_{\rm{L}} l \bar{N} \bar{q}  u     }(0,0)
		
		& i \left(\tau ^I\right)_j^i W_{\rm{L}}^{I}{}_{\mu }{}_{\nu } \left(\overline{q}_{s}^{a j} u_{ta}\right) \left(\overline{N}_{r} \sigma ^{\mu }{}^{\nu } l_{pi}\right)
		
	\end{array}\\
	
	&\begin{array}{c|l}
		
		\mathcal{O}_{W_{\rm{L}} l \bar{l}  \bar{N} N     }(0,0)
		
		& i \left(\tau ^I\right)_j^i W_{\rm{L}}^{I}{}_{\mu }{}_{\nu } \left(\overline{l}_{s}{}^{j} N_{t}\right) \left(\overline{N}_{r} \sigma ^{\mu }{}^{\nu } l_{pi}\right)
		
	\end{array}\\
	
	&\begin{array}{c|l}
		
		\mathcal{O}_{B_{\rm{L}} d  N  q^2    }(1,1)
		
		&\mathcal{Y}\left[\tiny{\young(p,r)}\right]i \epsilon ^{abc} \epsilon ^{ij} B_{\rm{L}}{}_{\mu }{}_{\nu } \left(d_{sc} C N_{t}\right) \left(q_{pa i} C \sigma ^{\mu }{}^{\nu } q_{rb j}\right)
		
	\end{array}\\
	
	&\begin{array}{c|l}
		
		\mathcal{O}_{B_{\rm{L}} \bar{l}  N  q \bar{u}    }(0,0)
		
		& i B_{\rm{L}}{}_{\mu }{}_{\nu } \left(\overline{l}_{s}{}^{i} N_{t}\right) \left(\overline{u}_{r}{}^{a} \sigma ^{\mu }{}^{\nu } q_{pa i}\right)
		
	\end{array}\\
	
	&\begin{array}{c|l}
		
		\mathcal{O}_{B_{\rm{L}} \bar{N} N  q \bar{q}     }(0,0)
		
		& i B_{\rm{L}}{}_{\mu }{}_{\nu } \left(\overline{q}_{t}^{a i} N_{s}\right) \left(\overline{N}_{p} \sigma ^{\mu }{}^{\nu } q_{ra i}\right)
		
	\end{array}\\
	
	&\begin{array}{c|l}
		
		\mathcal{O}_{B_{\rm{L}} d  \bar{e} N  \bar{u}    }(0,0)
		
		& i B_{\rm{L}}{}_{\mu }{}_{\nu } \left(d_{sa} C N_{t}\right) \left(\overline{e}_{p} \sigma ^{\mu }{}^{\nu } C \overline{u}_{r}{}^{a}\right)
		
	\end{array}\\
	
	&\begin{array}{c|l}
		
		\mathcal{O}_{B_{\rm{L}} \bar{N} N  \bar{u} u     }(0,0)
		
		& i B_{\rm{L}}{}_{\mu }{}_{\nu } \left(N_{s} C u_{ta}\right) \left(\overline{N}_{p} \sigma ^{\mu }{}^{\nu } C \overline{u}_{r}{}^{a}\right)
		
	\end{array}\\
	
	&\begin{array}{c|l}
		
		\mathcal{O}_{B_{\rm{L}} \bar{d} \bar{N} \bar{q} ^2    }(-1,-1)
		
		&\mathcal{Y}\left[\tiny{\young(st)}\right]i \epsilon _{abc} \epsilon _{ij} B_{\rm{L}}{}_{\mu }{}_{\nu } \left(\overline{q}_{s}^{b i} C \overline{q}_{t}^{c j}\right) \left(\overline{d}_{p}{}^{a} \sigma ^{\mu }{}^{\nu } C \overline{N}_{r}\right)
		
	\end{array}\\
	
	&\begin{array}{c|l}
		
		\mathcal{O}_{B_{\rm{L}} \bar{d} e  \bar{N} u     }(0,0)
		
		& i B_{\rm{L}}{}_{\mu }{}_{\nu } \left(e_{s} C u_{ta}\right) \left(\overline{d}_{p}{}^{a} \sigma ^{\mu }{}^{\nu } C \overline{N}_{r}\right)
		
	\end{array}\\
	
	&\begin{array}{c|l}
		
		\mathcal{O}_{B_{\rm{L}} \bar{d} d  \bar{N} N     }(0,0)
		
		& i B_{\rm{L}}{}_{\mu }{}_{\nu } \left(d_{sa} C N_{t}\right) \left(\overline{d}_{p}{}^{a} \sigma ^{\mu }{}^{\nu } C \overline{N}_{r}\right)
		
	\end{array}\\
	
	&\begin{array}{c|l}
		
		\mathcal{O}_{B_{\rm{L}} l \bar{N} \bar{q}  u     }(0,0)
		
		& i B_{\rm{L}}{}_{\mu }{}_{\nu } \left(\overline{q}_{s}^{a i} u_{ta}\right) \left(\overline{N}_{r} \sigma ^{\mu }{}^{\nu } l_{pi}\right)
		
	\end{array}\\
	
	&\begin{array}{c|l}
		
		\mathcal{O}_{B_{\rm{L}} l \bar{l}  \bar{N} N     }(0,0)
		
		& i B_{\rm{L}}{}_{\mu }{}_{\nu } \left(\overline{l}_{s}{}^{i} N_{t}\right) \left(\overline{N}_{r} \sigma ^{\mu }{}^{\nu } l_{pi}\right)
		
	\end{array}\\
	
	&\begin{array}{c|l}
		
		\mathcal{O}_{B_{\rm{L}} \bar{e} e  \bar{N} N     }(0,0)
		
		& i B_{\rm{L}}{}_{\mu }{}_{\nu } \left(e_{s} C N_{t}\right) \left(\overline{e}_{p} \sigma ^{\mu }{}^{\nu } C \overline{N}_{r}\right)
		
	\end{array}\\
	
	&\begin{array}{c|l}
		
		\mathcal{O}_{B_{\rm{L}} \bar{N}^2 N ^2    }(0,0)
		
		&\mathcal{Y}\left[\tiny{\young(p,r)},\tiny{\young(st)}\right]i B_{\rm{L}}{}_{\mu }{}_{\nu } \left(N_{s} C N_{t}\right) \left(\overline{N}_{p} \sigma ^{\mu }{}^{\nu } C \overline{N}_{r}\right)
		
	\end{array}\label{cl:FLpsiL2psiR2l}
	
\end{align}

\underline{Class $F_{\rm{L}}{}^2 \psi^\dagger{}^2 \phi  $}: 4 types

\begin{align}
	
	&\begin{array}{c|l}\label{cl:FL2psiR2hf}
		
		\mathcal{O}_{G_{\rm{L}}{}^2 \bar{l}  N  H^{\dagger}   }(0,0)
		
		& \epsilon _{ij} H^{\dagger}{}^{i} G_{\rm{L}}^{A}{}_{\mu }{}_{\nu } G_{\rm{L}}^{A}{}^{\mu }{}^{\nu } \left(\overline{l}_{p}{}^{j} N_{r}\right)
		
	\end{array}\\
	
	&\begin{array}{c|l}
		
		\mathcal{O}_{W_{\rm{L}}{}^2 \bar{l}  N  H^{\dagger}   }(0,0)
		
		& \epsilon _{km} \left(\tau ^I\right)_j^m \left(\tau ^J\right)_i^k H^{\dagger}{}^{i} W_{\rm{L}}^{I}{}_{\mu }{}_{\nu } W_{\rm{L}}^{J}{}^{\mu }{}^{\nu } \left(\overline{l}_{p}{}^{j} N_{r}\right)
		
	\end{array}\\
	
	&\begin{array}{c|l}
		
		\mathcal{O}_{B_{\rm{L}} W_{\rm{L}} \bar{l}  N  H^{\dagger}   }(0,0)
		
		& \epsilon _{jk} \left(\tau ^I\right)_i^k B_{\rm{L}}{}_{\mu }{}_{\nu } H^{\dagger}{}^{i} W_{\rm{L}}^{I}{}^{\mu }{}^{\nu } \left(\overline{l}_{p}{}^{j} N_{r}\right)
		
	\end{array}\\
	
	&\begin{array}{c|l}
		
		\mathcal{O}_{B_{\rm{L}}{}^2 \bar{l}  N  H^{\dagger}   }(0,0)
		
		& \epsilon _{ij} B_{\rm{L}}{}_{\mu }{}_{\nu } H^{\dagger}{}^{i} B_{\rm{L}}{}^{\mu }{}^{\nu } \left(\overline{l}_{p}{}^{j} N_{r}\right)
		
	\end{array}\label{cl:FL2psiR2hl}
	
\end{align}

\underline{Class $  \psi{}^3 \psi^\dagger \phi D$}: 16 types

\begin{align}
	
	&\begin{array}{c|l}\label{cl:psiL3psiRhDf}
		
		\multirow{4}*{$\mathcal{O}_{  N  q^3 H^{\dagger}  D}^{(1\sim 4)}(1,1)$}
		
		&\mathcal{Y}\left[\tiny{\young(pr,s)}\right]i \epsilon ^{abc} \epsilon ^{ik} D^{\mu } H^{\dagger}{}^{j} \left(q_{pa i} C q_{rb j}\right) \left(q_{sc k} C \gamma _{\mu } N_{t}\right)
		
		\\&\mathcal{Y}\left[\tiny{\young(pr,s)}\right]i \epsilon ^{abc} \epsilon ^{ik} H^{\dagger}{}^{j} \left(q_{pa i} C \gamma _{\mu } N_{t}\right) \left(q_{rb j} C D^{\mu } q_{sc k}\right)
		
		\\&\mathcal{Y}\left[\tiny{\young(prs)}\right]i \epsilon ^{abc} \epsilon ^{ik} D^{\mu } H^{\dagger}{}^{j} \left(q_{pa i} C q_{rb j}\right) \left(q_{sc k} C \gamma _{\mu } N_{t}\right)
		
		\\&\mathcal{Y}\left[\tiny{\young(p,r,s)}\right]i \epsilon ^{abc} \epsilon ^{ik} D^{\mu } H^{\dagger}{}^{j} \left(q_{pa i} C q_{rb j}\right) \left(q_{sc k} C \gamma _{\mu } N_{t}\right)
		
	\end{array}\\
	
	&\begin{array}{c|l}
		
		\multirow{3}*{$\mathcal{O}_{  \bar{e} N  q \bar{u} H^{\dagger}  D}^{(1\sim 3)}(0,0)$}
		
		& i D^{\mu } H^{\dagger}{}^{i} \left(\overline{e}_{p} q_{ra i}\right) \left(\overline{u}_{s}{}^{a} \gamma _{\mu } N_{t}\right)
		
		\\& i H^{\dagger}{}^{i} \left(\overline{e}_{p} \gamma _{\mu } N_{t}\right) \left(D^{\mu } \overline{u}_{s}{}^{a} q_{ra i}\right)
		
		\\& i D^{\mu } H^{\dagger}{}^{i} \left(\overline{e}_{p} C \overline{u}_{s}{}^{a}\right) \left(q_{ra i} C \gamma _{\mu } N_{t}\right)
		
	\end{array}\\
	
	&\begin{array}{c|l}
		
		\multirow{3}*{$\mathcal{O}_{  \bar{N} N  q \bar{u} H D}^{(1\sim 3)}(0,0)$}
		
		& i \epsilon ^{ij} D^{\mu } H_{j} \left(\overline{N}_{p} q_{ra i}\right) \left(\overline{u}_{s}{}^{a} \gamma _{\mu } N_{t}\right)
		
		\\& i \epsilon ^{ij} H_{j} \left(\overline{N}_{p} \gamma _{\mu } N_{t}\right) \left(D^{\mu } \overline{u}_{s}{}^{a} q_{ra i}\right)
		
		\\& i \epsilon ^{ij} D^{\mu } H_{j} \left(\overline{N}_{p} C \overline{u}_{s}{}^{a}\right) \left(q_{ra i} C \gamma _{\mu } N_{t}\right)
		
	\end{array}\\
	
	&\begin{array}{c|l}
		
		\multirow{3}*{$\mathcal{O}_{  \bar{d} e  \bar{N} q H D}^{(1\sim 3)}(0,0)$}
		
		& i \epsilon ^{ij} D^{\mu } H_{j} \left(\overline{d}_{p}{}^{a} C \overline{N}_{r}\right) \left(q_{sa i} C \gamma _{\mu } e_{t}\right)
		
		\\& i \epsilon ^{ij} H_{j} \left(\overline{d}_{p}{}^{a} \gamma _{\mu } e_{t}\right) \left(\overline{N}_{r} D^{\mu } q_{sa i}\right)
		
		\\& i \epsilon ^{ij} D^{\mu } H_{j} \left(\overline{N}_{r} \gamma _{\mu } e_{t}\right) \left(\overline{d}_{p}{}^{a} q_{sa i}\right)
		
	\end{array}\\
	
	&\begin{array}{c|l}
		
		\multirow{3}*{$\mathcal{O}_{  \bar{d} \bar{N} N  q H^{\dagger}  D}^{(1\sim 3)}(0,0)$}
		
		& i D^{\mu } H^{\dagger}{}^{i} \left(\overline{d}_{p}{}^{a} C \overline{N}_{r}\right) \left(q_{sa i} C \gamma _{\mu } N_{t}\right)
		
		\\& i H^{\dagger}{}^{i} \left(\overline{d}_{p}{}^{a} \gamma _{\mu } N_{t}\right) \left(\overline{N}_{r} D^{\mu } q_{sa i}\right)
		
		\\& i D^{\mu } H^{\dagger}{}^{i} \left(\overline{N}_{r} \gamma _{\mu } N_{t}\right) \left(\overline{d}_{p}{}^{a} q_{sa i}\right)
		
	\end{array}\\
	
	&\begin{array}{c|l}
		
		\multirow{6}*{$\mathcal{O}_{  l \bar{N} q \bar{q}  H D}^{(1\sim 6)}(0,0)$}
		
		& i \epsilon ^{ik} D^{\mu } H_{k} \left(\overline{N}_{r} l_{pi}\right) \left(\overline{q}_{t}^{a j} \gamma _{\mu } q_{sa j}\right)
		
		\\& i \epsilon ^{ij} D^{\mu } H_{k} \left(\overline{N}_{r} l_{pi}\right) \left(\overline{q}_{t}^{a k} \gamma _{\mu } q_{sa j}\right)
		
		\\& i \epsilon ^{ik} H_{k} \left(\overline{N}_{r} D^{\mu } q_{sa j}\right) \left(\overline{q}_{t}^{a j} \gamma _{\mu } l_{pi}\right)
		
		\\& i \epsilon ^{ij} H_{k} \left(\overline{N}_{r} D^{\mu } q_{sa j}\right) \left(\overline{q}_{t}^{a k} \gamma _{\mu } l_{pi}\right)
		
		\\& i \epsilon ^{ik} D^{\mu } H_{k} \left(l_{pi} C q_{sa j}\right) \left(\overline{N}_{r} \gamma _{\mu } C \overline{q}_{t}^{a j}\right)
		
		\\& i \epsilon ^{ij} D^{\mu } H_{k} \left(l_{pi} C q_{sa j}\right) \left(\overline{N}_{r} \gamma _{\mu } C \overline{q}_{t}^{a k}\right)
		
	\end{array}%
	
\end{align}

\subsection{Lists of the Dim-9 Operators}

\subsubsection{Classes involving Two-fermions}

\indent \indent \underline{Class $F_{\rm{L}}{}^2 \psi{}^2   D^2$}: 3 types

\begin{align}
	
	&%
	
\end{align}

\subsubsection{Classes involving Four-fermions}

\underline{Class $  \psi{}^3 \psi^\dagger   D^3$}: 12 types

\begin{align}
	
	&%
\label{cl:psiL2psiR2h3l}
	
\end{align}

\subsubsection{Classes involving Six-fermions}

\underline{Class $  \psi{}^4 \psi^\dagger{}^2    $}: 80 types

\begin{align}
	
	&%
\label{cl:psiL6l}
	
\end{align}

\section{Lists of Operators in $\nu$LEFT}\label{sec:listnuL}

In this section, we list the complete and independent operator basis in $\nu$LEFT from dimension 5 to dimension 9, and the statistic results of the operator basis are listed in table \ref{tab:sumdim5678L} and \ref{tab:sumdim9L}. We also give the relations of four-component Dirac spinors and two-component Weyl spinors here. 

\begin{align}
\nu_{\rm L}=,\quad
e_{\rm L}=\left(\begin{array}{c}e\\0\end{array}\right),\quad 
e_{\rm R}=\left(\begin{array}{c}0\\e_{_\mathbb{C}}^{\dagger}\end{array}%
\right), \quad \bar{N}_{\rm R}=\left(N_{_\mathbb{C}}\,,\,0\right)
\end{align}

In the following contents that list the operator basis, we should clarify that each $\psi$ ($\bar{\psi}$) in an operator class means a two-component left-handed (right-handed) spinor in this class. The baryon number and lepton number violation pattern of each operator type are presented next to the type as $(\Delta B, \Delta L)$. The subscripts and superscripts $\{p,r,s,t,u,v\}$, $\{a,b,c,d,e,f\}$ and $\{A,B,C,D\}$ denote flavor indices, $SU(3)_C$ group (anti)fundamental representation indices and $SU(3)_C$ group adjoint representation indices respectively. The number of up-type quark fields $u_{L,R}$ flavor in this effective field theory is 2 (excluding the heavy top quark which has been integrated out), and all other fermions flavor numbers are 3.

\begin{table}
	\begin{align*}
		\begin{array}{cc|c|c|c|c|c}
			\hline
			\multicolumn{7}{c}{\text{Dim-5 operators}}\\
			\hline
			N & (n,\tilde{n}) & \text{Classes} & \mathcal{N}_{\text{type}} & \mathcal{N}_{\text{term}} & \mathcal{N}_{\text{operator}} & \text{Equations}\\
			\hline
			3 & (2,0) & F_{\rm{L}} \psi^2    +h.c. & 2+0+2+0 & 4 & 24 & \multirow{1}*{(\ref{cl:FLpsiL2fL}-\ref{cl:FLpsiL2lL})} \\
			\hline\hline
			\multicolumn{7}{c}{\text{Dim-6 operators}}\\
			\hline
			N & (n,\tilde{n}) & \text{Classes} & \mathcal{N}_{\text{type}} & \mathcal{N}_{\text{term}} & \mathcal{N}_{\text{operator}} & \text{Equations}\\
			\hline
			4 & (2,0) & \psi^4    +h.c. & 10+4+12+2 & 50 & 2034 & \multirow{1}*{(\ref{cl:psiL4fL}-\ref{cl:psiL4lL})} \\
			\cline{2-7}
			& (1,1) & \psi^2 \psi^\dagger{}^2 & 20+8+28+2 & 58 & 3189 & \multirow{1}*{(\ref{cl:psiL2psiR2fL}-\ref{cl:psiL2psiR2lL})} \\
			\hline
			\multicolumn{2}{c|}{\text{Total}} & 5 & 30+12+40+4 & 108 & 5223 & \\
			\hline\hline
			\multicolumn{7}{c}{\text{Dim-7 operators}}\\
			\hline
			N & (n,\tilde{n}) & \text{Classes} & \mathcal{N}_{\text{type}} & \mathcal{N}_{\text{term}} & \mathcal{N}_{\text{operator}} & \text{Equations}\\
			\hline
			4 & (3,0) & F_{\rm{L}}^2 \psi^2    +h.c. & 4+0+4+0 & 8 & 60 & \multirow{1}*{(\ref{cl:FL2psiL2fL}-\ref{cl:FL2psiL2lL})} \\
			\cline{2-7}
			& (2,1) & F_{\rm{L}}^2 \psi^\dagger{}^2    +h.c. & 4+0+4+0 & 8 & 60 & \multirow{1}*{(\ref{cl:FL2psiR2fL}-\ref{cl:FL2psiR2lL})} \\
			& & \psi^3 \psi^\dagger{} D   +h.c. & 30+12+38+4 & 84 & 3846 & \multirow{1}*{(\ref{cl:psiL3psiRDfL}-\ref{cl:psiL3psiRDlL})} \\
			\hline
			\multicolumn{2}{c|}{\text{Total}} & 6 & 38+12+46+4 & 100 & 3966 & \\
			\hline\hline
			\multicolumn{7}{c}{\text{Dim-8 operators}}\\
			\hline
			N & (n,\tilde{n}) & \text{Classes} & \mathcal{N}_{\text{type}} & \mathcal{N}_{\text{term}} & \mathcal{N}_{\text{operator}} & \text{Equations}\\
			\hline
			4 & (3,1) & \psi^4 D^2   +h.c. & 10+4+12+2 & 78 & 3204 & \multirow{1}*{(\ref{cl:psiL4D2fL}-\ref{cl:psiL4D2lL})} \\
			\cline{2-7}
			& (2,2) & F_{\rm{L}} F_{\rm{R}} \psi \psi^\dagger{} D & 2+0+4+0 & 6 & 54 & \multirow{1}*{(\ref{cl:FLFRpsiLpsiRDfL}-\ref{cl:FLFRpsiLpsiRDlL})} \\
			& & \psi^2 \psi^\dagger{}^2   D^2 & 20+8+28+2 & 116 & 6129 & \multirow{1}*{(\ref{cl:psiL2psiR2D2fL}-\ref{cl:psiL2psiR2D2lL})} \\
			\hline
			5 & (3,0) & F_{\rm{L}} \psi^4   +h.c. & 16+8+18+2 & 136 & 5238 & \multirow{1}*{(\ref{cl:FLpsiL4fL}-\ref{cl:FLpsiL4lL})} \\
			\cline{2-7}
			& (2,1) & F_{\rm{L}} \psi^2 \psi^\dagger{}{}^2   +h.c. & 68+32+92+4 & 212 & 10800 & \multirow{1}*{(\ref{cl:FLpsiL2psiR2fL}-\ref{cl:FLpsiL2psiR2lL})} \\
			\hline
			\multicolumn{2}{c|}{\text{Total}} & 8 & 116+52+154+10 & 548 & 25425 & \\
			\hline
		\end{array}
	\end{align*}
	\caption{The complete statistics of dimension 5, 6, 7, 8 $\nu$LEFT operators.
		$N$ in the leftmost column shows the number of particles. $(n,\tilde{n})$ are the numbers of $\epsilon$
		and $\tilde{\epsilon}$ in the Lorentz structure. $\mathcal{N}_{\rm type}$, $\mathcal{N}_{\rm term}$, and $\mathcal{N}_{\rm operator}$
		show the number of types, terms and Hermitian operators respectively (independent conjugates
		are counted), while the numbers under $\mathcal{N}_{\rm type}$ describe the sum of each possible $|\Delta L|$
		types/operators with $\mathcal{N}=\mathcal{N}(|\Delta L|=0)+\mathcal{N}(|\Delta L|=1)+\mathcal{N}(|\Delta L|=2)+\mathcal{N}(|\Delta L|=4)$. The links in the rightmost column refer to the list(s) of the terms in given
		classes.
	}
	\label{tab:sumdim5678L}
\end{table}

\begin{table}
	\begin{align*}
		\begin{array}{cc|c|c|c|c|c}
			\hline
			\multicolumn{7}{c}{\text{Dim-9 operators}}\\
			\hline
			N & (n,\tilde{n}) & \text{Classes} & \mathcal{N}_{\text{type}} & \mathcal{N}_{\text{term}} & \mathcal{N}_{\text{operator}} & \text{Equations}\\
			\hline
			4 & (3,2) & F_{\rm{L}}{}^2 \psi^2   D^2    +h.c. & 4+0+4+0+0+0 & 16 & 120 & \multirow{1}*{(\ref{cl:FL2psiL2D2fL}-\ref{cl:FL2psiL2D2lL})} \\
			& & F_{\rm{L}} F_{\rm{R}} \psi^2   D^2    +h.c. & 4+0+4+0+0+0 & 8 & 60 & \multirow{1}*{(\ref{cl:FLFRpsiL2D2fL}-\ref{cl:FLFRpsiL2D2lL})} \\
			& & F_{\rm{L}}{}^2 \psi^\dagger{}{}^2   D^2    +h.c. & 4+0+4+0+0+0 & 8 & 60 & \multirow{1}*{(\ref{cl:FL2psiR2D2fL}-\ref{cl:FL2psiR2D2lL})} \\
			& & \psi^3 \psi^\dagger{}   D^3    +h.c. & 30+12+38+0+4+0 & 162 & 8136 & \multirow{1}*{(\ref{cl:psiL3psiRD3fL}-\ref{cl:psiL3psiRD3lL})} \\
			\hline
			5 & (4,0) & F_{\rm{L}}{}^3 \psi^2    +h.c. & 6+0+6+0+0+0 & 20 & 126 & \multirow{1}*{(\ref{cl:FL3psiL2fL}-\ref{cl:FL3psiL2lL})} \\
			\cline{2-7}
			& (3,1) & F_{\rm{L}}{}^3 \psi^\dagger{}^2    +h.c. & 2+0+2+0+0+0 & 4 & 30 & \multirow{1}*{(\ref{cl:FL3psiR2fL}-\ref{cl:FL3psiR2lL})} \\
			& & F_{\rm{L}}{} \psi^3 \psi^\dagger D   +h.c. & 50+24+62+0+4+0 & 602 & 28854 & \multirow{1}*{(\ref{cl:FLpsiL3psiRDfL}-\ref{cl:FLpsiL3psiRDlL})} \\
			\cline{2-7}
			& (2,2) & F_{\rm{L}} F_{\rm{R}}{}^2 \psi^2    +h.c. & 8+0+8+0+0+0 & 16 & 96 & \multirow{1}*{(\ref{cl:FLFR2psiL2fL}-\ref{cl:FLFR2psiL2lL})} \\
			& & F_{\rm{R}} \psi^3 \psi^\dagger   D    +h.c. & 50+24+62+0+4+0 & 450 & 21300 & \multirow{1}*{(\ref{cl:FRpsiL3psiRDfL}-\ref{cl:FRpsiL3psiRDlL})} \\
			\hline
			6 & (3,0) & \psi^6    +h.c. & 30+34+40+10+12+2 & 676 & 100782 & \multirow{1}*{(\ref{cl:psiL6fL}-\ref{cl:psiL6lL})} \\
			\cline{2-7}
			& (2,1) & \psi^4 \psi^\dagger{}^2    +h.c. & 252+254+344+78+72+4 & 2492 & 629862 & \multirow{1}*{(\ref{cl:psiL4psiR2fL}-\ref{cl:psiL4psiR2lL})} \\
			\hline
			\multicolumn{2}{c|}{\text{Total}} & 22 & 440+348+574+88+96+6 & 4454 & 789426 & \\
			\hline
		\end{array}
	\end{align*}
	\caption{The complete statistics of dimension 9 $\nu$LEFT operators. The numbers under $\mathcal{N}_{\rm type}$ describe the sum of each possible $|\Delta L|$ types/operators with $\mathcal{N}=\mathcal{N}(|\Delta L|=0)+\mathcal{N}(|\Delta L|=1)+\mathcal{N}(|\Delta L|=2)+\mathcal{N}(|\Delta L|=3)+\mathcal{N}(|\Delta L|=4)+\mathcal{N}(|\Delta L|=6)$.
	}
	\label{tab:sumdim9L}
\end{table}

\subsection{Lists of the Dim-5 Operators}

\underline{Class $F_{\rm{L}} \psi{}^2    $}: 2 types

\begin{align}
	
	&\begin{array}{c|l}\label{cl:FLpsiL2fL}
		
		\mathcal{O}_{F_{\rm{L}} \nu _{\rm L} \bar{N}_{\rm R}    }(0,0)
		
		& i F_{\rm{L}}{}^{\mu }{}^{\nu } \left(\overline{N}_{\rm R}{}_{p} \sigma _{\mu }{}_{\nu } \nu _{\rm L}{}_{r}\right)
		
	\end{array}\\
	
	&\begin{array}{c|l}\label{cl:FLpsiL2lL}
		
		\mathcal{O}_{F_{\rm{L}} \bar{N}_{\rm R}^2    }(0,-2)
		
		&\mathcal{Y}\left[\tiny{\young(p,r)}\right]i F_{\rm{L}}{}^{\mu }{}^{\nu } \left(\overline{N}_{\rm R}{}_{p} \sigma _{\mu }{}_{\nu } C \overline{N}_{\rm R}{}_{r}\right)
		
	\end{array}
	
\end{align}

These operators are listed and discussed in Ref.~\cite{Aparici:2009fh,Chala:2020vqp,Li:2020lba}. Here the $(0,0)$ and $(0,-2)$ denote the baryon and lepton numbers of the operators and each of the $\psi$s or $\psi^{\dagger}$s in the class name means a left-handed two-component spinor or a right-handed two-component spinor in the class respectively. The definition of the Dirac charge conjugation matrix $C$ and its conversions can be found near eq.~(\ref{eq:psiC}), along with the conversion between $F$, $\tilde{F}$, $F_L$ and $F_R$ eq.~(\ref{eq:FLRtilde1},\ref{eq:FLRtilde2}) , the conversion between two-component spinors and four-component spinors eq.~(\ref{eq:bilinear}) and Hermitian conjugation of operators eq.~(\ref{eq:hermit1},\ref{eq:hermit2}).

\subsection{Lists of the Dim-6 Operators}

\underline{Class $  \psi{}^4    $}: 14 types

\begin{align}
	
	&

	
\end{align}

\subsection{Lists of the Dim-7 Operators}

\underline{Class $F_{\rm{L}}{}^2 \psi{}^2    $}: 4 types

\begin{align}
	
	&%
	
\end{align}

\subsection{Lists of the Dim-8 Operators}

\underline{Class $  \psi{}^4   D^2$}: 14 types

\begin{align}
	
	&%
	
\end{align}

\subsection{Lists of the Dim-9 Operators}

\subsubsection{Classes involving Two-fermions}

\underline{Class $F_{\rm{L}}{}^2 \psi{}^2   D^2$}: 4 types

\begin{align}
	
	&%
	
\end{align}

\subsubsection{Classes involving Four-fermions}

\underline{Class $  \psi{}^3 \psi^{\dagger}   D^3$}: 42 types

\begin{align}
	
	&%
	
\end{align}

\subsubsection{Classes involving Six-fermions: $\psi^6$}

\underline{Class $  \psi{}^6    $}: 64 types

\begin{align}
	
	&%
	
\end{align}

\subsubsection{Classes involving Six-fermions: $\psi^4 \psi^{\dagger2}$}

\input{nuLEFT_operators_psi4psidagger2}

\section{Conclusion}\label{sec:con}

In this paper, we investigated the operator basis in the $\nu$SMEFT and $\nu$LEFT frameworks, in which the SMEFT is extended with light sterile right-handed neutrinos, its mass ranging from KeV to tens of GeV.  We briefly review the Young tensor method introduced in a series of our work~\cite{Li:2020gnx,Li:2020xlh,Li:2020tsi}: starting from
the spinor-helicity formalism  and the amplitude-operator correspondence, and then exhibiting the Lorentz and gauge structure construction of operators using Young tableaux. We utilize this method to obtain the complete and independent operator basis in $\nu$SMEFT and $\nu$LEFT up to dimension 9, and illustrate the whole procedure with  a specific example of a dimension 9 operator type $L N_{_\mathbb{C}}^3 H D^2$. Following the same procedure, we listed all the explicit  operators in section~\ref{sec:listnuSM} and section~\ref{sec:listnuL}. We also discussed the differences between the Dirac neutrinos and the Majorana neutrinos, which directly relates to the lepton number $U(1)_L$. If the lepton number $U(1)_L$ violating operators are turned on in the $\nu$SMEFT and $\nu$LEFT, the neutrinos should be recognized as Majorana neutrinos, otherwise the neutrinos are Dirac.

At the mass dimension 5, 6, 7, 8, 9 in the $\nu$SMEFT, we found that there are 2, 29, 80, 323, 1358 independent operators involving the right-handed neutrino for one generation of fermions, and 18, 1614, 4206, 20400, 243944 flavor-specified independent operators involving the right-handed neutrinos for three generation of fermions, as shown in table \ref{tab:sumdim567}, \ref{tab:sumdim8} and \ref{tab:sumdim9}.
In the $\nu$LEFT with the top quark integrated out, at the mass dimension 5, 6, 7, 8, 9, we found that there are 24, 5223, 3966, 25425, 789426 independent operators involving the right-handed neutrinos as shown in table \ref{tab:sumdim5678L} and \ref{tab:sumdim9L}.

The operator bases in the $\nu$SMEFT and $\nu$LEFT can be applied to various physical processes and give rise to new contributions in addition to those of SMEFT and LEFT. The $\nu$SMEFT and $\nu$LEFT have been applied to study various processes involving light sterile neutrinos at the future colliders and various precision experiments. With the complete operator bases written down, more physical processes are expected to be explored within the framework of $\nu$SMEFT and $\nu$LEFT, such as various lepton and baryon number violating processes with sterile neutrinos. Furthermore, the renormalization group (RG) running equations in the $\nu$SMEFT and $\nu$LEFT~\cite{Chala:2020pbn,Datta:2020ocb,Datta:2021akg}, and the matching between the $\nu$SMEFT and the $\nu$LEFT~\cite{Chala:2020vqp,Li:2020lba} has been investigated up to the dimension 6, although the full RG running equations are still under construction. After all the above theoretical setup equipped, it would be ready to explore light sterile neutrino related processes involving several scales in the EFT framework.    


\section*{Acknowledgments} 
HLL, ZR and JHY were supported by the National Natural Science Foundation of China (NSFC) under Grants No. 12022514 and No. 11875003. MLX was supported by the NSFC under grant No.2019M650856 and the 2019 International Postdoctoral Exchange Fellowship Program. JHY was also supported by the NSFC under Grants No. 12047503 and National Key Research and Development Program of China under Grant No. 2020YFC2201501.

\bibliographystyle{JHEP}
\bibliography{nuSMEFTref}

\end{document}